\PassOptionsToPackage{table}{xcolor}
\documentclass{SciPost}

\hypersetup{
    colorlinks,
    linkcolor={red!50!black},
    citecolor={blue!50!black},
    urlcolor={blue!80!black}
}

\usepackage[table]{xcolor}

\usepackage{enumitem}
\usepackage{tcolorbox}
\usepackage{graphicx}
\usepackage{subcaption}

\usepackage[bitstream-charter]{mathdesign}
\DeclareSymbolFont{usualmathcal}{OMS}{cmsy}{m}{n}
\DeclareSymbolFontAlphabet{\mathcal}{usualmathcal}

\fancypagestyle{SPstyle}{
\fancyhf{}
\lhead{\colorbox{scipostblue}{\bf \color{white} ~SciPost Physics Lecture Notes }}
\rhead{{\bf \color{scipostdeepblue} ~Submission }}

\fancyfoot[C]{\textbf{\thepage}}
}

\usepackage{xcolor}
\definecolor{SciPostPhysicsLectureNotes}{HTML}{000000} 

\newcommand{\beq}{\begin{equation}}
\newcommand{\eeq}{\end{equation}}
\newcommand{\exclude}[1]{}

\usepackage{xcolor}

\begin{document}

\pagestyle{SPstyle}

\begin{center}
{\Large\textbf{\textcolor{SciPostPhysicsLectureNotes}{%
QCD-driven dark matter: AQNs formation and observational tests\\
}}}
\end{center}

\begin{center}\textbf{
Ludovic Van Waerbeke$\star$
}\end{center}

\begin{center}
Department of physics and Astronomy, The University of British Columbia, 6224 Agricultural Road, V6T 1Z1, Vancouver, Canada

$\star$ \href{mailto:email1}{\small waerbeke@phas.ubc.ca}\,

\bigskip
\today
\end{center}

\section*{\color{scipostdeepblue}{Abstract}}
\textbf{\boldmath{%
The nature of dark matter remains a central problem in cosmology. A compelling possibility is that dark matter is macroscopic, consisting of composite objects formed in the early Universe. We introduce the QCD-AQN framework, a well-motivated scenario in which dark matter is composed of dense aggregates of quark  and antiquark matter stabilised by axion domain walls. The framework proposes a unified explanation for both dark matter and the observed matter–antimatter asymmetry. Particular emphasis is placed on existing observational constraints and on observational tests. Finally, we  explore a possible QCD-based scenario for dark energy.
}}

\vspace{\baselineskip}


\vspace{1cm}

\begin{figure}[htbp]
\centering
\includegraphics[width=0.9\textwidth]{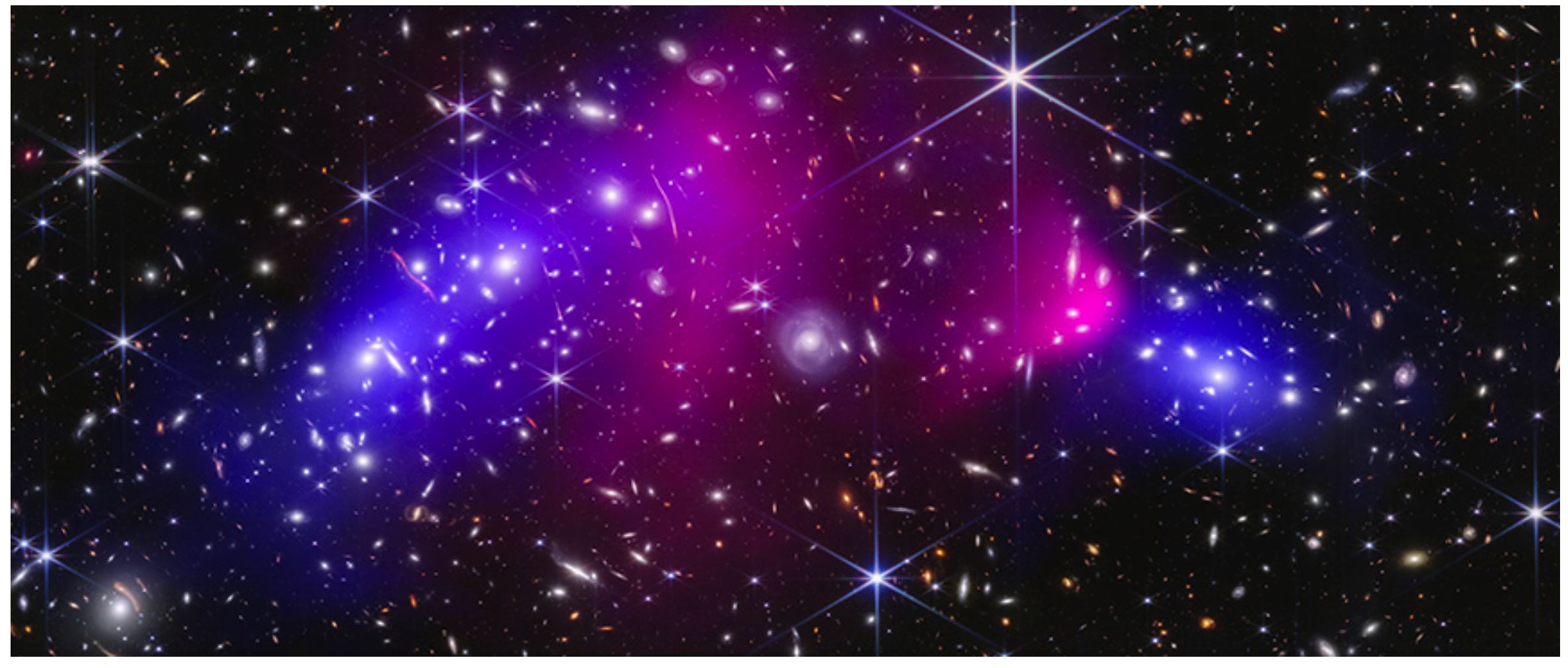}
\caption{Recent data from NASA’s James Webb Space Telescope (JWST), combined with observations from the Chandra X-ray Observatory, have produced a new high-resolution image of the Bullet Cluster \cite{2006ApJ...648L.109C}. This system is historically important because it provided the first direct empirical evidence for dark matter, based on earlier analyses combining Chandra, the Hubble Space Telescope, and ground-based observations in 2006. The X-ray emission detected by Chandra traces the hot gas in the two merging clusters (displayed in pink), whereas the mass distribution, shown in blue, is inferred from gravitational lensing measurements made possible by the detailed imaging from JWST, and reveals the presence of a dominant dark matter component \cite{2025ApJ...987L..15C}.  Adapted from \url{http://chandra.harvard.edu/photo/2025/bullet/} }
\label{fig:bullet}
\end{figure}

\newpage


\vspace{10pt}
\noindent\rule{\textwidth}{1pt}
\tableofcontents
\noindent\rule{\textwidth}{1pt}
\vspace{10pt}

\section{Introduction}
\label{sec:intro}

Since the pioneering work of Fritz Zwicky in the 1930s \cite{1933AcHPh...6..110Z}, the quest to understand the nature of dark matter (DM) has stood at the centre of modern cosmology. Subsequent observations -- from the rotation curves measured by Vera Rubin \cite{1970ApJ...159..379R,1980ApJ...238..471R} to the precision cosmology of Planck -- have transformed DM into a cornerstone of the standard cosmological model. Yet despite overwhelming evidence for its existence, the nature of DM remains one of the most profound open questions at the intersection of cosmology and particle physics \cite{2024arXiv240601705C,2025PDU....4901965D}.
One of the striking facts is that its abundance is so close to that of ordinary matter:
\[
\Omega_{\rm DM} \approx 5\,\Omega_b.        
\] 
Within the standard paradigm, there is no structural reason for the two densities to be comparable: The visible baryon\footnote{When dealing with condensed matter, baryon (B) will be understood stricto sensu as a composite hadron made of three quarks bound together by the strong interaction ($B=+1$, e.g. protons and neutrons). In cosmology, baryons (b) are defined in contrast to dark matter and radiation, and refer to all ordinary matter composed of hadrons. This includes protons and neutrons, the light nuclei produced during Big Bang nucleosynthesis, as well as gas, dust, stars, etc.} population is generated through a high-scale baryogenesis mechanism, while dark matter is produced independently.

The QCD--AQN framework, developed by Ariel R.~Zhitnitsky \cite{2003JCAP...10..010Z}, proposes a different interpretation rooted in quantum chromodynamics (QCD) and matter--antimatter symmetry. The foundational idea emerged in part from earlier work by \cite{PhysRevD.30.272,1986PhRvD..34.2947M,2004PhRvD..69f3509L}: aggregates of condensed quark matter carrying a large baryon number in the colour superconducting phase may become stable against decay into ordinary hadronic matter at temperatures $T \lesssim 60\,\mathrm{MeV}$.

In this framework, the Universe may remain globally symmetric,
\[
    B_{\text{universe}} = B_{\text{matter}} + {\bar {B}_{\text{antimatter}}} = 0 ,
\]
where $B$ is the baryon number.

During the QCD transition at $T\sim 170\,{\rm MeV}$, a baryon number separation mechanism gives rise to the formation of quark and antiquark aggregates, known as Axion Quark Nuggets (AQNs), with a dynamical imbalance favouring the antimatter AQNs. The visible sector consists of the leftover unbound baryons.  Because both AQN dark matter and the visible baryons emerge from the same underlying QCD dynamics, their present-day mass densities are correlated. Being remarkably stable,  AQNs survive to the present epoch and, owing to their macroscopic masses and consequent rarity, have so far evaded direct detection.

The conceptual shift is significant:
\begin{itemize}[noitemsep, topsep=0pt]
    \item DM abundance is a consequence of QCD-scale physics;

    \item   The photon-to-baryon ratio is controlled by the QCD binding scale of the Color Superconducting phase;
    \item Matter--antimatter symmetry is preserved globally;
    \item    One additional fundamental particle is involved, the QCD axion.

\end{itemize}
A major strength of the framework lies in its multi-wavelength testability and its clear falsifiability. Subsequent studies by several authors have examined how AQN–baryon interactions may produce observable signatures across a wide range of astrophysical environments, through electromagnetic emission and/or localised energy deposition.

This lecture begins with an overview of the current observational constraints on dark matter, with a strong emphasis on distinguishing firmly established results from model-dependent interpretations (Section \ref{sec:model_independant}), and on highlighting regions of the DM parameter space to which additional theoretical and observational effort could profitably be devoted. Before discussing the condensed-matter aspects of the QCD-AQN framework, we first review the observational evidence that motivates the study of AQNs (Section \ref{sec:AQN_and_Obs}). We then present a concise overview of the theoretical foundations of the QCD-AQN model (Section~\ref{sec: formation}). This is followed by a very brief review of axion physics and a discussion of the possibility that DM consists of AQNs together with a subdominant component of QCD axions (Section~\ref{section:axion}). Finally, to complement this excursion into applications of QCD in cosmology, we outline a QCD-based approach to the dark-energy problem (Section~\ref{sec:QCDDE}).

\section{A critical perspective on model-independent constraints}
\label{sec:model_independant}

Cosmology is currently at a turning point, with the community actively diversifying both theoretical and experimental strategies to address the dark matter problem \cite{2018Natur.562...51B}.
This section reviews the set of model-independent observational constraints that any viable DM candidate must satisfy. It is important to distinguish between \textit{direct }empirical constraints and conclusions that rely on additional theoretical assumptions about the nature of DM. As we shall see, some commonly stated “properties” of dark matter are not direct observations, but interpretations that depend on specific modelling choices.

\subsection{Where is dark matter ?}
Observations of galaxy rotation curves, galaxy velocity dispersions, hot intracluster gas, gravitational lensing, and anisotropies in the Cosmic Microwave Background (CMB) leave little doubt that dark matter exists and constitutes a substantial fraction of the Universe’s total energy density.

\begin{figure}[ht]
\centering
\includegraphics[width=14 cm]{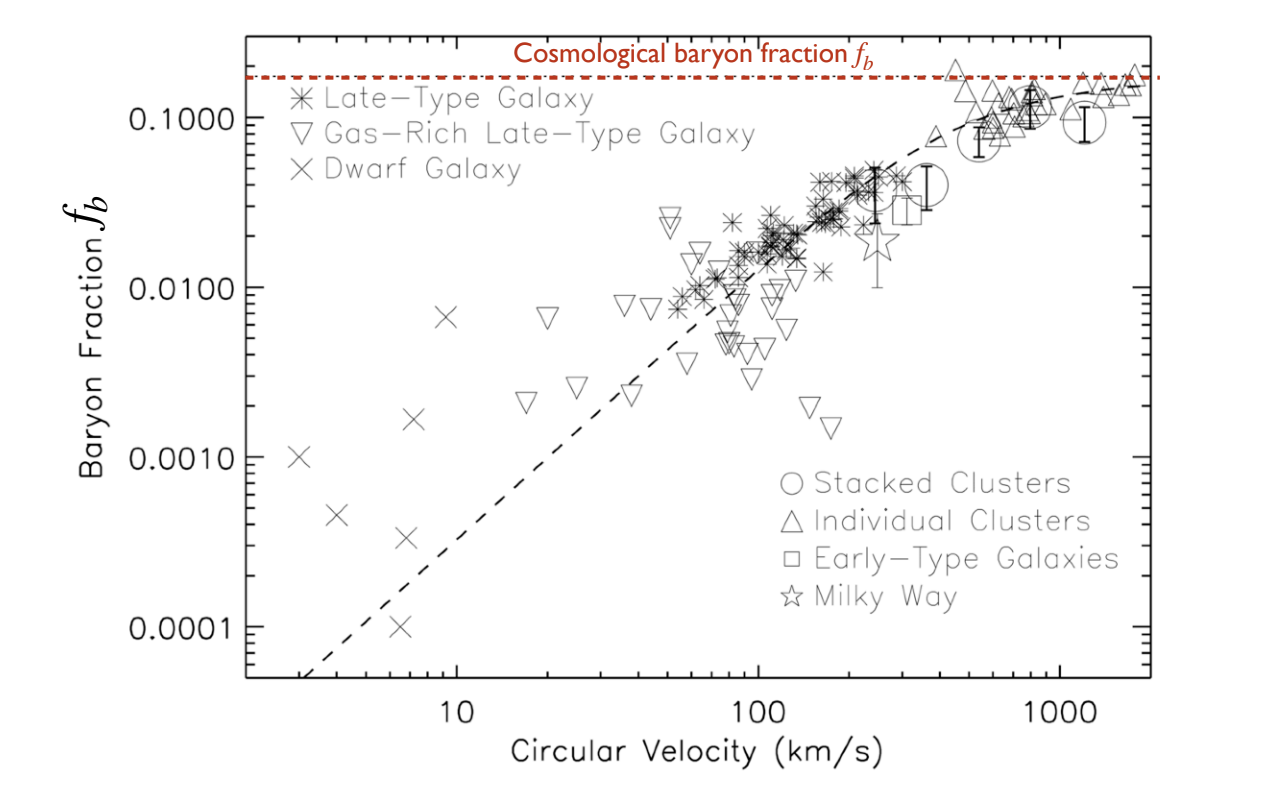}
  \vspace{-0cm}
  \captionof{figure}{Fraction of baryon mass relative to total mass (circular velocity) in various systems and environments (from \cite{2010ApJ...719..119D}).
  }
  \label{fig:B_2_Dm_ratio}
\end{figure}

From the latest Planck results \cite{2020A&A...641A...6P} the baryon density today is $\Omega_bh^2\simeq 0.0224\pm 0.0001$ and the DM density is $\Omega_c h^2\simeq 0.120\pm 0.001$. The cosmological baryon fraction is therefore given by

\begin{equation}
 f_{b,\rm cosmo} \equiv \frac{\Omega_b}{{\Omega_b+\Omega_{c}}} \simeq 0.156
 \label{eq:baryon_fraction}
\end{equation}

Although DM is present throughout the Universe, its spatial distribution is highly biased towards lower mass systems. Fig.~\ref{fig:B_2_Dm_ratio} shows the baryon fraction $f_b(V_c)$ as a function of circular velocity $V_c$, which serves as a proxy for the depth of the gravitational potential well. Observations indicate that dwarf galaxies are at least two orders of magnitude more DM dominated than massive galaxy clusters. 

\subsection{How dark is dark matter ?}

\begin{figure}[t]
\centering
  \includegraphics[width=12 cm]{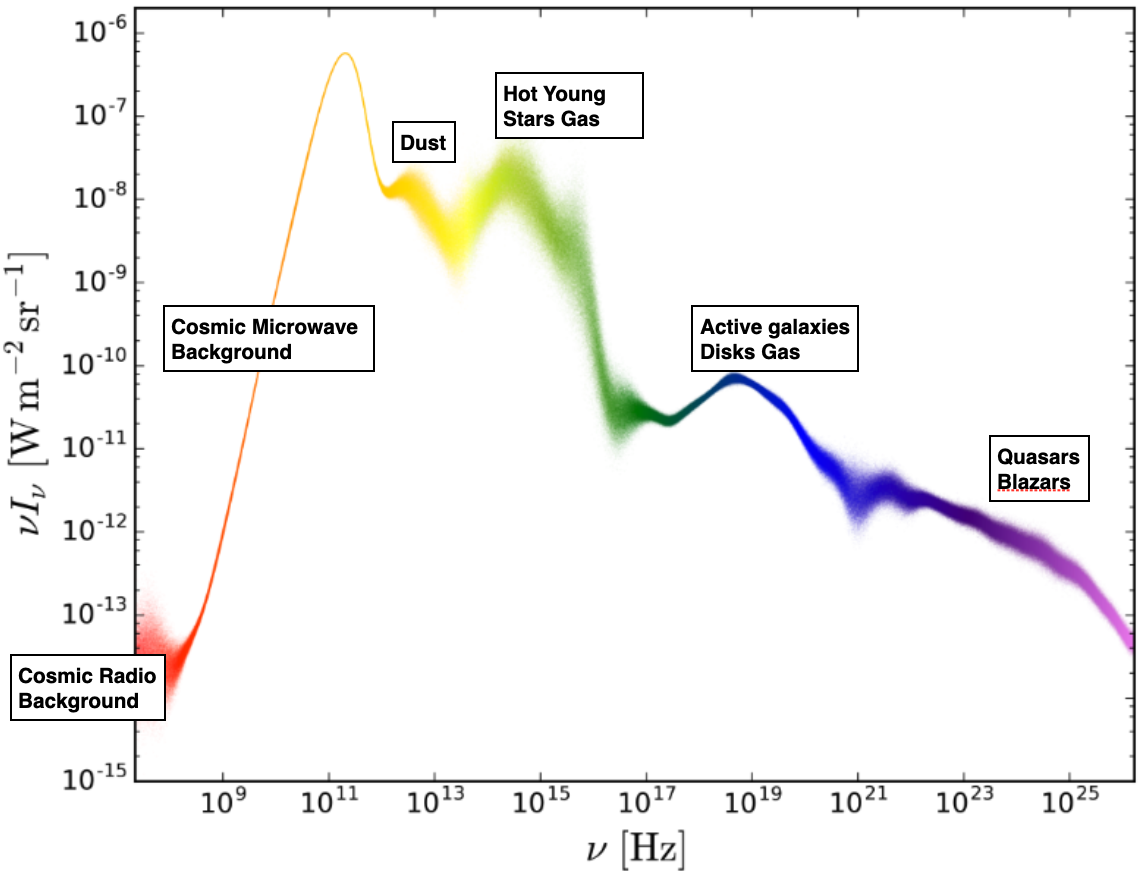}
  \vspace{-0cm}
  \captionof{figure}{Intensity of the sky monopole as a function frequency \cite{2018ApSpe..72..663H}. The width of scattered points along the vertical direction indicate the measurement errors.
  }
  \label{fig:cosmic_backgrounds}
\end{figure}

The qualifier {\it dark} is understood to mean that DM does not emit electromagnetic radiation above the sensitivity thresholds of current instruments; it should not be conflated with {\it black}, which would imply the complete absence of any emission. Fig \ref{fig:cosmic_backgrounds} show the measured sky intensity frequency spectrum, together with the dominant sources responsible for this emission \cite{2018ApSpe..72..663H}. The integrated emission from known sources over the whole sky is in good agreement with the measurement, which means that no significant excess has been seen relative to the expectations. One exception is the ARCADE excess at $\nu < 1\,{\rm GHz}$ which remains significant, even after all known galactic and extragalactic sources have been taken into account \cite{2011ApJ...734....5F}.

In principle, DM could emit radiation at any frequency, provided that its intensity remains below the error bars.
Interestingly, it is considerably easier for rare, macroscopic DM candidates to produce such an emission without violating observational constraints than for DM composed of elementary particles. The reason is that the very high number density of elementary-particle DM would lead to a substantial cumulative emission, which is not observed. By contrast, a rare population of massive DM objects could emit significantly on an individual basis, yet still produce a negligibly small total luminosity owing to their low number density.

{\bf Key takeaway.} Dark matter is dark, yet this does not preclude the possibility that it emits a faint glow below current detection thresholds.

\subsection{Cold dark matter}

Dark matter is termed \emph{cold} if its typical velocity $v_{\rm eq}$ at the epoch of matter--radiation equality ($z_{\rm eq} \sim 3400$) is much smaller than the speed of light $c$, such that its free-streaming length $\lambda_{\rm FS}$ remains below galactic scales\footnote{The definition of a galactic scale is somewhat arbitrary, but it is generally accepted that a typical halo mass of a dwarf galaxy, $M \sim 10^8\,{\rm M}_\odot$, provides a reasonable reference value.}. $\lambda_{\rm FS}$ is defined as:

\begin{equation}
    \lambda_{\rm FS}=\int_0^{t_0}\frac{v(t)}{a(t)}{\rm d}t = \int_0^{a_{\rm nr}}\frac{1}{a^2H(a)}{\rm d}a+\int_{a_{\rm nr}}^1\frac{v(a)}{a^2H(a)}{\rm d}a,
\end{equation}
where $v(t)$ is the DM velocity and $a(t)$ is the cosmological scale factor. $a_{\rm nr}$ is the scale factor when DM becomes non-relativistic. Although the free-streaming integral formally depends on the epoch at which dark matter becomes non-relativistic, the classification into cold, warm, and hot dark matter is set by whether the relativistic–to–non-relativistic transition occurs well before the time of matter–radiation equality. If $a_{\rm nr}\ll a_{\rm eq}$, the relativistic contribution to free streaming is negligible and dark matter behaves as cold.

The existence of an observational upper bound on the free-streaming length is largely model independent: structure formation data imply that dark matter cannot erase density fluctuations on scales comparable to galaxies. However, the exact \emph{calculation} of the free-streaming length $\lambda_{\rm FS}$ is model dependent, since it depends on the detailed velocity history of the DM particles. For example, consider a thermal relic of mass $m_{\rm DM}$. While it remains in thermal equilibrium, its typical momentum scales as $p \sim T$. The particle becomes non-relativistic when the temperature drops to $T \sim m_{\rm DM}$. In this case, there is a relatively direct relation between the particle mass and its velocity at matter--radiation equality, which can be translated into a lower bound on $m_{\rm DM}$. By contrast, if dark matter is produced through a non-thermal mechanism, its momentum distribution can be significantly different. 

The key point is that free-streaming constraint limit the DM \textit{velocity} at matter--radiation equality. It does not directly impose a universal bound on the particle mass, since the specifics of the production mechanism plays a decisive role in determining the velocity distribution. What observations constrain in a truly model-independent way is the relative transfer function with respect to CDM:

\[
T(k) \equiv \frac{P(k)}{P_{\rm CDM}(k)} \, ,
\]
where $P(k)$ is the three-dimensional power spectrum of a given model. $P_{\rm CDM}(k)$ is the Cold Dark matter (CDM) reference spectrum corresponding to a negligible free-streaming length ($\lambda_{\rm FS} \ll 1\,\mathrm{kpc}$).

\begin{figure}[ht]
\centering
  \includegraphics[width=12 cm]{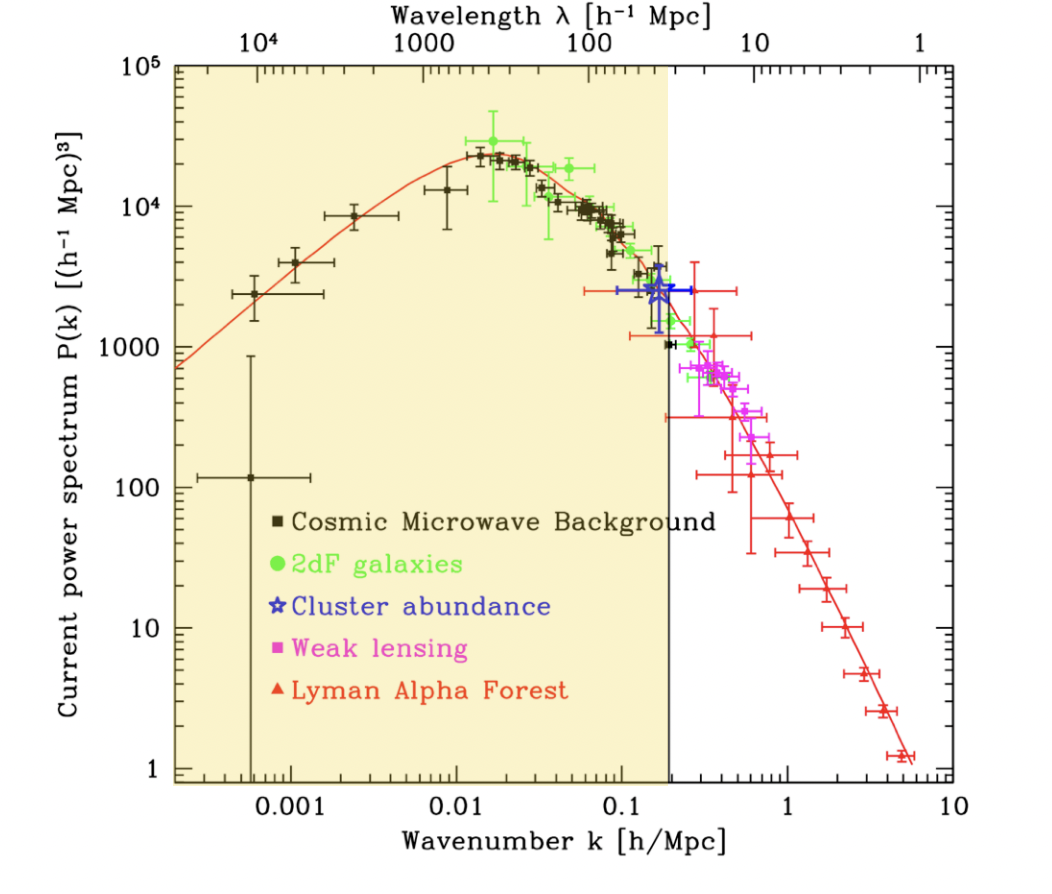}
  \vspace{-0cm}
  \captionof{figure}{ Power spectrum from a combination of observational probes, by  \cite{2002PhRvD..66j3508T}. 
  The red line shows the CDM power spectrum fitted to the multiple datasets.
  }
  \label{fig:CDM_lyman_alpha}
\end{figure}

The solid red line in Fig.~\ref{fig:CDM_lyman_alpha} shows the CDM power spectrum fitted to multiple datasets. One can see that it retains significant power on very small scales, extending to $k \gg 10\,h\,\mathrm{Mpc}^{-1}$. Not shown on Fig.~\ref{fig:CDM_lyman_alpha}, is that hot DM models predict a strong suppression for $k \gtrsim 0.01\text{--}0.1\,h\,\mathrm{Mpc}^{-1}$, while warm DM leads to suppression at intermediate scales, typically $k \gtrsim 5\text{--}30\,h\,\mathrm{Mpc}^{-1}$, depending on the specific model parameters.

Observations of the CMB, represented by the yellow region in Fig.~2, probe $P(k)$ on large scales ($k \lesssim 0.2\,h\,\mathrm{Mpc}^{-1}$) and show no significant deviation from the CDM prediction, thereby excluding hot dark matter and strongly constraining warm dark matter scenarios in this regime.
On sufficiently large scales, DM behaves effectively as cold and has already begun to cluster gravitationally by the epoch of recombination.

The Lyman-$\alpha$ forest provides access to intermediate scales, $ k \sim 0.5 \text{--} 5\,h\,\mathrm{Mpc}^{-1} $. Galaxy redshift surveys and counts of dwarf galaxies probe substantially smaller scales, reaching $ k \gtrsim 10\,h\,\mathrm{Mpc}^{-1} $. The absence of any suppression of power on these scales provides strong empirical support for the CDM paradigm and rules out hot DM. 
\medskip

{\bf Key takeaways.} Cold dark matter is empiricially supported. Warm dark matter scenarios are now tightly constrained  (e.g. \cite{PalanqueDelabrouille2020,Nadler2021}). Hot dark matter is incompatible with current data.

\subsection{Collisionless dark matter}
The interaction rate of DM is strongly constrained by observations. Consider DM-DM collisions. The mean free path is given by $\lambda=(\sigma n)^{-1}$, where $n$ is the DM number density and $\sigma$ the DM-DM cross-section. Dark matter is described as collisionless within a system of characteristic size $L$ when its mean free path $\lambda$ is much larger than $L$, i.e.\ $\lambda \gg L$. This can be rewritten as:

\[
    \left(\frac{\sigma}{m}\right)\rho L \ll 1,
\]
where $\rho=n\,m$ is the DM mass density. Hence $\Sigma_{\rm obs}=\rho L$ is the surface mass density of the system. Observations indicate that $\Sigma_{\rm obs} \sim 0.05\text{--}0.3\,\mathrm{g\,cm^{-2}}$ across a broad range of systems, including galaxy clusters, groups, and individual galaxies.

\begin{table}[t]
\centering
\begin{tabular}{l c c l}
\hline
\textbf{Constraints} & $\sigma/m$ & Velocity & Observable / Method \\
\hline
Halo shapes / ellipticity 
    & $\lesssim 1\,\mathrm{cm^2\,g^{-1}}$ 
    & $1300\,\mathrm{km\,s^{-1}}$ 
    & Cluster lensing surveys \\

Substructure mergers 
    & $\lesssim 2\,\mathrm{cm^2\,g^{-1}}$ 
    & $\sim 500$--$4000\,\mathrm{km\,s^{-1}}$ 
    & DM--galaxy offset \\

Merging clusters 
    & $\lesssim \text{few}\,\mathrm{cm^2\,g^{-1}}$ 
    & $2000$--$4000\,\mathrm{km\,s^{-1}}$ 
    & Post--merger halo survival \\

\emph{Bullet Cluster} 
    & $\lesssim 0.7\,\mathrm{cm^2\,g^{-1}}$ 
    & $4000\,\mathrm{km\,s^{-1}}$ 
    & Mass--to--light ratio \\
\hline
\end{tabular}
\caption{Astrophysical constraints on the dark matter self--interaction cross-section per unit mass $\sigma/m$ from cluster-scale observations. From \cite{2018PhR...730....1T}}
\label{sigma_table}
\end{table}

At cluster and larger scales, observations show that (see Table ~\ref{sigma_table}):
\[
\frac{\sigma}{m} \lesssim 1\,{\rm cm^2\,g^{-1}},
\]
which is $\ll \Sigma_{\rm obs}^{-1}$, indicating that dark matter is collisionless at these scales. This constraint is model-independent, as it follows directly from gravitational lensing measurements and basic dynamical arguments, without invoking assumptions about baryonic physics or halo modelling.

On smaller scales, such as dwarf galaxies, galaxy groups, and cluster cores, larger values are favoured:
\[
\frac{\sigma}{m} \sim 1\text{--}10\,{\rm cm^2\,g^{-1}},
\]
and may even provide an attractive feature of self-interacting DM models \cite{2000PhRvL..84.3760S}. However, unlike the cluster-scale upper bound, such interpretation relies on a complete understanding of the complex astrophysical processes, such as stellar and active galactic nuclei feedback, gas stripping, gas outflows and gas cooling to name a few \cite{2012MNRAS.421.3464P,2018PhR...730....1T}.
No observations currently exclude collisionless DM at any scales.

\medskip
{\bf Key takeaway.} Although observational bounds on $\sigma/m$ are model-independent in nature, their theoretical interpretation and calculation depend on the specific DM scenario under consideration.

\subsection{Dark matter interactions with baryons and photons}

Observational data impose limits on the strength of DM interactions with baryons and photons. If DM scatters with baryons then we can define the interaction rate $\Gamma_{\rm DM-b}$:

\[
    \Gamma_{{\rm X}-b}=\frac{\rho_b}{m_{\rm X}}\langle \sigma_{{\rm X}-b} v\rangle,
\]
where $\langle \sigma_{{\rm X}-b}\, v\rangle$is the velocity dependent cross-section of the DM-baryon interaction. Momentum exchange between the two fluids will modify the acoustic oscillations in the primordial plasma. This will alter the heights and phases of the CMB acoustic peaks, suppress the growth of small-scale density perturbations, and consequently distort the CMB temperature and polarisation power spectra. Such deviations will accumulate prior to recombination. They will also affect the subsequent formation of structure, leaving observable imprints in large-scale structure probes such as the Lyman-$\alpha$ forest.

Since neither Planck CMB measurements nor redshift surveys detect such departures from the standard cosmological model, one obtains the robust requirement that $\Gamma_{X-b} \ll H$ around recombination, a condition which is independent of the DM model. However, the corresponding constraint on the cross-section per unit mass, $\sigma_{X-b}/m_X$, depends on the velocity dependence of the DM–baryon interaction, and more generally on the DM model under consideration. For different velocity scalings of the cross-section, the resulting bounds are given by \cite{2014PhRvD..89b3519D,2018PhRvL.121h1301G,2018PhRvD..98h3510B,2018PhRvD..98l3506B}:

\[
    \frac{\sigma_{X-b}}{m_X}\lesssim 10^{-9}-10^{-6}\,{\rm cm^2g^{-1}}
\]
The interaction between DM and photons would lead to similar effects. The absence of such signature also leads to an upper bound for the DM-photon cross-section \cite{2014JCAP...04..026W,2018JCAP...10..009S}:

\[
    \frac{\sigma_{X-\gamma}}{m_X}\lesssim 10^{-6}-10^{-4}\,{\rm cm^2g^{-1}}
\]
These are robust bounds on DM–baryon and DM–photon scattering, obtained from the requirement that dark matter must have decoupled from the primordial plasma before recombination. However, unlike the cluster-scale self-interaction limit, these constraints stem from early-Universe drag effects, and their translation into limits on the cross-section depends on the specific DM model. By contrast, the optical-depth constraint on DM–DM interactions is largely geometric in nature and therefore model independent.

\begin{figure}[t]
\centering
  \includegraphics[width=13 cm]{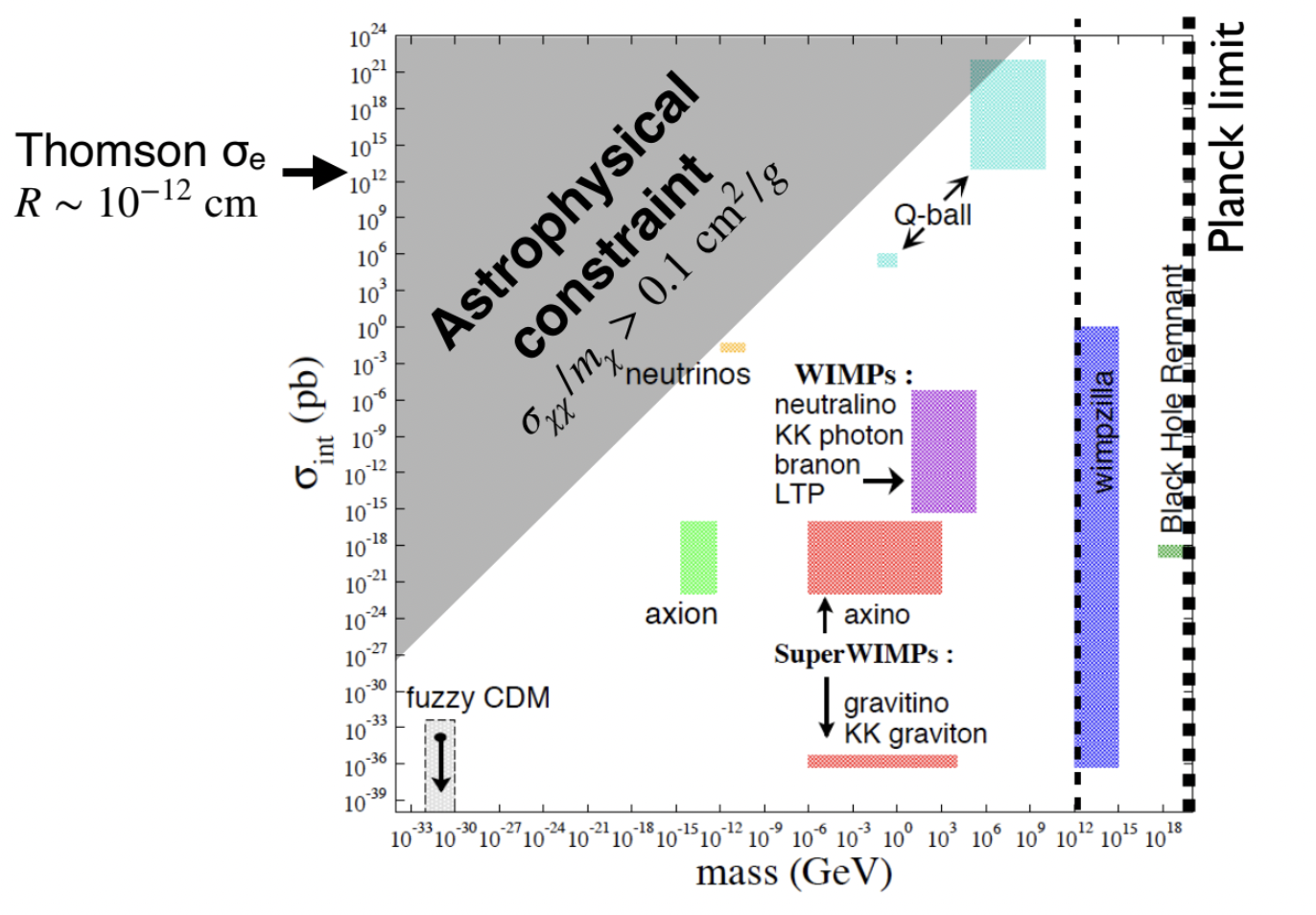}
  \vspace{-0cm}
  \captionof{figure}{Limits on the annihilation cross-section of DM as a function of DM mass \cite{2009arXiv0901.4732B}. The top-left gray region is excluded because DM must be collisionless.
  }
  \label{fig:sigma M parameter space}
\end{figure}

By comparison, for WIMPs the self-interaction cross-section cannot be inferred directly from observations and must instead be estimated within a specific particle-physics framework. In the minimal weak-scale scenario, i.e. point-like WIMPs without a light mediator\footnote{ This model is closely aligned with the historical WIMP paradigm}, one typically obtains $\sigma_{XX}/m_X \sim 10^{-17}-10^{-13}\ {\rm cm^2\,g}^{-1}$. Direct-detection experiments constrain the WIMP–nucleon interaction cross-section to approximately $\sigma_{Xb}/m_X \sim 10^{-26}-10^{-24}\ {\rm cm^2\,g}^{-1}$ for weak-scale masses (see for example \cite{2023PhRvL.131d1002A}).
Finally, still for minimal WIMP models, the particle does not couple to photons at tree level, so the WIMP–photon cross-section is expected to be extremely small (i.e. suppressed at the loop level).
These estimates demonstrate that more than ten orders of magnitude in parameter space separate the canonical WIMP expectation for self-interactions from the current astrophysical upper limits. A substantial region of the parameter space remains largely unexplored: This wide open regime is illustrated in Fig.~\ref{fig:sigma M parameter space}, which displays only the self-interaction cross-section.

\medskip

{\bf Key takeaway.} A substantial region of the $\sigma$--mass parameter space for DM candidates remains largely unexplored.

\subsection{Observational constrains from Big Bang Nucleosynthesis}
Big Bang Nucleosynthesis (BBN) provides a powerful model-independent probe of DM, as it occurs during a particularly well-controlled epoch in cosmic history ($t \sim 1-10^{3}\,\mathrm{s}$, $
T \sim 0.1-10\,\mathrm{MeV}$), during which both the expansion rate of the Universe and energy injection processes are tightly restricted by observations.
The constraints come from the primordial abundances of light elements (D, $^3$He, $^4$He, and $^7$Li): an additional energy injection from DM and/or an increase of relativistic species $\Delta N_{\rm eff}$ (e.g. sterile neutrino as DM) at this epoch, would leave an imprint on the light elements abundances. Energy injection will generally photodissociate light elements, over producing deuterium and ${}^3\mathrm{He}$ and decrease ${}^4\mathrm{He}$. If DM produces hadrons instead, e.g. from decay or annihilation, the $n/p$ ratio will differ, which can subsequently modify the abundance of certain light elements. Letting the ${}^7\mathrm{Li}$ problem asside, the current constraints on $D/H$ and ${}^4\mathrm{He}/H$, combined with CMB, set a strong upper bound on $\Delta N_{\rm eff} \lesssim 0.2-0.3$ \cite{Fields_2020,Pitrou_2018}.

Let's emphasise that the baryon–to–photon ratio,

\begin{equation}
    \eta\equiv n_b/n_\gamma \simeq 6.1\times 10^{-10}
\end{equation}
is determined independently by BBN and by the CMB. The agreement between these two measurements -- spanning the period from $\sim 3 $ min after the Big Bang to about 380{,}000 years later -- provides one of the most compelling consistency tests of the standard cosmological model.
Since $\eta$ fixes the baryon density, one infers that:

\begin{equation}
    \Omega_{\rm DM}\sim 5\Omega_b.
\end{equation}
This fact is often invoked as evidence that dark matter must be non-baryonic, since a dominant baryonic-DM component would have significantly altered the primordial light-element abundances. However, it is important to clarify what BBN really implies. BBN does not rule out baryonic dark matter in a model-independent way; rather, it constrains the abundance of baryons that were thermally coupled and participated in nucleosynthesis at the MeV epoch. In other words, DM must not have been overwhelmingly coupled to the baryon–photon plasma during BBN.

If dark matter consists of elementary particles interacting with baryons, the simplest way to satisfy this requirement is indeed for it to be non-baryonic. However, another possibility is that DM is composed of rare, compact, stable  \textit{macros}: even if such objects were of baryonic origin, their rarity would prevent them from significantly modifying primordial nucleosynthesis.
\medskip

{\bf Key takeaways.} 
The observed light-element abundances impose stringent constraints: DM must be cold, non-relativistic, and non-disruptive at MeV temperatures \cite{2016RvMP...88a5004C}. Dark matter must interact only weakly with ordinary matter. This requirement may be satisfied either by abundant non-baryonic particles with very small interaction cross-sections, or by massive but extremely rare objects. BBN does not exclude dark-matter candidates composed of quark matter.

\subsection{An overlooked hypothesis: rare dark matter}

An observational upper bound on the ratio $\sigma/m_X$ does not, by itself, place independent constraints on the interaction cross-section $\sigma$ or on the number density $n_X$ separately. Let us rewrite the upper bound constraints in terms of $n_X$. For the self-interaction rate in a halo:

\[
    \sigma_{XX} n_X L_{\rm halo}\ll 1.
\]
For DM-baryon, we obtain:

\[
\sigma_{X-b} n_X (v_{\rm rel}/H)\ll 1,
\]
and for DM-photon:
\[
\sigma_{X-\gamma} n_X (v_{\rm rel}/H)\ll 1.
\]
One can see that the same upper bound can be realised either with a small cross-section and a large number density $n_X$, or with a large cross-section and a small $n_X$.

It is common to hypothesize that DM is composed of abundant subatomic particles. In the Solar System environment, where the DM mass density is $\rho_{x}\simeq 0.3\,{\rm GeV\,cm^{-3}}$, their number density is extremely large, $n_X\simeq 3\times 10^5 ({\rm GeV/m_X})\,m^{-3}$. Given the current absence of direct detection of subatomic DM, such particles must interact only very weakly with baryons, hence their very small cross-sections $\sigma_{XX}, \sigma_{X-b}$ and $\sigma_{X-\gamma}$.

An alternative possibility is that DM consists of macroscopic objects with mass $m_X\gg M_{\rm Planck}$, characterised by extremely low number densities $n_X \ll 10^{-14}\,\mathrm{m}^{-3}$. In this case, even large values of $\sigma_{XX}, \sigma_{X-b} $ and $\sigma_{X-\gamma}$ can produce a low total emission rate to have so far evaded detection. Moreover, for $n_X \ll 1\,\mathrm{m}^{-3}$, direct detection on Earth would be unlikely.
\medskip

{\bf Key takeaway.} Massive, rare DM particles with a large interaction cross-section are not ruled out by observations.

\subsection{Bayesian perspectives on dark-matter models}

\begin{figure}[t]
\centering
  \includegraphics[width=13 cm]{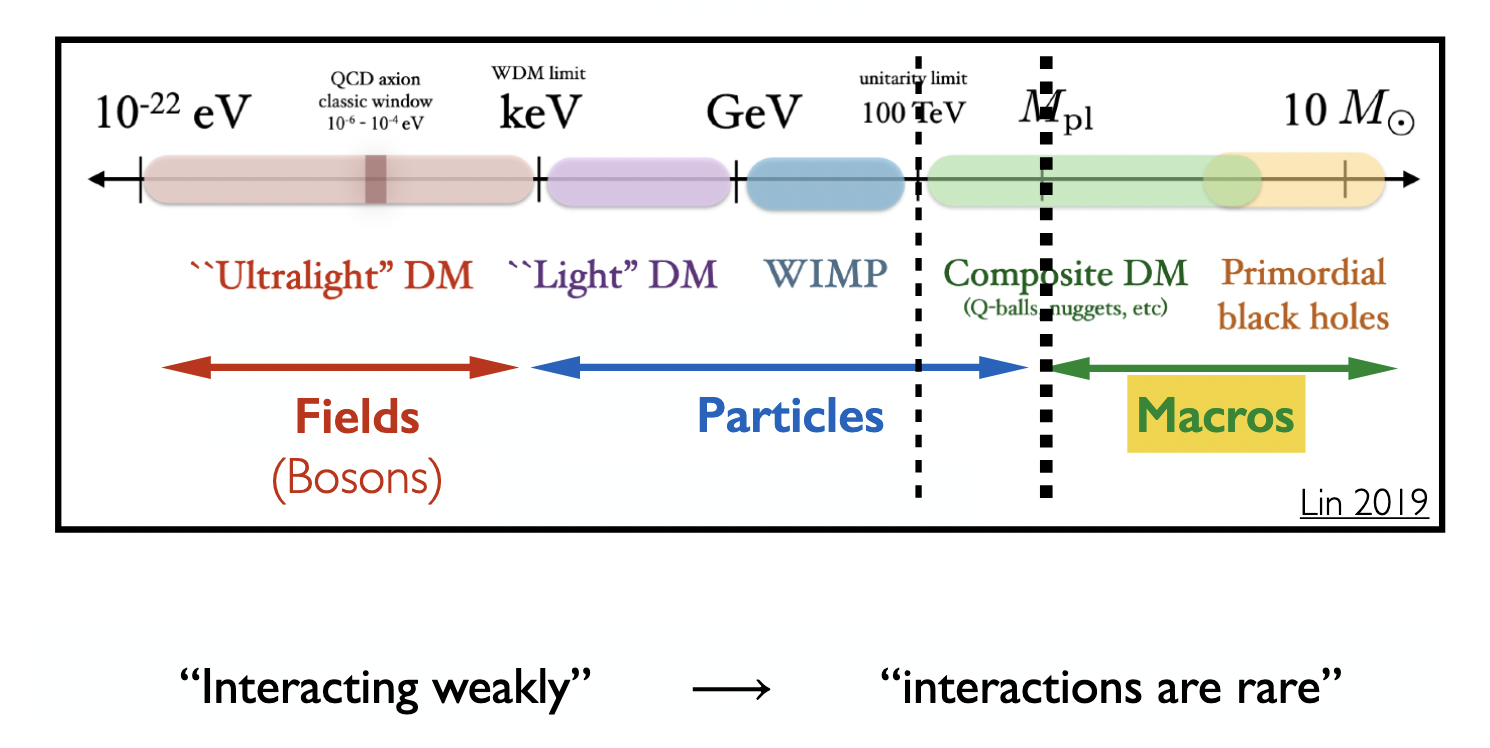}
  \vspace{-0cm}
  \captionof{figure}{Mass range for various DM candidates. The thicker dashed line indicates the Planck mass. AQNs belong to the composite DM region, with $m_{\rm AQN} \sim 10^{20}\text{--}10^{23}\,\mathrm{TeV}$. From \cite{2019arXiv190407915L}.
  }
  \label{fig:DM_mass_scale}
\end{figure}

Astrophysical measurements establish that a dark, cold, and collisionless matter component exists, but no observation to date demonstrates that dark matter is composed of elementary particles. We are left with two hypotheses: Either DM is composed of abundant, weakly interacting particles or it is made of rare, strongly interacting macroscopic objects. The latter has received comparatively little attention and remains the largest unexplored region of parameter space among DM models \cite{2015MNRAS.450.3418J}. Fig.~\ref{fig:DM_mass_scale} illustrates the wide range of viable DM mass scales. With these conclusions in mind, we now proceed to examine the Bayesian evidence associated with the leading DM candidates.

Alternative dark matter models refer to theoretical proposals beyond the Weakly Interacting Massive Particle (WIMP) scenario. Examples include axions, sterile neutrinos, primordial black holes (PBHs), self-interacting DM (SIDM) and macroscopic composite objects such as AQNs. Each of these models makes different assumptions about the mass, interactions, and cosmological behaviour of dark matter. Given the variety of DM models and the size of the parameter space, a quantitative comparison based on Bayesian inference provides a systematic framework for assessing their relative plausibility.

A common concept in evaluating these models is falsifiability: does the model make clear predictions that could, in principle, be proven wrong? For example, WIMPs predict signals in underground direct detection experiments, axions predict narrow-band signals in resonant cavity experiments. These are examples of observational tests, where theory is confronted with independent measurements. In a Bayesian framework, we do not ask if a model is falsifiable, rather we consider the \textbf{posterior probability} $P(M|D)$ that model $M$ is correct after taking the data $D$ into account:

\[
    P(M|D)\propto P(D|M)P(M)
    \label{bayes}
\]
where $P(D|M)$ is the \textbf{likelihood}, which quantifies how well the model explains the data. $P(M)$ is the \textbf{prior probability}, i.e. the degree of plausibility assigned to the model before considering the current data. It encodes theoretical motivation, simplicity, and previous knowledge.

Now consider that model $M$ has a parameter space $\theta$, then by marginalizing over the parameter space, the Bayesian evidence can be written as:

\[
    P(D|M)=\int_{\rm Priors} P(D|\theta,M)P(\theta|M){\rm d}\theta
\]
One can see that for models in which only a small fraction of the parameter space $\theta$ yields a high likelihood, while most of the space gives a poor likelihood, the Bayesian evidence $P(D|M)$ is small. This is known as the Occam penalty.
It arises when a model has a large parameter space, but only a small region of that space produces predictions consistent with the data. 
In such cases, most of the prior volume contributes little to the likelihood, which reduces the overall Bayesian evidence. Importantly, this penalty is not imposed by hand; it appears automatically when the likelihood is marginalised over the full parameter space.

To illustrate these definitions, let us consider the example of WIMPs:
\begin{itemize}[noitemsep, topsep=0pt]
    \item With the original WIMPs miracle, the prior $P(M_{\rm WIMPs})$ was high for several reasons: WIMPs were naturally linked to the electroweak scale; the thermal freeze-out explained $\Omega_{\rm DM}\sim \Omega_b$; the model had very few parameters and appeared natural. Its Occam penalty was low.
        \item The null results from direct, indirect, and collider searches progressively reduced the likelihood $P(D|M_{\rm WIMPs})$, since an increasing fraction of the parameter space became incompatible with the data;
    \item The absence of supersymmetry at the LHC, together with the lack of evidence for new weak-scale physics, has led to a substantial reduction of the prior plausibility $P(M_{\rm WIMPs}$).
\end{itemize}
As a result, the posterior $P(M_{\rm WIMPs}|D)$ has significantly reduced  over the last twenty years.

\bigskip
In contrast, many DM alternatives (see Table. \ref{tab:dm_candidates_comparison}) are characterized by broad prior parameter spaces and have limited, often single-channel, phenomenological predictions. These models distribute likelihood over a wide range of possible datasets and incur only a mild Occam penalty, even as observational constraints accumulate. In other words,  \textit{the field now tolerates less predictive models, broader parameter spaces, and weaker experimental commitments. }

\smallskip
\textbf{What about QCD-AQN ?}
\medskip

The QCD–AQN framework is rooted in QCD-scale physics and axion dynamics. It belongs to a well-motivated Standard Model extension, the QCD axion. Its extensive explanatory scope will be explained in Section \ref{sec: formation}.  Sociogically, such a broad scope elicits scepticism within the scientific community, not because Bayesian reasoning penalises it, but because of historical factors. Past experience  has fostered a culture of caution. Unified frameworks that claim to explain many seemingly unrelated phenomena can appear too ambitious. Scientists tend to apply an \textit{informal epistemic prior}: the broader the claim, the more rigorous the required evidence. 
      
QCD-AQN carries a historical burden stemming from an earlier proposal. In 1984, Edward Witten \cite{PhysRevD.30.272} introduced the idea that stable lumps of strange quark matter  could have formed during a first-order QCD transition, carry a large baryon number and behave as DM: the so-called \textit{quark nuggets}.  However, later lattice QCD studies indicated that, at zero baryon chemical potential, the cosmological QCD transition is a smooth crossover \cite{2002LNP...583..209K} rather than a strong first-order phase transition, thereby weakening Witten's nuggets formation mechanism.  The QCD-AQN scenario, which introduces new ingredients to provide a radically different formation pathway (see Section \ref{sec: formation}), has sometimes mistakenly been viewed through the lens of Witten's quark nuggets fall out.

QCD-AQN makes unavoidable, multi-channel predictions arising from the coupling of AQNs with  a large variety of baryonic environments, which results in a heavy Occam penalty. These astrophysical environments span several fields of study: cosmology, high-energy astrophysics, stellar physics, interstellar medium, planetary science etc. Modern research is highly specialised, with distinct subfields operating semi-independently. From a sociological perspective, a model that connects disparate domains challenges disciplinary boundaries and may face resistance simply for crossing them. There is concern that wide explanatory scope might conceal hidden fine-tuning or selective interpretation of data, thereby requiring stronger empirical support to achieve comparable evidence.

On the positive side, QCD-AQN is governed by a single fundamental parameter, the mean AQN mass, $\langle m_{AQN}\rangle$, which is restricted to a relatively narrow window by independent observational and theoretical considerations (see Section~\ref{sec:synthesis_mass}).
\medskip

{\bf Key takeaways.}  QCD-AQN is high-risk but potentially high-reward framework. It is neither easily excluded nor easily accommodated. Its ultimate viability
hinges on the emergence of distinctive, environment-specific signatures that cannot be replicated by more weakly predictive DM scenarios.
\bigskip

\begin{table*}[htbp]
\centering
\small
\begin{tabular}{p{2.6cm} p{3.6cm} p{4.0cm} p{4.0cm}}
\hline
\textbf{Candidate} & \textbf{Prior theoretical motivation} & \textbf{Explanatory scope beyond DM} & \textbf{Main penalties / tensions} \\
\hline

\textbf{QCD axions} &
Very high: solves the strong CP problem within a well-motivated Standard Model extension; no dark sector &
High: DM + resolution of the strong CP problem &
No confirmed detection; broad but narrowing parameter space \\

\hline

\textbf{QCD-AQNs} &
Moderate--high: QCD and axion physics with additional dynamical assumptions; no dark sector &
Very high:  matter-antimatter asymmetry; natural link $\Omega_{\rm DM} \sim \Omega_b$, baryon-to-photon ratio $\eta$, possible lithium-7 tension, possibility of avoiding baryogenesis; potential multi-messenger signatures &
 No decisive detection; historical
scepticism from earlier
nugget models; structural complexity\\

\hline

\textbf{WIMPs} &
Historically very high (“WIMP miracle”); electroweak-scale new physics &
Moderate: primarily relic abundance via thermal freeze-out &
Strong null results from direct detection, indirect searches, and LHC; shrinking parameter space \\

\hline

\textbf{ALPs} &
Moderate: generic prediction of string compactifications and hidden sectors &
Moderate: DM candidate; possible astrophysical transparency anomalies &
Highly flexible parameter space; reduced predictivity; no confirmed detection \\

\hline

\textbf{Sterile neutrinos} &
Moderate: minimal extension of the neutrino sector; motivated by neutrino mass generation &
Moderate: DM + possible links to neutrino physics &
X-ray constraints; structure formation limits; uncertainties in production mechanisms \\

\hline

\textbf{Primordial Black Holes (PBH)} &
Moderate: no new particle physics required; arises from early-universe collapse &
Moderate: DM; possible connection to gravitational-wave events &
Strong constraints across many mass ranges (microlensing, CMB, evaporation, dynamics) \\

\hline

\textbf{Self-Interacting DM (SIDM)} &
Moderate: phenomenological extension of CDM addressing small-scale tensions &
Moderate--high: DM + potential resolution of core--cusp and diversity problems &
Requires tuned interaction cross sections; not a unique underlying particle model \\

\hline
\end{tabular}
\caption{Qualitative comparison of major dark matter candidates, highlighting theoretical motivation, explanatory scope, and principal tensions.}
\label{tab:dm_candidates_comparison}
\end{table*}

\section{Confronting QCD-AQN to observations}
\label{sec:AQN_and_Obs}

Having reviewed the observational constraints that any viable dark matter model must satisfy, we now turn to the observational aspects of the QCD-AQN framework, which constitute the principal motivation for its investigation. In this scenario, when the QCD transition begins, at $T\sim 170\,{\rm MeV}$, the pre-existing small asymmetric coupling between quarks (antiquarks) to the axion field, combined with the collapsing of the axion domain walls, leads to the formation of aggregates with large baryon numbers ($ |B| \gtrsim 10^{24}$), the so called Axion Quark Nuggets (AQNs). The formation of AQNs, the baryon number separation process and the AQNs survival in the early Universe will be discussed in Section \ref{sec: formation}. For the purpose of observational tests, it is sufficient to focus on the structure of the aggregates. In this section, after reviewing their  fundamental properties, we explore a range of potential AQN signatures, considering first their behaviour in dilute media and then in dense environments.

\subsection{AQNs structure and properties}

\begin{figure}[t]
\centering
  \includegraphics[width=13
  cm]{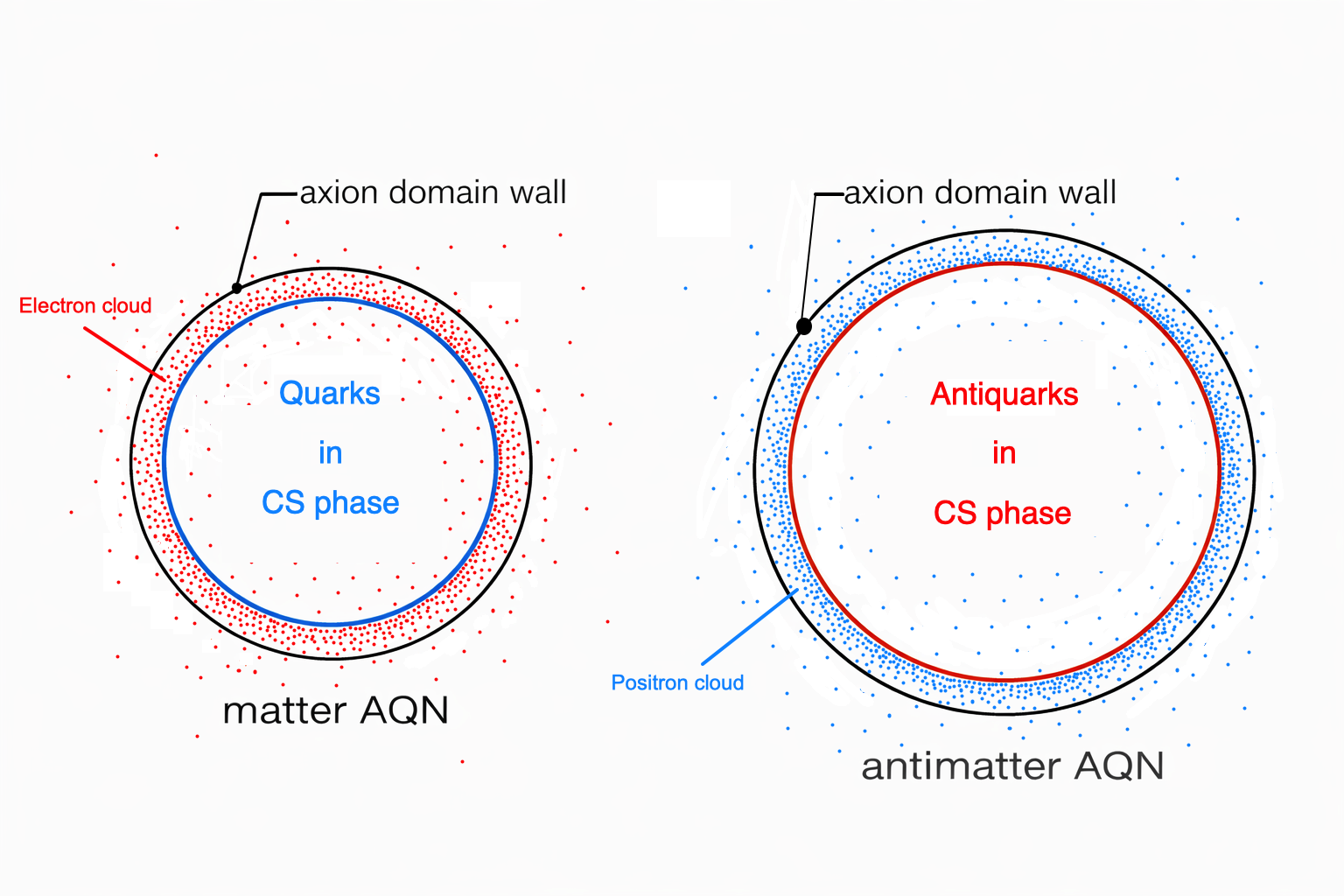}
  \vspace{-0cm}
  \captionof{figure}{The quarks (antiquarks) in an AQN core are bound by an axion domain wall and stabilised by a superconducting energy gap, $\Delta_{\rm gap} \sim 100\,\mathrm{MeV}$. Matter AQNs are surrounded by an electron cloud. Antimatter $\bar{\rm AQN}$s are composed of an antiquark core  surrounded by a positron cloud. The cosmological distribution of matter and antimatter AQNs is asymmetric (see \ref{subsection:asymmetry}). }.
  \label{fig:AQN structure}
\end{figure}

Matter (antimatter) AQNs  are macroscopic aggregates of quarks (antiquarks) in a Colour-Superconducting (CS) phase, stabilised by an axion domain wall that acts as a confining shell. The quark (antiquark) core is surrounded by an electron (positron) cloud that ensures overall electric neutrality (Fig.~\ref{fig:AQN structure}).

The baryon number of AQNs is not a freely adjustable parameter but emerges from QCD dynamics at the axion domain-wall formation epoch; theoretical estimates based on the balance between domain-wall tension, Fermi pressure, and colour-superconducting energy gaps lead to typical values in the range $|B| \sim 10^{22}$--$10^{28}$ (see \ref{sec:AQN_mass_range}), corresponding to macroscopic masses $\sim 10^{-2}$ -- $10^4$ grams.

Matter AQNs do not produce observable annihilation signatures: Because of the low abundance of antimatter cosmic rays, quark aggregates remain dark. Their interactions can be studied mainly through their gravitational effects as a CDM component. They are not the primary focus of this lecture.

Of particular interest is the population of antimatter AQNs which collide with ambient baryons, thereby generating observable signatures across a broad range of wavelengths and astrophysical environments.

\begin{tcolorbox}[colback=white, colframe=black, boxrule=0.5pt]
\emph{Convention: Throughout the remainder of this lecture, $\bar{\rm {AQN}}$ will denote antimatter aggregates.} 
\end{tcolorbox}

\smallskip

The dense CS core of the $\bar{\rm {AQN}}$ generates an electric field that binds positrons in an extended structure outside the sharp surface. This positron cloud, also know as positron electrosphere, behaves like a Thomas-Fermi gas \cite{1927PCPS...23..542T,1928ZPhy...48...73F} and has a stratified profile\cite{2010PhRvD..82h3510F}: 
\begin{itemize}[noitemsep, topsep=0pt]
    \item degenerate inside the antiquark core;
    \item distributed as a thin classical relativistic gas layer at the core surface;
    \item non-relativistic Boltzmann gas above the surface .
\end{itemize}
\begin{table}[ht]
\centering
\label{tab:AQN_properties}
\begin{tabular}{l c}
\hline
Physical quantity  & Characteristic value\\[1ex]
\hline
Geometric cross section &
$\displaystyle {\sigma}/{M} \sim 10^{-10}\ \mathrm{cm^2\,g^{-1}}$ \\[1ex]

Radius &
$\displaystyle R \sim 10^{-5}\ \mathrm{cm}$ \\[1ex]

Mass &
$\displaystyle m \gtrsim 10^{24}\,m_p \sim 1\ \mathrm{g}$\\

Density &
$\displaystyle \rho \gtrsim \rho_{\rm nucl}
= 3.5\times10^{14} \mathrm{g .cm^{-3}}\ $\\[1ex]

Number density (Earth environment) &
$\displaystyle (10^{6}\,\mathrm{km})^{-3} \lesssim n_{\rm AQN} \lesssim (10^{3}\,\mathrm{km})^{-3}$
\\
\hline
\end{tabular}
\caption{Main physical characteristics of AQNs.}
\end{table}

\subsection{Cosmological behaviour of AQNs: cold and collisionless}

Because the QCD-AQN framework lies at the intersection of cosmology and condensed matter physics, differences in terminology can obscure the discussion. It is therefore useful to clarify what is meant by baryonic matter.

From a particle physics perspective, AQNs are baryons because they are made of quarks bound into dense quark matter and therefore $B \neq 0 $. Their mass originates from QCD-scale strong interactions, just like ordinary protons and neutrons.

In cosmology, baryonic matter is defined by its role in early-Universe rather than by its quark content. Here, a baryon refers to ordinary matter that carries baryon number and participates in the thermal and nuclear processes of the early Universe, i.e. the matter component that was tightly coupled to photons before recombination. In this context, AQNs are not baryons because their baryon number is confined within compact objects before BBN, preventing them from participating in primordial element formation or in the photon–baryon plasma that shapes the CMB acoustic peaks.

In short, AQNs are baryonic in their microscopic composition yet non-baryonic in their cosmological behaviour.
\bigskip

AQNs are non-relativistic very early on. Being macroscopic objects with gram-scale masses, their thermal velocities after formation at the QCD transition are negligible compared to the speed of light. As the Universe expands, their velocities redshift further, so they behave as non-relativistic matter well before matter--radiation equality. This gives them an equation of state
\[
w \equiv \frac{P}{\rho} \approx 0,
\]
just like standard cold DM.

AQNs are collisionless on large scales. Although annihilation events happen when an $\bar{\rm AQN}$ collides with baryons, their geometric cross section per unit mass is extremely small. In the dilute intergalactic medium, interaction rates are negligible compared to the Hubble expansion rate. Finally, the free-streaming length of AQNs is negligible, as these objects move with very small, non-relativistic velocities. They cluster gravitationally and seed structure formation as expected for cold DM.
\medskip

{\bf Key takeaways.} Despite being baryonic in composition, AQNs behave cosmologically as non-baryonic cold, collisionless DM. They are massive, slow-moving, and dynamically decoupled objects.
\medskip

Having reviewed the large-scale cosmological behaviour of AQNs, we now focus on the thermal emission produced locally as they pass through baryon-rich environments.

\subsection{Radiative signatures of $\bar{\rm{AQN}}$s in dilute environments}
\label{subsec:radiation}

In this section, we will discuss the electromagnetic radiation arising from the interaction of $\bar{\rm {AQN}}$s with ordinary baryons in a dilute medium, which was first described by \cite{2010PhRvD..82h3510F}. Dense media will be discussed in \ref{subsec:section:observations_dense} and prospects for future tests in various astrophysical environments are detailed in \ref{subsec:futuretests} and \ref{sec:dark_glow}.

\smallskip

\begin{figure}[ht]
\centering
  \includegraphics[width=13 cm]{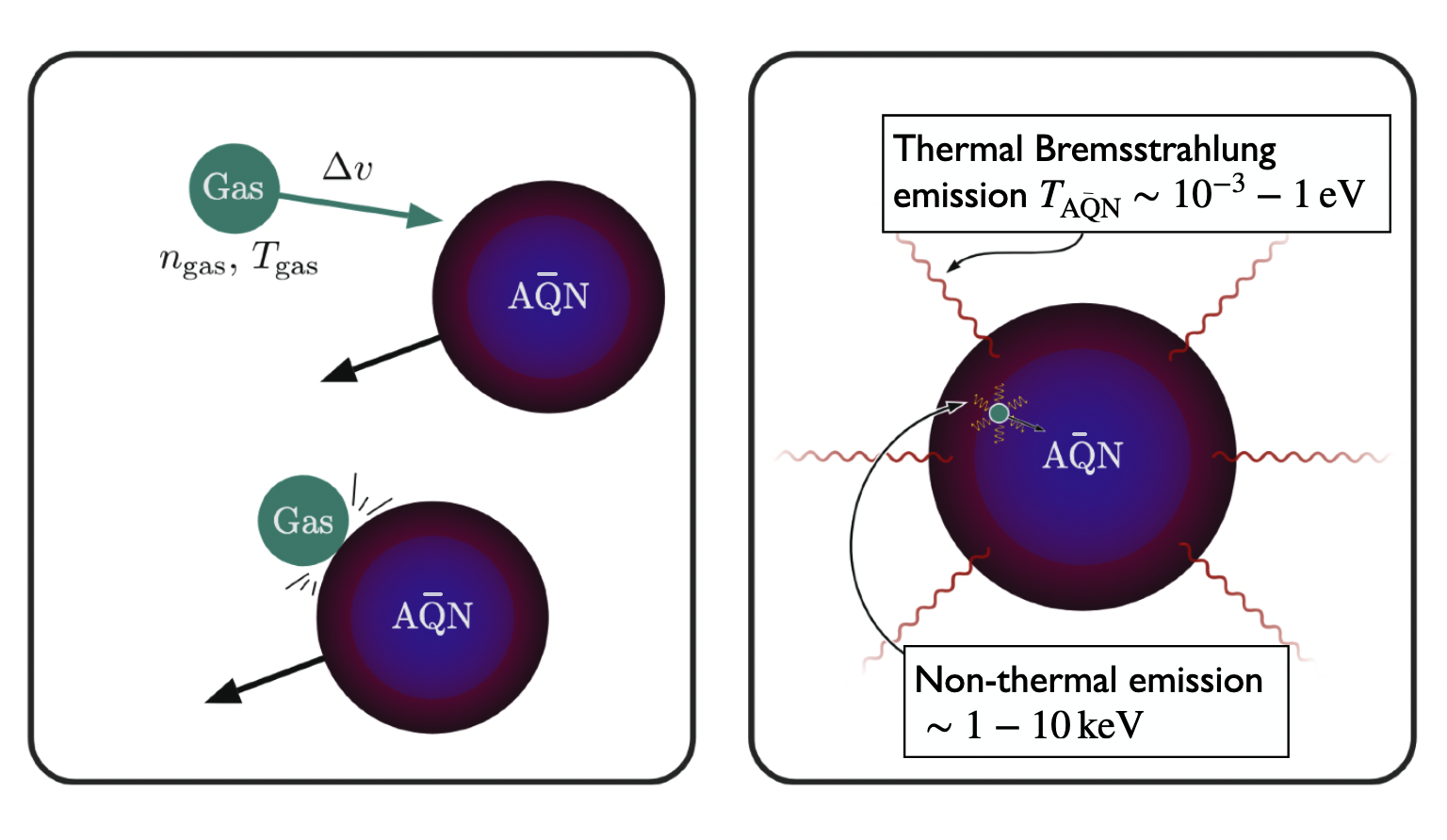}
  \vspace{-0cm}
  \captionof{figure}{Left: A baryon (i.e. proton) collides with an $\bar{\rm {AQN}}$ [not to scale]. Right: The matter-antimatter annihilation emits $\sim 2$ GeV of energy in the form of thermal Bremsstrahlung and a secondary beam in the 1-10 keV range \cite{2024A&A...691A..38S}
  .}
  \label{fig:AQN_B_collision}
\end{figure}

\begin{figure}[t]
    \centering
    \includegraphics[width=13cm]{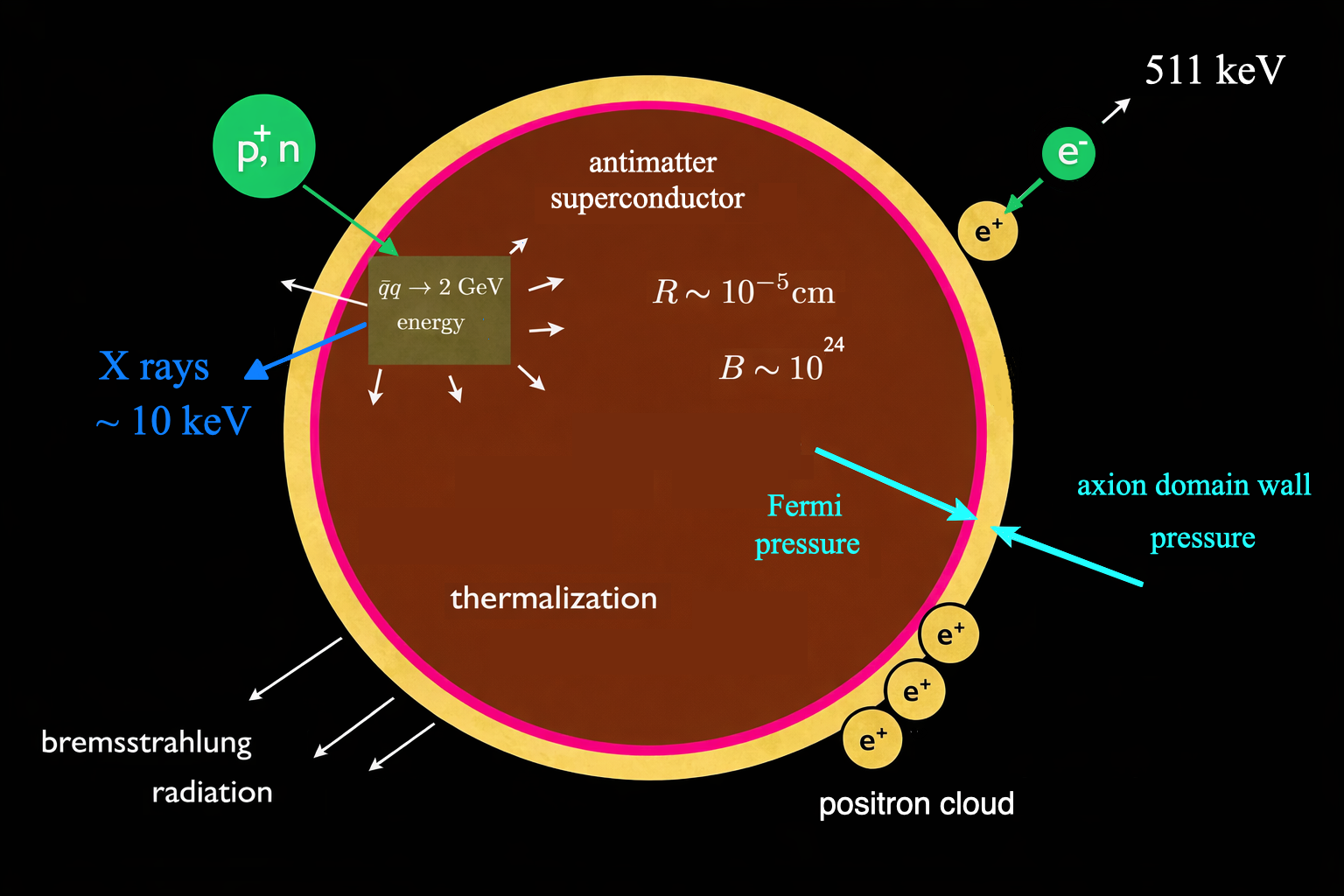}
    \caption{Summary of the structure and emission sources of an $\bar{\rm {AQN}}$. From \cite{2013arXiv1305.6318L}.}
    \label{fig:AQN_summary}
\end{figure}

The collision of an $\bar{\rm {AQN}}$ with a proton might lead to an annihilation releasing $\sim 2\,{\rm GeV}$ in the system (see Fig.~\ref{fig:AQN_B_collision}). A fraction of this energy is promptly emitted as X-ray at the point of impact. A dominant fraction is deposited in the positron cloud and subsequently radiated away as a thermal Bremsstrahlung emission. A negligible fraction, which is ignored, is emitted in the form of axions, neutrinos and other subatomic particles.
Fig.~\ref{fig:AQN_summary} summarises the structure and emission sources of an $\bar{\rm {AQN}}$. 

\subsubsection{$\bar{\rm {AQN}}$-baryon capture cross-section in dilute media}

The $\bar{\rm {AQN}}$ temperature   $T_{\rm \bar{\rm {AQN}}}$ is defined as the thermal temperature of the positron cloud. In the absence of $\bar{\rm {AQN}}$-baryon collision:
\begin{itemize}[noitemsep, topsep=0pt]
 \item no energy is injected in the system and $T_{\rm \bar{\rm {AQN}}}=0$;
    \item the positron gas in the $\bar{\rm {AQN}}$ core is fully degenerate;
    \item no radiation is emitted;
    \item the $\bar{\rm {AQN}}$ is electrically neutral: $Q=0$.
\end{itemize}
In a dilute medium, when a baryon collides with an $\bar{\rm {AQN}}$, the capture cross-section depends on the ionisation state of the gas.

 \begin{itemize}
     \item {\bf Geometrical cross-section in a predomintaly neutral  medium}
 \end{itemize}

 When ordinary matter collides with an $\bar{\rm {AQN}}$, annihilation deposits energy into the $\bar{\rm {AQN}}$ core. A fraction of this energy is thermalised, increasing the core temperature and heating the surrounding positron cloud.

Two effects follow.  First, the hotter positron cloud emits optically thin bremsstrahlung radiation. Second, a fraction of the least tightly bound positrons in the Boltzmann layer can escape the system. As a result, the $\bar{\rm {AQN}}$ acquires a net electric charge $Q<0$, whose value depends on $T_{\rm \bar{\rm {AQN}}}$. 

If the medium is predominantly neutral, the $\bar{\rm AQN}$s electric field has no effect on the capture rate of baryons and the effective capture cross-section is not significantly enhanced. As a result, although the $\bar{\rm {AQN}}$ heats and becomes mildly charged, its interaction rate remains controlled by the geometrical cross-section 

\begin{equation}
\sigma_{\mathrm{geo}}=\pi R_{\rm \bar{\rm {AQN}}}^2.
\end{equation}

\begin{itemize}
    \item {\bf{Effective capture cross-section in an ionised medium} }
\end{itemize}

If the medium is ionised, the charged $\bar{\rm {AQN}}$  exerts a Coulomb force that attracts and captures additional baryons. These subsequently annihilate, depositing energy into the $\bar{\rm {AQN}}$ core and  heating the system further. 

This enhanced heating increases the radiative output and drives the charge $Q$ to more negative values. A thermal equilibrium is eventually established, determined by the balance between the cooling timescale, the collision rate, the effective impact parameter, and the annihilation timescale of captured baryons. Upon reaching thermal equilibrium, the effective cross-section of the charged $\bar{\rm {AQN}}$ depends on several factors related to the astrophysical environment, such as the ionization level, the gas temperature and the density of the medium. We therefore define an effective cross-section:

\begin{equation}
\sigma_{\mathrm{eff}}={\sigma_{\mathrm{geo}}}
\left(\frac{R_{\mathrm{eff}}}{R_{\mathrm{\bar{AQN}}}}\right)^{2},
\end{equation}
where $R_{\mathrm{eff}}$ is the effective radius of the effective capture volume.

\vspace{0.2cm}

\subsubsection{$\bar{\rm {AQN}}$s emission in a predominatly neutral, dilute  medium }

In the derivation that follows, we consider an $\bar{\rm {AQN}}$ evolving in a dilute medium such that its equilibrium temperature satisfies 
$T_{\bar{\rm {AQN}}} \lesssim 1\,\mathrm{eV}$. 
At such temperatures, the interaction cross-section is effectively equal to the geometrical cross-section, 
as long-range Coulomb effects associated with the $\bar{\rm {AQN}}$'s electric charge can be neglected.\footnote{Owing to its large mass, a charged AQN experiences negligible deflection in astrophysical magnetic fields. 
Its Larmor radius is typically of order tens of megaparsecs. This estimate remains valid even in extreme scenarios where the charge reaches $|Q| \sim 10^{8}\!-\!10^{9}\,e$.
Consequently, AQNs cannot be detected through synchrotron radiation.}

At thermal equilibrium, the thermal emission spectrum (in $\rm erg \,s^{-1}\,cm^{-2}\,Hz^{-1}$) of an $\bar{\rm {AQN}}$ at temperature $T_{\bar{\rm {AQN}}}$ and frequency $\nu$ is given by:

\vspace{0.2cm}

\[
    dF(\nu,T_{\bar{\rm {AQN}}})
= \frac{8\pi}{45}\,
\frac{T_{\bar{\rm {AQN}}}^3\,\alpha^{5/2}}{\pi}
\left(\frac{T_{\bar{\rm {AQN}}}}{m_e}\right)^{1/4}
\left(1+\frac{2\pi\nu}{T_{\bar{\rm {AQN}}}}\right)
h\!\left(\frac{2\pi\nu}{T_{\bar{\rm {AQN}}}}\right)
\exp\!\left(-\frac{2\pi\nu}{T_{\bar{\rm {AQN}}}}\right)d\nu,
\label{eq:thermal_emission_spectrum}
\]
where $\alpha$ is the fine structure constant, $m_e$ the electron mass and $h(x)$ is a slowly varying "Coulomb-log" type factor which results from the exact QED Bremsstrahlung calculation of the positron cloud. It is given by:

\[
h(x)=
\begin{cases}
17 - 12\,\ln\!\left(\dfrac{x}{2}\right), & x < 1, \\[6pt]
17 + 12\,\ln 2, & x \ge 1 .
\end{cases}
\]

\begin{figure}[t]
\centering
  \includegraphics[width=12 cm]{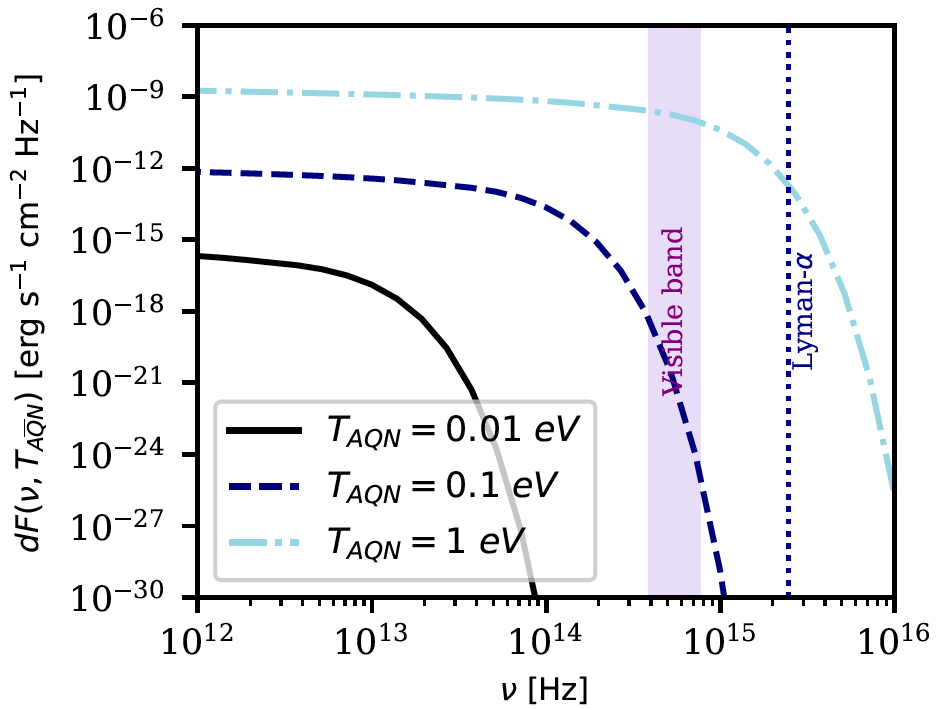}
  \vspace{-0cm}
  \captionof{figure}{$\bar{\rm {AQN}}$ thermal emission spectrum for various temperatures of the aggregates. The purple vertical band indicates the visible spectrum range and the Lyman-$\alpha$ line corresponds to $10.2\,{\rm eV}$.}
  \label{fig:AQN_Thermal_emission}
\end{figure}

\vspace{0.3cm}
Fig.~\ref{fig:AQN_Thermal_emission} shows the thermal emission spectrum $dF(\nu,T_{\bar{\rm {AQN}}})$ for several values of $T_{\bar{\rm {AQN}}}$. One notices a sharp emission drop around the visible band, which motives searches for a DM glow in the radio-visible range (see ~\ref{sec:dark_glow}).

\vspace{0.2cm}
The total bolometric flux (in $\rm erg \,s^{-1}\,cm^{-2}$) emitted by an $\bar{\rm {AQN}}$ is obtained by integrating over frequency,

\begin{equation}
F_{\rm bol}(T_{\bar{\rm {AQN}}}) = \int \, dF(\nu,T_{\bar{\rm {AQN}}}) \, .
\label{eq:Fbol}
\end{equation}

For simplicity, we now consider the collision with protons only (the generalisation to a more complex environment is straightforward). For each collision between an $\bar{\rm {AQN}}$ and a proton, the total energy transferred as heat to the positron cloud is:

\[
dE_{\rm ann} = ( 2~\mathrm{GeV}) \, f\,(1-g),
\]
where $f\sim 0.1$ is the quantum reflexion factor which accounts for the fact that only $10\%$ of the colliding protons will be annihilated while the others will bounce back. The factor $(1-g)$ represents the fraction of the annihilation energy that is thermalised within the $\bar{\rm {AQN}}$, while the remaining fraction, $g$, is emitted as X-rays from the point of impact. $2\,{\rm GeV}\simeq 2 m_p$ is the energy available for annihilation, if successful.

Over a time interval ${\rm d}t$, the $\bar{\rm {AQN}}$ will collide with a number of protons ${\rm d}N_p=\pi R_{\bar{\rm {AQN}}}^2 \, n_b \, \Delta{\rm v}\,{\rm d}t$, where $n_b$ is the number density of protons and $\Delta{\rm v}$ is the average relative speed between the $\bar{\rm {AQN}}$ and the protons. The total power (in $\rm erg \,s^{-1}$) injected into the $\bar{\rm {AQN}}$ is therefore

\begin{equation}
\frac{\mathrm{d}E_{\rm ann}}{\mathrm{d}t}
= ( 2~\mathrm{GeV}) \, f\,(1-g)\,
\pi R_{\bar{\rm {AQN}}}^2   \, \Delta {\rm v}  \, n{_\mathrm{bar}}
\label{eq:Eann_inj}
\end{equation}
The equilibrium temperature $T_{\bar{\rm {AQN}}}$ is determined by equating Eq.(\ref{eq:Eann_inj}) with Eq.(\ref{eq:Fbol}):

\begin{equation}
\frac{\mathrm{d}E_{\rm ann}}{\mathrm{d}t}
=
4\pi R_{\bar{\rm {AQN}}}^{2}\,
F_{\rm bol}(T_{\bar{\rm {AQN}}}) \, .
\label{eq:T_AQN_def}
\end{equation}
Consider a unit volume in which $\bar{\rm {AQN}}$s are characterised by a number density $n_{\bar{\rm {AQN}}}$ and a temperature $T_{\bar{\rm {AQN}}}$. 
The resulting spectral emissivity, $d\varepsilon(\nu, T_{\bar{\rm {AQN}}})$, in units of $\mathrm{erg\,s^{-1}\,Hz^{-1}\,cm^{-3}}$, is given by:

\[
\mathrm{d}\epsilon(\nu, T_{\bar{\rm {AQN}}})
=
4\pi R_{\bar{\rm {AQN}}}^{2}\,
\mathrm{d}F(\nu, T_{\bar{\rm {AQN}}})\,
n_{\bar{\rm {AQN}}} \,
\]
where $T_{\bar{\rm {AQN}}}$ is determined by the thermal equilibrium condition Eq.(\ref{eq:T_AQN_def}):

\begin{equation}
\boxed{
T_{\bar{\rm {AQN}}}
=
\left[
\frac{3\pi}{4}\,
\frac{2\,\mathrm{GeV}}{16\,\alpha^{5/2}}\,
f\,(1-g)\,m_e^{1/4}\,\Delta {\rm v}\,n_b
\right]^{4/17}
}
\label{eq:T_AQN}
\end{equation}

Since the aggregate temperature depends on the local baryon density $n_b$, the emissivity scales as:
\begin{equation}
\mathrm{d}\epsilon(\nu, T_{\bar{\rm {AQN}}})
\;\propto\;
n_{\bar{\rm {AQN}}}\, n_{\rm b}^{13/17} 
\label{eq: emissivity_scale_no_ions}
\end{equation}

\vspace{0.2cm}
{\bf Key takeaway.} An emissivity scaling with both $n_{\bar{\rm {AQN}}}$ and $n_b$ is unique to the QCD-AQN framework. 
Standard DM models do not predict emission that correlates with baryons. 
Decaying DM produces a signal proportional to $n_X$, and annihilating DM scales as $n_X^2$ : 
neither mechanism intrinsically follows the baryonic distribution.

\subsubsection{$\bar{\rm {AQN}}$s emission in an ionised dilute medium}
\label{sec:ion_dilute}

As discussed above, when $T_{\bar{\rm {AQN}}} > 0$, the $\bar{\rm {AQN}}$ acquires a net electric charge, $Q < 0$. 
If $|Q|$ exceeds a certain threshold and the surrounding medium is ionised, 
Eq.(\ref{eq:T_AQN}) no longer applies. 
In this regime, the long-range Coulomb attraction must be taken into account. 
This enhances the effective interaction cross-section, increases the baryon capture rate,  and consequently leads to a higher annihilation rate and a larger equilibrium temperature $T_{\bar{\rm {AQN}}}$. The equilibrium will still be reached, but it now explicitly depends on the gas temperature $T_{\rm gas}$. The reason is that the impact cross-section of the Coulomb interaction depends on the speed of the incoming baryon: for a given impact parameter, faster baryons will be less likely to be captured by the charged $\bar{\rm {AQN}}$ than slower baryons. Therefore, {\bf a higher $T_{\rm gas}$ will not heat an $\bar{\rm {AQN}}$ as much as a lower $T_{\rm gas}$}, everything else being equal. Taking this effect into account, the $\bar{\rm {AQN}}$ temperature is given by \cite{2024JCAP...09..045M}:

\begin{equation}
\boxed{
T_{\bar{\rm {AQN}}}
=
\left[
\frac{3\pi}{4}\,
\frac{2\,\mathrm{GeV}}{16\,\alpha^{5/2}}\,
f\,(1-g)\,m_e^{1/4}\,\Delta {\rm v}\,n_{\rm ion}
\left(\frac{\sigma_{\mathrm{eff}}}{\sigma_{\mathrm{geo}}}\right)
\right]^{4/17},
}
\label{eq:temperature_AQN_ionised_medium}
\end{equation}
where $n_b$ has been replaced by $n_{\rm ion}$, the number density of ions. This equation is similar to Eq.(\ref{eq:T_AQN}), but with an enhancement factor set by the ratio of the effective to geometrical cross-sections:

\begin{equation}
\frac{\sigma_{\mathrm{eff}}}{\sigma_{\mathrm{geo}}}
= \left(\frac{R_{\mathrm{eff}}}{R_{\bar{\rm {AQN}}}}\right)^{2} =
8\,\alpha\,m_e^{2}\,R_{\bar{\rm {AQN}}}^{2}
\left(\frac{T_{\bar{\rm {AQN}}}}{T_{\mathrm{gas}}}\right)^{2}
\sqrt{\frac{T_{\bar{\rm {AQN}}}}{m_e}} \, .
\label{eq:ratio_of_sigmaeff_sigmageo}
\end{equation}

By definition, the ratio $\sigma_{\rm eff}/\sigma_{\rm geo}$ cannot be smaller than unity 
(with the exception of negatively charged ions in the medium, for which the Coulomb interaction would be repulsive). 
In practice, this means that whenever the gas temperature $T_{\rm gas}$ implies 
$\sigma_{\rm eff} < \sigma_{\rm geo}$, the geometrical cross-section must be adopted.

Eq.(\ref{eq:temperature_AQN_ionised_medium}) and Eq.( \ref{eq:ratio_of_sigmaeff_sigmageo}) are not valid in dense or semi-dense environments.  In these regimes, ion capture becomes efficient: large numbers of ions accumulate around the $\bar{\rm {AQN}}$, partially shielding its electric charge and consequently reducing $|Q|$.

\medskip
{\bf Key takeaways.}  In a fully ionised medium, the emission properties of $\bar{\rm {AQN}}$s are determined by a
\textbf{self-consistent feedback loop} involving charge build-up, baryon capture
and annihilation, and cooling via thermal emission. The resulting equilibrium temperature $T_{\bar{\rm {AQN}}}$ is given by Eq.~(\ref{eq:temperature_AQN_ionised_medium}).
In a partially ionised medium, the temperature must be computed from a combination of Eqs.~(\ref{eq:T_AQN}) and (\ref{eq:temperature_AQN_ionised_medium}).

\medskip
In the following three subsections, we use hydrodynamical simulations to evaluate the $\bar{\rm {AQN}}$ emission in two distinct environments, large-scale structures and the Milky Way, dilute enough for Eq.(\ref{eq:T_AQN}) and Eq.(\ref{eq:temperature_AQN_ionised_medium}) to apply. 

\begin{figure}[t]
\centering
  \includegraphics[width=14 cm]{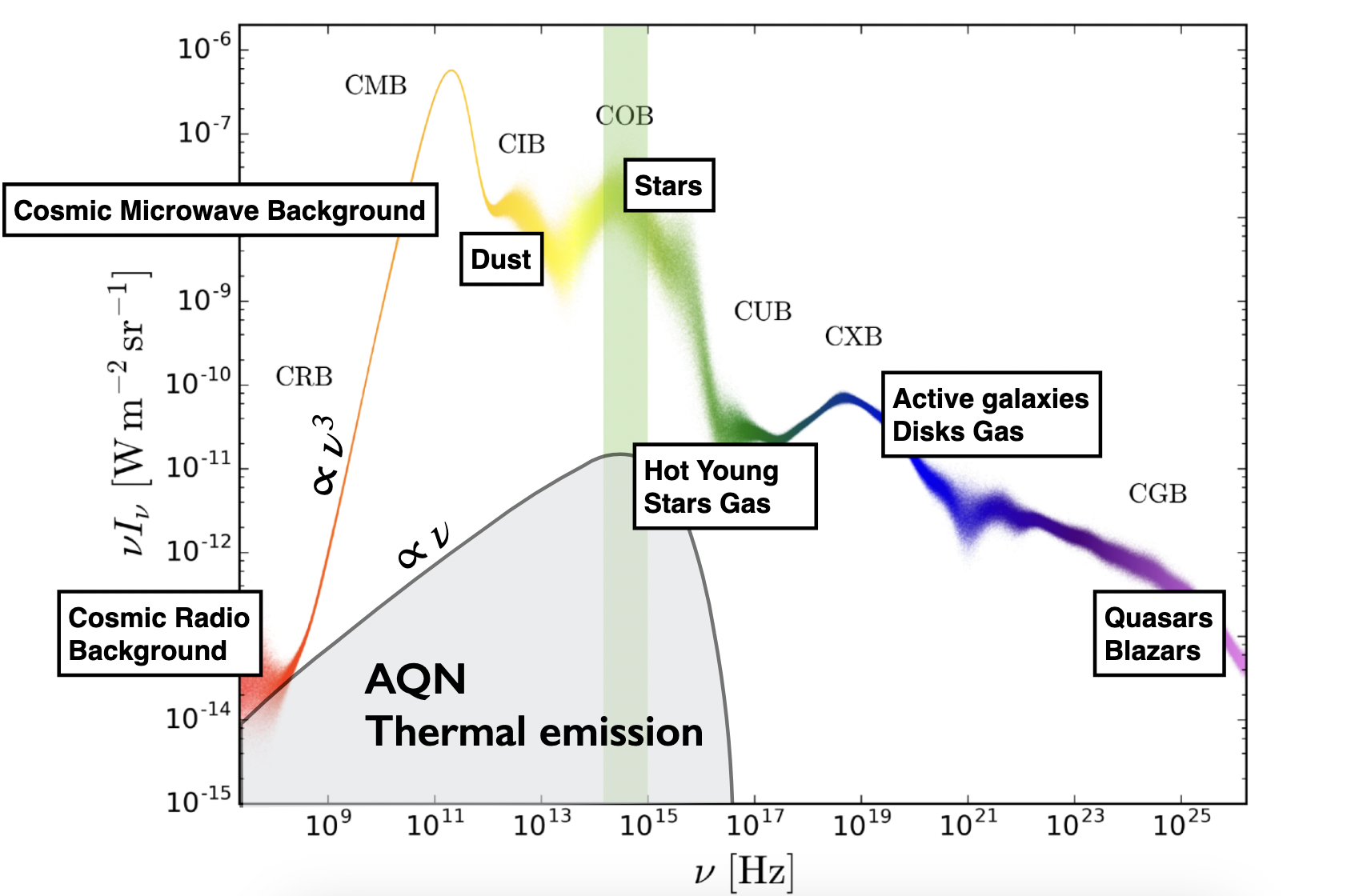}
  \vspace{-0cm}
  \captionof{figure}{Comparison of the $\bar{\rm {AQN}}$ thermal emission calculated for $\langle m_{\bar{\rm {AQN}}} \rangle = 100\,\mathrm{g}$ to the sky monopole calculated by \cite{2018ApSpe..72..663H}.}
  \label{fig:sky_monopole_with_AQN}
\end{figure}
\subsection{Observations in dilute environments}
\label{subsec:observations_dilute}

\subsubsection{$\bar{\rm {AQN}}$s contribution to the sky monopole}
\label{sec:monopole3}

Could the $\bar{\rm {AQN}}$ thermal emission contribute substantially to the sky monopole at any wavelength, potentially leading to an excess that would invalidate the QCD-AQN framework? Using the Magneticum simulation\footnote{Magneticum denotes the \textit{Magneticum Pathfinder} cosmological simulation suite, a set of large-volume hydrodynamical $\Lambda$CDM simulations including dark matter, baryons, star formation, black-hole feedback, and magnetic fields, commonly used to model galaxy and cluster formation and the baryonic content of haloes.}, \cite{2024arXiv240612122M} constructed light cones and computed the cumulative $\bar{\rm {AQN}}$ emission from $z \sim 6$ to $z = 0$. The simulation outputs include the gas temperature and the velocity fields for both DM and baryons, but the ionisation fraction is not computed. However, given that by $z \sim 7\text{–}8$ the Universe is fully reionised, and large-scale hydrogen has not subsequently recombined, it is reasonable to assume a fully ionised medium.

Fig.~\ref{fig:sky_monopole_with_AQN} presents the sky-averaged thermal emission from $\bar{\rm {AQN}}$s, compared with the measured sky monopole. The sole free parameter of the framework, the mean $\bar{\rm {AQN}}$ mass, $\langle m_{\bar{\rm {AQN}}} \rangle$, was set to $100\,\mathrm{g}$. We see that the $\bar{\rm {AQN}}$ thermal component contributes a non-negligible fraction of the sky monopole only at very low frequencies, $\nu \lesssim 1\,\mathrm{GHz}$, where the ARCADE 2 excess was reported (\cite{2011ApJ...734....5F}). Across the remainder of the electromagnetic spectrum, the contribution remains subdominant. It should be noted that increasing $\langle m_{\bar{\rm {AQN}}} \rangle$ would enhance the $\bar{\rm {AQN}}$ overall emissivity. Consequently, the observed sky intensity imposes an upper bound on $\langle m_{\bar{\rm {AQN}}} \rangle$, as discussed in the following subsection.

\subsubsection{$\bar{\rm {AQN}}$s and South Pole Telescope observations}
\label{sec:SPT_AQN}

The sky monopole serves as a consistency check to ensure that any DM emission does not exceed current observational limits, but the anisotropies of the mass distribution provide a far more informative testing ground: Model predictions can be confronted with data through their dependence on redshift, physical scale, environment, and observing frequency.

The QCD-AQN framework predicts an emission signal that correlates with both dark matter and baryons, and can therefore be tested using cosmological surveys that map the sky at specific wavelengths. At present, the most sensitive high-resolution, wide-area maps suitable for this purpose are found in the radio and far-infrared bands, notably from the South Pole Telescope (SPT) \cite{2023PhRvD.108b3510B}, the Atacama Cosmology Telescope \cite{2024ApJ...962..113M} and Planck \cite{2020A&A...641A...6P}.

\begin{figure}[t]
\centering
  \includegraphics[width=13 cm]{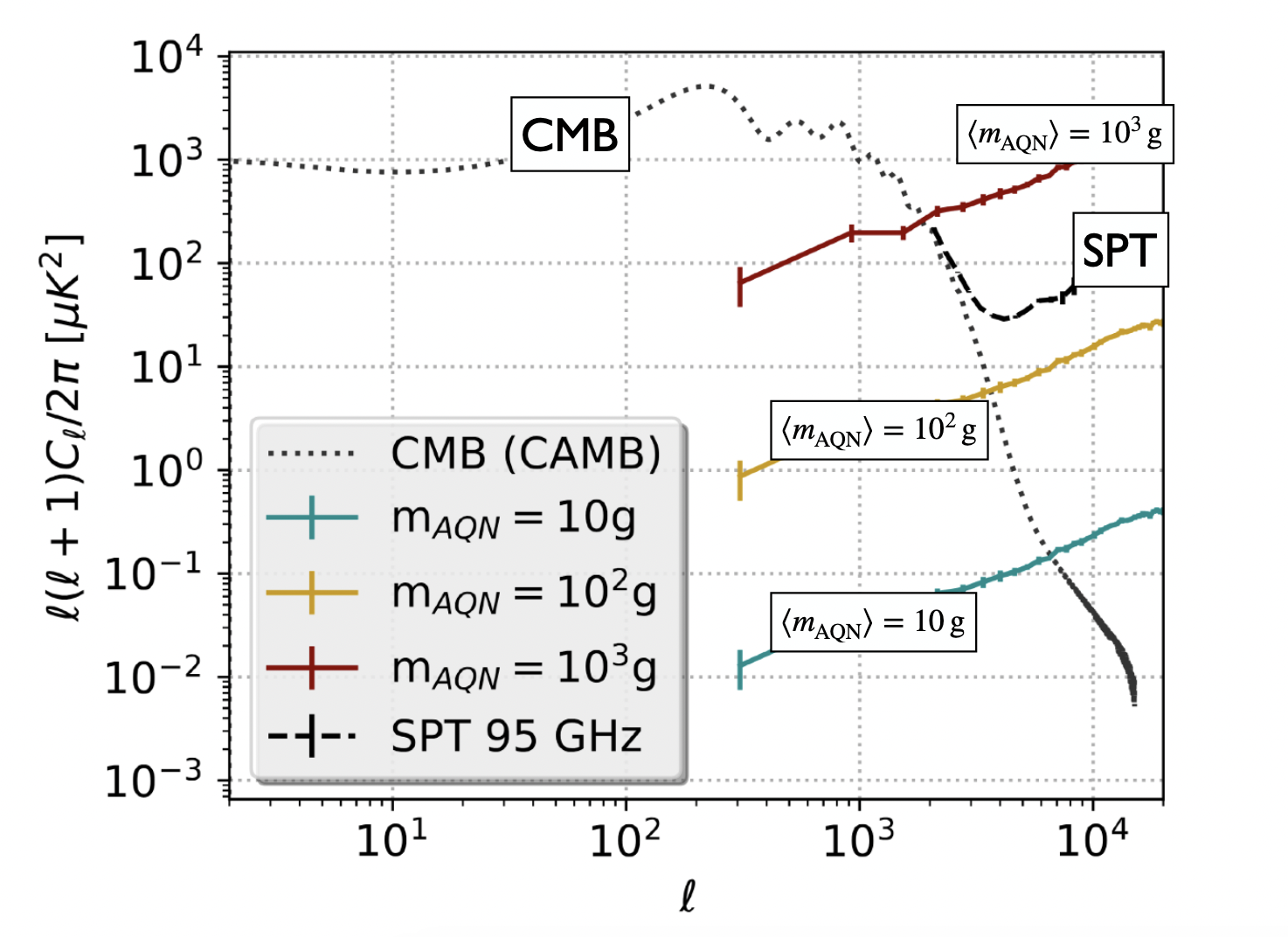}
  \vspace{-0cm}
  \captionof{figure}{$\bar{\rm {AQN}}$ emission fluctuations compared to the South Pole Telescope measurements for various values of the mean $\bar{\rm {AQN}}$ mass \cite{2024arXiv240612122M}.}
  \label{fig:SPT_upper_bound}
\end{figure}

Using the same light-cone simulations as in (\ref{sec:monopole3}), \cite{2024arXiv240612122M} computed the $\bar{\rm {AQN}}$ fluctuation spectrum, $C_\ell$. Fig.~\ref{fig:SPT_upper_bound} compares the predicted angular power spectrum of millimetre-wave fluctuations from $\bar{\rm {AQN}}$s with measurements from SPT for $\langle m_{\bar{\rm {AQN}}} \rangle = [10\,{\rm g},10^2\,{\rm g}, 10^3\,{\rm g}]\,$.

Above $\ell \sim 3000$, the signal measured by SPT is dominated by conventional extragalactic and galactic foregrounds, including dusty star-forming galaxies and the thermal and kinetic Sunyaev–Zel’dovich effects \cite{2021ApJ...908..199R}. Any $\bar{\rm {AQN}}$ contribution must therefore act as a subdominant component.
Consequently, Fig.~\ref{fig:SPT_upper_bound} may be interpreted as placing an upper limit on the amplitude of $\bar{\rm {AQN}}$-induced emission fluctuations, which translates into an upper bound on the mean $\bar{\rm {AQN}}$ mass:
\begin{equation}
\langle m_{\bar{\rm {AQN}}} \rangle \lesssim 2 \times 10^{2}\ \mathrm{g} .
\label{eq:mass_SPT}
\end{equation}
Provided that the $\bar{\rm {AQN}}$ mass remains below this upper bound, the predicted $\bar{\rm {AQN}}$ signal lies systematically below the power measured by SPT across the full multipole range considered. Moreover, the upper bound $\langle m_{\bar{\rm {AQN}}} \rangle = 200\,\mathrm{g}$ does not lead to an overproduction of optical emission that would be inconsistent with the sky monopole. This is illustrated in Fig.~\ref{fig:sky_monopole_with_AQN}, where the $\bar{\rm {AQN}}$ signal is shown for $\langle m_{\bar{\rm {AQN}}} \rangle = 100\,\mathrm{g}$.
\medskip

{\bf Key takeaway.} The thermal emission associated with $\bar{\rm {AQN}}$s does \emph{not} significantly overproduce small-scale anisotropies and is fully compatible with current SPT measurements. It sets an upper limit on the mean $\bar{\rm {AQN}}$ mass compatible with observations of the all-sky mean intensity.

\subsubsection{$\bar{\rm {AQN}}$s and Galactic UV background observations}
\label{subsec: GALEX}

The ultraviolet window offers a complementary probe of the QCD-AQN framework, since temperatures in the range $T_{\bar{\rm {AQN}}} \sim 1\text{--}10\,\mathrm{eV}$ produce emission extending into the UV  (see Fig.\ref{fig:AQN_Thermal_emission}). Observationally, the NUV-FUV sky ($\sim 2800-1300\,$\AA) is largely dominated by low redshift sources $z\lesssim 0.3-0.8$, with a strong contribution from the Milky-Way and UV emitting stars. 

\begin{figure}[t]
\centering
  \includegraphics[width=12 cm]{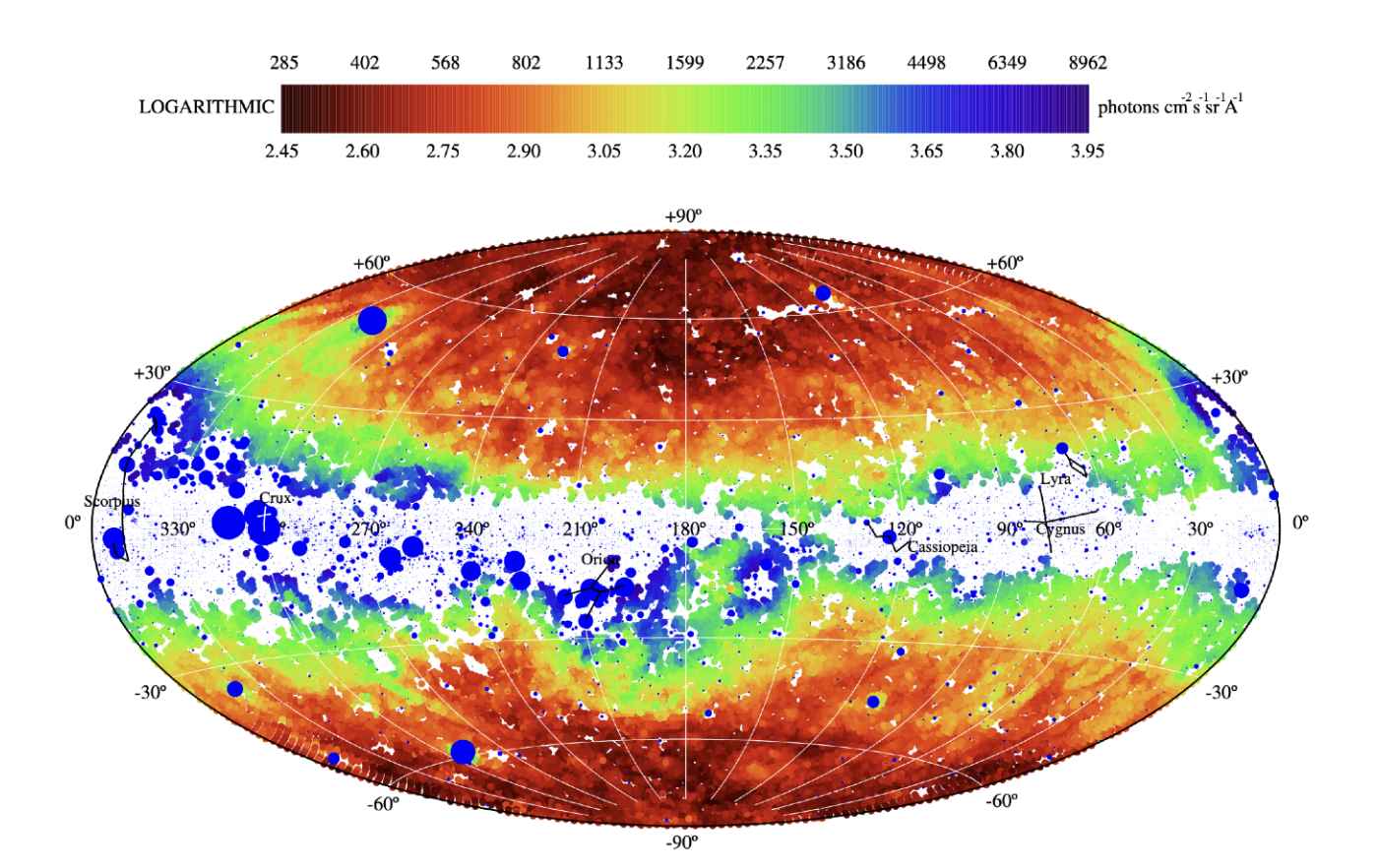}
  \vspace{-0cm}
  \captionof{figure}{FUV galactic background map measured by GALEX (from \cite{2015ApJ...798...14H}). Notice the isotropic excess at the galactic poles.    
}
  \label{fig:galactic_UV_BCKG}
\end{figure}

\cite{2015ApJ...798...14H} analysed diffuse far-ultraviolet (FUV) maps obtained with GALEX (see Fig.\ref{fig:galactic_UV_BCKG}). A Galactic dust-scattered component was modelled using the FUV–FIR ($\sim 100,\mu{\rm m}$) cross-correlation and subtracted from the GALEX maps. The remaining emission is nearly isotropic, with an intensity of $\sim 300$–$500$ photon counts. Using deep GALEX and HST galaxy surveys and the Quasars/Active Galactic Nuclei UV luminosity function, they argue that the total extragalactic component intensity is $I_{\rm ext}\sim 100-150$ photon counts. A substantial residual component, $I_{\rm excess} \sim 200\text{--}300$ photon counts, remains of unknown origin. 
It is characterised by a remarkable degree of uniformity and symmetry at high Galactic latitudes, shows no correlation with Galactic dust nor stellar UV sources and is present in both the northern and southern Galactic poles.
To evaluate a dark matter origin for the excess, \cite{2015ApJ...798...14H} calculated the UV emission expected from WIMPs and PBHs, and found that the predicted intensity falls short of the observed signal by roughly ten orders of magnitude.

\begin{figure}[t]
\centering
  \includegraphics[width=12 cm]{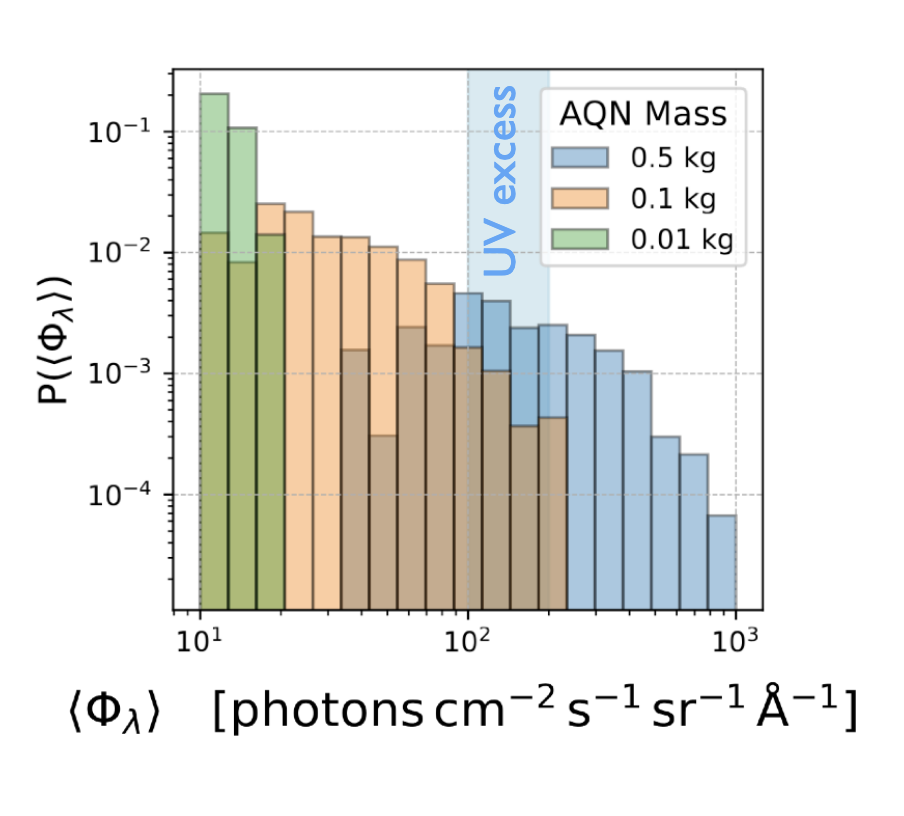}
  \vspace{-0cm}
  \captionof{figure}{Probability distributions $P(\langle \Phi_\lambda \rangle)$ of the $\bar{\rm AQN}$ FUV signal from voxels 
surrounding the candidate solar neighbourhood regions for three values of the $\bar{\rm AQN}$ masse, with 
$\Delta v$ modelled with a Maxwell--Boltzmann distribution \cite{2025arXiv250415382S}.}
  \label{fig:Sekatchev_mass_bound}
\end{figure}

In the Solar System environment, FUV photons have a relatively short mean free path, $L \sim 0.5\text{--}1\,\mathrm{kpc}$. The FUV band is therefore particularly well suited to testing the QCD-AQN framework within the $\sim \mathrm{kpc}^3$ volume surrounding the Sun, where the distributions of DM and baryons are reasonably well constrained. \cite{2025arXiv250415382S} computed the FUV intensity produced by baryon–$\bar{\rm {AQN}}$ annihilation within a region of radius $ \sim 1\,\mathrm{kpc} $ around the Sun. In order to obtain a realistic distribution of DM and baryons they used state-of-the-art FIRE-2 Latte simulations of individual Milky Way–like galaxies and they selected 281 sites whose DM density, baryon content, and ionisation fraction closely resemble those of the Solar System. The distribution of photon units counts for $\langle m_{\bar{\rm {AQN}}} \rangle = [10\,{\rm g},10^2\,{\rm g},5\times 10^2\,{\rm g}]\,$ is shown on Fig.\ref{fig:Sekatchev_mass_bound}. The authors demonstrated that the $\bar{\rm AQN}$ annihilation emissivity produces a FUV intensity similar to the excess observed by GALEX. Their results show that
\begin{equation}
\langle m_{\bar{AQN}} \rangle \sim 10^{1}\text{--}10^{3}\ {\rm g}
\label{eq:mass_GALEX}
\end{equation}
produces both the amplitude and the band-averaged intensity of the signal measured by \cite{2015ApJ...798...14H}. Although not constituting proof, the agreement between this mass range and those discussed in Sections 3.2.1 and 3.2.2 provides evidence for the internal consistency of the QCD-AQN framework. 

Critics of the analysis by \cite{2015ApJ...798...14H} argued that, GALEX being a low-Earth orbit instrument, the reported UV excess could arise from, or be biased by, residual systematics in the zodiacal light subtraction. However, \cite{2025AJ....169..103M} extended and refined GALEX's UV measurements with the Alice ultraviolet spectrograph aboard the New Horizons spacecraft (2014–2023). Thirty-six spectra from independent lines-of-sights were taken at $5\times 10^9\,{\rm km}$ away from the Sun, near Pluton's orbit where the zodiacal light is negligible. The New Horizons measurements are in perfect agreement with GALEX, confirming  the presence of a diffuse Galactic FUV component in excess of the predictions from standard dust-scattered light models. The detected signal exhibits only weak correlations with dust tracers and shows consistency across multiple independent sky directions.

\subsubsection{$\bar{\rm AQN}$s and CMB observations}

\cite{2025arXiv251205401M}  investigated whether energy injection from $\bar{\rm AQN}$s into the primordial plasma prior to recombination could generate CMB anisotropies or spectral distortions at a level sufficient to challenge the viability of the QCD-AQN framework. They concluded that $\bar{\rm AQN}$s do not produce any detectable signature in the TT, TE, or EE anisotropy power spectra and that $\mu$- and $y$-type distortions  remain below the sensitivity threshold of COBE/FIRAS.

A distinctive prediction of the QCD-AQN framework is that continuous annihilation results in $\mu$-type distortion exceeding the $y$-type component by a factor of  $\sim 3$. Such a $\mu$-to-$y$ ratio is not expected in other DM energy-injection scenarios. Detection of such distortions would require a next-generation CMB spectral distortion mission, such as the proposed PIXIE experiment. ESA’s Voyage 2050 programme provides a long-term strategic framework within which such missions could potentially emerge. 

\subsection{$\bar{\rm AQN}$s interactions with dense objects}
\label{subsec:section:observations_dense}

Having examined some aspects of the thermal emission of $\bar{\rm AQN}$s and the resulting diffuse glow produced through their interaction with ambient baryons in dilute media, it is now important to explore the encounters of $\bar{\rm AQN}$s with dense astrophysical bodies (stars and planets), as it offers an independent avenue for testing the AQN hypothesis beyond cosmology.

\subsubsection{Example 1: the Sun}
\label{sec:corona}

While matter AQNs cross stars unaffected, $\bar{\rm AQN}$s may imprint observable signatures in stellar plasmas. The most natural laboratory in which to search for such effects is our own Sun.

Following the initial 2017 paper by \cite{2017JCAP...10..050Z}, \cite{2018PhRvD..98j3527R} examined whether $\bar{\rm AQN}$s could contribute to the heating of the solar corona. The solar  corona emits a continuous EUV radiation that persists in quiet regions of the Sun and remains present even during the minima of the solar activity cycle. Maintaining the corona requires a sustained energy input of approximately

\[
L_{\rm corona} \sim 10^{27}\,\mathrm{erg.\,s^{-1}}.
\]

The solar corona is thought to reside near a self-organised criticality (SOC) state, in which even minor perturbations can initiate magnetic reconnection within pre-existing current sheets. However, the precise physical trigger responsible for the ubiquitous small-scale heating events is still under debate. While small-scale impulsive brightenings are routinely observed in the solar atmosphere, the even smaller nanoflares required to account for coronal heating are expected to lie below the resolution limits of current instruments.

Using shock-wave theory and without introducing additional free parameters,
\cite{2018PhRvD..98j3527R} evaluate the energy deposition resulting from $\bar{\rm AQN}$ interactions
with the solar plasma. Collisions between $\bar{\rm AQN}$s and solar baryons produce
photons, leptons, and hadrons that rapidly thermalise in the tenuous coronal
environment, generating localised heating events in the
$10^{21}$--$10^{26}\,\mathrm{erg}$ range, comparable to nanoflares.

$\bar{\rm AQN}$s enter the solar atmosphere with a typical Galactic halo velocity of 
${\rm v} \simeq 220\,\mathrm{km\,s^{-1}}$, 
which sets the impact parameter relevant for gravitational capture by the Sun. As they descend, the ambient plasma density increases, and the annihilation rate  rises with local density:

\begin{itemize} [noitemsep, topsep=0pt]
\item In very low-density regions, annihilation is inefficient.
\item
In very high-density regions (deeper layers), $\bar{\rm AQN}$s rapidly lose kinetic energy and the annihilation rate increases.
\end{itemize}

By combining the solar density profile, the DM velocity, the annihilation cross-section, and the plasma response, \cite{2018PhRvD..98j3527R} found that the peak energy deposition occurs within the transition region between the chromosphere and the corona (a narrow layer where plasma temperature abruptly rises from $\sim 10^4\,\mathrm{K}$  to $\sim 10^6\,\mathrm{K}$ ).  

The total power deposited onto the Sun depends on the DM flux and is largely independent of the detailed $\bar{\rm AQN}$ mass distribution. Interestingly, inserting independently measured DM Galactic parameters yields an energy deposition rate of $\sim 10^{27}\,{\rm erg\,s^{-1}}$, of the same order of magnitude as $L_{\rm corona}$, without tuning solar inputs. 

The annihilation of $\bar{\rm AQN}$ could thus contribute substantially to the heating of the solar corona. The advantage of this mechanism
is that it reproduces nanoflare-like statistics without invoking unresolved magnetic reconnection. It also provides a heating channel that operates even in magnetically quiet regions of
the solar atmosphere. 

\textbf{Path Forward} A first-principles MHD simulation demonstrating that $\bar{\rm AQN}$ injection can initiate magnetic reconnection is still required to establish $\bar{\rm AQN}$s as a viable nanoflare trigger mechanism.

The observed nanoflare energy distribution follows
\[
P(E) \propto E^{-\alpha},
\]
with $\alpha \sim 1.5$--$2$, consistent with SOC phenomenology. In such systems, the avalanche slope is largely insensitive to the microscopic origin of the trigger.  Here, the critical observable is not the precise value of $\alpha$, but instead the event rate and temporal statistics, which can be extracted from high-cadence EUV and X-ray nanoflare catalogues (for example, from SDO/AIA \cite{2012SoPh..275....3P}).

\subsubsection{Example 2 : the Earth }
\label{sec:seasonal_variations}

The terrestrial environment provides the next natural setting in which to search for $\bar{\rm AQN}$ signatures. Adopting a local DM density in the Solar System of $\rho_{\rm DM} \simeq 0.3\,\mathrm{GeV\,cm^{-3}}$, and assuming a mean mass $ \langle m_{\bar{AQN}} \rangle \sim 10\,\mathrm{g} $, one can estimate the terrestrial encounter rate. Under these assumptions, approximately $4 \times 10^{7}$ $\bar{\rm AQN}$s intersect the Earth’s surface each year. This corresponds to an average flux of roughly $\sim 10$ $\bar{\rm AQN}$s per $3{,}000\,\mathrm{km}^{2}$ per year, with characteristic velocities of order $v \sim 220\,\mathrm{km\,s^{-1}}$. An $\bar{\rm AQN}$ traversing the Earth is heated through interactions with terrestrial matter \cite{2022PDU....3601031G}, emerges largely intact and exits the Earth as a hot $\bar{\rm AQN}$ ($T_{\bar{AQN}}\sim 100-200\,{\rm keV}$), subsequently emitting X-rays in the $\sim 1-10\,\mathrm{keV}$ range over distances of up to $\sim 10\,R_\oplus$. At ground level, the associated radiation dose would be less hazardous than that of a typical medical CT scan.

Seasonal variations in the DM detection rate are a standard and important consistency check for any Earth-based detection strategy. Their study was motivated by the claim of \cite{Fraser:2014wja} that XMM-Newton observations of the diffuse X-ray background in the $6\,\mathrm{keV}$ band exhibit a statistically significant annual modulation of $\sim 20\%$, which exceeds the typical modulation of $\sim 5-10\%$ expected by direct DM detection experiments. 

\cite{2022PDU....3601031G} investigated the seasonal modulation of near-Earth X-ray emission arising from the orbital motion of the Earth around the Sun, while the Solar System moves through the Galaxy (at $\sim 220\,{\rm km\,s^{-1}}$). They found that the X-rays most relevant to an orbiting observatory are produced near the “exit” point of an $\bar{\rm AQN}$ trajectory after it has traversed the Earth. The resulting emission is therefore associated with $\bar{\rm AQN}$ Earth-transit and depends on the changing orientation of the Earth relative to the incoming DM flow. Because this mechanism relies not only on the seasonal variations in the Earth’s speed through the Galactic halo, but also on directional effects, the predicted annual modulation can be substantially larger than that expected in conventional particle DM models. \cite{2022PDU....3601031G} estimated that the resulting seasonal variation could reach the $ \sim 20
 \text{--} 25\% $ level, an amplitude broadly compatible with the modulation reported in the XMM-Newton analysis. They further argued that the radiation spectrum produced by $\bar{\rm AQN}$ interactions during transit through the Earth can account for emission in the $ 6\,\mathrm{keV} $ band without the need for extreme fine-tuning, hence presented this signal as a generic consequence of $\bar{\rm AQN}$–matter interactions\footnote{No variations associated with the Earth’s seasons are expected when measurements are conducted far from the Earth and with cameras that never point toward the planet. This constitutes a distinctive feature of the QCD-AQN framework, as the $\rm A\bar QN$'s emission can transmit information about the Earth’s seasonal cycle to distances as large as $\sim 10\,{\rm R_{Earth}}$.}.

\cite{2019PhRvD.100d3531L,2020PhRvD.101d3512L} examined the dynamical, thermal, and observational consequences of an $\bar{\rm AQN}$ traversing the Earth. They showed that $\bar{\rm AQN}$s can pass through the planet without undergoing catastrophic disruption, thereby demonstrating the effective transparency of the Earth to such composite objects. Their analysis indicates that no large-scale melting, shock formation, or structural damage is produced, explaining why $\bar{\rm AQN}$s are not excluded by existing geological constraints. \cite{2022PDU....3601031G} further argued that, under favourable conditions, current seismic networks could in principle be sensitive to the passage of $\bar{\rm AQN}$s, offering a potential -- albeit challenging -- avenue for direct detection.
\medskip

{\bf Key takeaways.} The Earth should be continuously crossed by AQNs, producing X-ray with a large seasonal modulation, while remaining consistent with geological and observational constraints, making the terrestrial environment a viable — though challenging — detection channel.

\subsubsection{Example 3: neutron-stars}
\label{sec:magnetars}

Neutron stars (NSs) are compact stellar remnants with masses of approximately $1.2\text{–}2\,M_\odot$ and radii $R_{\rm NS} \sim 10\text{–}15\,\mathrm{km}$. The magnetic field strength of an ordinary neutron star is typically ${\cal B} \simeq 10^{10}\text{–}10^{13}\,\mathrm{G}$, whereas magnetars host the strongest magnetic fields known in the Universe, reaching ${\cal B} \simeq 10^{14}\text{–}10^{15}\,\mathrm{G}$.
Given the enormous energy stored in such fields, an important question is what impact infalling AQNs might have on these systems.

In the same manner as the Sun, a neutron star will gravitationally capture AQNs, but it will do it with a large impact parameter $b_{\rm cap}$:

\[
\frac{b_{\rm cap}}{R_{\rm NS}}\simeq \frac{c}{v_\infty}
\left( \frac{10^{6}\,\mathrm{cm}}{R_{\rm NS}} \right)^{1/2}
\left( \frac{M_{\rm NS}}{2\,M_\odot} \right)^{1/2} \gg 1.
\]
A number of AQNs comparable to the number accreted by the Sun will be captured and fall onto a single NS. This is a substantial flux of particles, particularly in view of the very small physical size of a neutron star.

\cite{2019PhRvD..99d3535V} analyse the interaction of AQNs with neutron-star magnetospheres. Using simple dynamical arguments and elementary shock wave physics in plasma, they show that both matter and antimatter AQNs can considerably speed up the magnetic reconnection in a NS magnetosphere. The argument is based on the fact that AQNs will reach the speed of light, $\rm v_{AQN}\sim c$, larger than the speed of sound, leading to a Mach number of the order:

\[
M\simeq 20\,
\left( \frac{1\,\mathrm{MeV}}{k_B T} \right)^{1/2}
\left( \frac{10^{6}\,\mathrm{cm}}{R_{\rm NS}} \right)^{1/2}
\left( \frac{M_{\rm NS}}{2\,M_\odot} \right)^{1/2}.
\]
Consequently, the matter flow can reach  ${\rm v}\sim c$ locally, leading to an Alfvén Mach number $M_A\sim 1$. This will considerably accelerate the magnetic reconnection and release vast amounts of energy. From this key point,  \cite{2019PhRvD..99d3535V} calculated the amount of energy released by reconnection events triggered by the shock wave generated by AQNs falling on a magnetar. They found that it is consistent with the coherent plasma radiation mechanism proposed by \cite{2017MNRAS.468.2726K,2018MNRAS.477.2470L} that could be at the origin of Fast Radio Bursts (FRBs).

The appeal of this work lies in the absence of ad hoc parameter adjustment: the burst frequency, duration, and energy scale are set by independently motivated quantities (axion mass, DM density, magnetar magnetic field strength and magnetar abundance). Moreover, the prediction that FRBs should originate preferentially in strongly magnetised neutron stars preceded and is consistent with subsequent observational evidence linking FRBs to magnetars \cite{2020Natur.587...59B}. \cite{2024PhRvD.109f3018Z} discussed other proposals that could link AQNs to NS.

\subsubsection{Additional constrains on the mean $\bar{\rm AQN}$ mass }
\label{additional_constrains}

{\bf IceCube constrains}

Considering the extremely low number density of $\bar{\rm AQN}$s, direct detection is exceedingly unlikely. IceCube’s non-detection of $\bar{\rm AQN}$-like events implies that $\bar{\rm AQN}$s must be sufficiently massive for their flux through the detector to be extremely small \cite{2019PhRvD.100d3531L}. This places a lower bound on the $\bar{\rm AQN}$ mean mass:
\begin{equation}
\langle m_{\bar{AQN}}\rangle  \gtrsim 5\,\mathrm{g},
\label{eq:ICECUBE}
\end{equation}
For smaller masses, $\bar{\rm AQN}$s would be too numerous and would deposit enough energy through annihilation, ionisation, and secondary particle production to generate detectable, highly penetrating tracks. 

\vspace{0.2cm}

{\bf Geothermal constrains}

The Earth’s geothermal heat flow is well measured, which strongly constrains any dark-matter population that deposits energy while passing through or being captured by the Earth. As a result, geothermal data place robust limits on the mass, interaction cross-section, and flux of macroscopic DM candidates. According to \cite{2012PhRvD..86l3005G}, macros can contribute no more than $\sim$ a quarter of Earth’s heat flow, impling that, for an AQN population making up the local DM density:  

\begin{equation}
\langle m_{\bar{AQN}}\rangle \gtrsim 4\,\mathrm{g}.
\label{eq:geothermal}
\end{equation}

{\bf Antartic Impulsive Transient Antenna}

The Antartic Impulsive Transient Antenna (ANITA) is a long-duration balloon experiment that flies above Antarctica to detect nanosecond-scale radio pulses from ultra-high-energy particle interactions in the Antarctic ice. During its 2014 flight, the \textit{ANITA} experiment detected an $\sim 10^{18}\,\mathrm{eV}$ upward-going, cosmic-ray--like radio event emerging at roughly $30^\circ$ above the horizontal, implying a chord length through Earth of roughly 5,000–7,000 km \cite{2018PhRvL.121p1102G}. During its 2016 flight, the \textit{ANITA} experiment reported a second upward-going cosmic-ray--like radio event with an energy of $\sim 1-4 \times  10^{18}\, \mathrm{eV}$, emerging at approximately $35^\circ$ above the horizontal, corresponding to an even longer Earth-crossing distance \cite{2020arXiv200805690A}. 

These observations present a striking challenge to conventional ultra-high-energy neutrino interpretations:  At \textit{ANITA} energies ($E_\nu \sim 10^{17}\text{--}10^{19}\,\mathrm{eV}$), the neutrino--nucleon scattering cross section grows to $\sim 10^{-33}\,\mathrm{cm}^2$, so the Earth becomes effectively opaque to such neutrinos. \cite{2022PhRvD.106f3022L} interpreted the upward-going events as $\bar{\rm AQN}$s exiting the surface. The heated aggregate and its ionisation trail can generate a coherent radio pulse that mimics an upward-going air shower. A key distinction is that, unlike a point-like neutrino, an $\bar{\rm AQN}$ deposits energy continuously along a track, leading to characteristic polarisation, duration, and angular features. 

{\bf $\bar{\rm AQN}$ mass bound} 

A mean mass $\langle m_{\bar{AQN}} \rangle
 \gtrsim   1\,\text{-} 10 \, \mathrm{g}$ is required for  an $\bar{\rm AQN}$ to survive propagation through a substantial chord of the Earth, and for the emerging shower to match \textit{ANITA}’s observed energy scale. On the other hand,  for $\langle m_{\bar{AQN}} \rangle
 \gg 100\,\mathrm{g}$, the flux would be too small to explain two \textit{ANITA} events within the exposure time, leading to an approximate mass window:

\begin{equation}
1 \text{-} 10 \,  \mathrm{g} \lesssim \langle m_{\bar{AQN}} \rangle
 \lesssim 1000\,\mathrm{g}
 \label{eq:ANITA}
\end{equation}
This window is consistent with mass ranges that have been derived previously from independent considerations.

\subsubsection{Synthesis of observational constraints}

\label{sec:synthesis_mass}
To summarise \ref{subsec:observations_dilute}  and \ref{subsec:section:observations_dense}, Tables~\ref{tab:AQN_mass_constraints} and~\ref{tab:AQN_energy_injection} compile the principal mass-range constraints and the consistency of $\bar{\rm AQN}$-induced energy injection across the relevant astrophysical and terrestrial environments.

\begin{table}[h!]
\centering
\begin{tabular}{|>{\centering \arraybackslash}p{0.55\linewidth}|>{\centering\arraybackslash}p{0.25\linewidth}|>{\centering \arraybackslash}p{0.1\linewidth}|}
\hline
\textbf{Theory} & \textbf{$[m_{\min}, m_{\max}]$}  &\textbf{Ref.}\\
\hline
AQNs survival/Limited formation time  & $[10^{-2}, 10^{4}]\mathrm{g}$ & \ref{sec:formation_mass_range}\\
\hline
\textbf{Independent observations} &  {\bf Mean mass $\langle m_{\bar{AQN}} \rangle$} &\\
\hline
SPT small-scale anisotropies & $\langle m_{\bar{AQN}} \rangle \lesssim 2 \times 10^2\,\mathrm{g}$& Eq.(\ref{eq:mass_SPT})\\
GALEX \& ALICE UV background constraints &  $ 1\,{\rm g} \lesssim \langle m_{\bar{AQN}} \rangle \lesssim 10^3\,\mathrm{g}$& Eq.(\ref{eq:mass_GALEX})\\
IceCube non-detection constraints & $ 5\,\mathrm{g} \lesssim \langle m_{\bar{AQN}} \rangle $  &Eq.(\ref{eq:ICECUBE})\\
Geothermal heat flow & $ 4\,\mathrm{g} \lesssim \langle m_{\bar{AQN}} \rangle $  &Eq.(\ref{eq:geothermal})\\
ANITA up-going events (rate + energetics) & $\sim 1\text{--}10^3\,\mathrm{g}$ & Eq.(\ref{eq:ANITA})\\
\hline
\end{tabular}
\caption{Summary of observational and theoretical constraints on the mean $\bar{\rm AQN}$ mass.}
\label{tab:AQN_mass_constraints}
\end{table}

\begin{table}[h!]
\centering
\begin{tabular}{|p{5.5cm}|p{8.5cm}|c|}

\hline
\textbf{Environments} & \textbf{$\bar{\rm AQN}$ energy injection } & \textbf{Ref.} \\
\hline
Sky monopole (multi-wavelength) 
& Compatible with observed sky-averaged intensity 
& \ref{sec:monopole3} \\
CMB 
& Compatible with anisotropies and spectral distortions 
& \ref{sec:SPT_AQN} \\
Solar corona heating 
& Compatible with nanoflare energetics 
& \ref{sec:corona} \\
Magnetars (burst energetics) 
& Compatible with FRB rates and energetics 
& \ref{sec:magnetars} \\
 Seasonal variations  of the diffuse X-ray background
& Compatible with XMM-Newton observations
& \ref{sec:seasonal_variations} \\
\hline
\end{tabular}
\caption{Consistency of $\bar{\rm AQN}$-induced energy injection with selected astrophysical environments.}
\label{tab:AQN_energy_injection}
\end{table}

\subsection{Upcoming tests for the QCD-AQN framework}
\label{subsec:futuretests}

The QCD-AQN framework must be confronted with observations in any astrophysical environment where $\bar{\rm AQN}$s encounter a sufficiently high baryon density to trigger annihilation. We outline below the principal observational tests that will be pursued in the near future.

\subsubsection{511 keV emission in the Galactic bulge}

\begin{figure}[t]
    \centering
    \includegraphics[width=\linewidth]{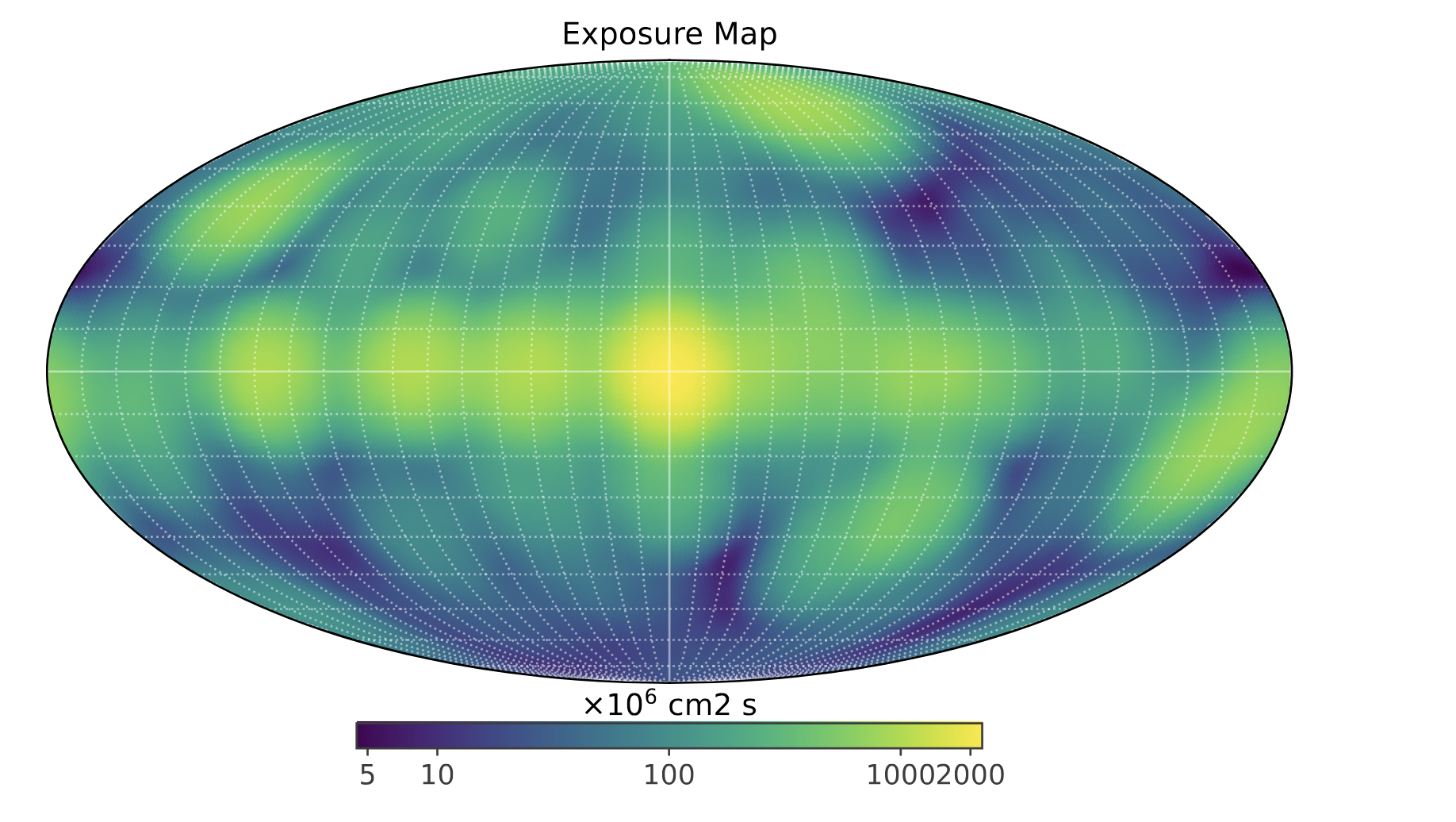}
\caption{Exposure map for the 511~keV line observations with INTEGRAL/SPI over 20 years in Galactic coordinates. Image adapted from \texttt{www.aanda.org/articles/aa/pdf/2025/10/aa55895-25.pdf}.}
    \label{fig:integral_spi_511kev_exposure_map}
\end{figure}

Observations, notably by INTEGRAL/SPI (Fig.\ref{fig:integral_spi_511kev_exposure_map}), reveal an intense and spatially extended emission from the Galactic bulge, corresponding to a positron injection rate of order $\dot N_{e^+} \sim 10^{43}\,\mathrm{s^{-1}}$. While several astrophysical sources are capable of producing positrons -- including $\beta^+$-decaying radioactive isotopes synthesised in supernovae, pulsars, microquasars, and low-mass X-ray binaries -- most of these populations trace the Galactic disc rather than the observed bulge.

As $\bar{\rm AQN}$s traverse the dense bulge environment, baryons from the interstellar medium annihilate on their surface, releasing energy that heats the surrounding positron cloud and leads to the emission of low-energy positrons. These positrons subsequently thermalise and annihilate in the interstellar medium, producing the observed 511\,keV line. This scenario was suggested by \cite{2005PhRvL..94j1301O} in order to simultaneously explain the bulge-dominated morphology of the emission and the relatively low injection energies implied by the narrow line width, while avoiding the overproduction of high-energy gamma rays.  
Another distinctive feature of the QCD-AQN framework is that the emitted signal is correlated with baryonic matter, as demonstrated by Eq. (\ref{eq: emissivity_scale_no_ions}), in contrast to annihilating or decaying DM scenarios. It remains to be seen if the framework can predict a 511\,keV intensity map -- obtained from realistic Galactic DM and baryon density distributions -- consistent with the observed spatial extent, symmetry, and surface-brightness profile of the bulge emission.

\subsubsection{Cosmic dawn and 21-cm cosmology}

The energy difference $\Delta E$ between the $1s$ hyperfine levels of neutral hydrogen defines the spin temperature $T_S$.{\footnote{$n_1$ is the number density of neutral hydrogen atoms in the excited (triplet) hyperfine level of the 1s ground state, while $n_0$ is the number density of atoms in the lower-energy (singlet) hyperfine level.}

\[
    \frac{n_1}{n_0}=\frac{g_1}{g_0}e^{-\frac{h\nu}{k_B T_S}}=3 e^{-\frac{\Delta E}{k_B T_S}}
\]

where $\Delta E$ corresponds to $\lambda=21\,{\rm cm}$ ($1420\,{\rm MHz}$). The 21-cm signal is observed either in absorption or emission against the CMB and can only be observed when $T_S \neq T_{\rm CMB}$. Throughout the history of the Universe:
\begin{itemize}[noitemsep, topsep=0pt]
    \item Between $1100 \gtrsim z \gtrsim 200$ the gas temperature remains equal to the CMB temperature because of Compton scattering between CMB photons and residual free electrons. This process keeps the baryon gas thermally coupled to the radiation even after recombination, hence $T_S= T_{\rm gas} = T_{\rm CMB}$.
    \item Between $200\gtrsim z\gtrsim 40$, the gas cools faster than the CMB but collisional coupling maintains $T_S\simeq T_{\rm gas}<T_{\rm CMB}$, hence the 21cm is seen in absorption against the CMB.
    \item As the universe continues expanding, at $z\sim 30$--$40$ collisional coupling becomes less effective than CMB photons scattering and $T_S\simeq T_{\rm CMB}>T_{\rm gas}$ : the 21cm signal disappears.
    \item Then the first stars form and inject UV and X-ray photons. The mechanism known as the Wouthuysen-Field effect (Wouthuysen 1952, Field 1958) redistributes electrons between the two hyperfine levels and results in $T_S\sim T_{\rm gas}$: the 21-cm signal is seen in absorption again.
    \item At the beginning of the reionisation era  ( $z\sim 15$) the gas temperature rises well above $T_{\rm CMB}$ and the 21-cm signal is seen in emission again.

\end{itemize}
    \smallskip
There has been no confirmed detection of the global 21-cm signal to date\footnote{The EDGES collaboration reported the detection of an unexpectedly deep 21-cm absorption feature at redshift $z \sim 17$ \cite{2018Natur.555...67B}. It is still debated.}, although it remains an area of intensive research. The interaction between $\bar{\rm AQN}$s and baryons provides several mechanisms by which the 21-cm signal could be modified, either enhancing absorption or driving emission. In particular, the injection of low-energy photons would contribute to the radio background in addition to the CMB, while higher-energy UV and X-ray photons could alter the thermal and ionisation history of the intergalactic medium. In this context, precision 21-cm cosmology may constitute a decisive empirical test of the QCD-AQN scenario.

\subsubsection{ARCADE~2 and the cosmic radio background}
\label{subsection:ARCADE}

As shown in Fig.~\ref{fig:sky_monopole_with_AQN}, the $\bar{\rm AQN}$ emission can make an appreciable contribution to the sky monopole only at low frequencies, specifically below $1 \mathrm{GHz}$. One potential origin of such low-frequency emission is the cumulative low-energy radiation produced during the cosmic dark ages.

The ARCADE 2 excess refers to an unexpectedly bright, diffuse radio background detected at frequencies below approximately $10\,\mathrm{GHz}$ \cite{2011ApJ...734....5F}. This signal is nearly isotropic, significantly exceeds the contribution anticipated from known galactic and extragalactic radio sources, and follows a smooth power-law spectrum with a steep spectral index ($T \propto \nu^{-2.6}$).

Any potential connection between the ARCADE 2 excess and a cosmic radio background generated within the QCD-AQN framework (as suggested in \cite{2013PhLB..724...17L}) must, however, be treated with caution. A substantial radio background at high redshift would fundamentally alter the reference against which the 21-cm line is observed. Rather than hydrogen interacting solely with the CMB radiation bath, it would experience a brighter effective radio background, thereby enhancing the depth of the absorption signal without modifying the underlying spin physics \cite{2018ApJ...858L..17F}. Note that this effect can be reversed by heat injection into the gas, for example through soft photons \cite{2024MNRAS.534..738C}. The gas can also be heated by ionising photons. As discussed in Section 3.4.2, $\bar{\rm AQN}$s also inject UV and X-ray photons. This additional energy input heats the gas and reduces the 21-cm absorption amplitude, potentially offsetting or even cancelling the enhanced absorption caused by an excess radio background.

If the $\bar{\rm AQN}$ radio emission is indeed related to the ARCADE excess, it would necessarily also be linked non-trivially to the global 21-cm signal. Taken together, measurements of the radio background and the 21-cm line could provide a decisive test of the QCD-AQN framework.

\subsubsection{Where and how to search for $\bar{\rm AQN}$ signatures}
\label{subsec:otherprobes}

The detectability of $\bar{\rm AQN}$-induced phenomena depends sensitively on the physical properties of the surrounding medium. Because $\bar{\rm AQN}$s interact predominantly through annihilation with baryons, the strength and character of any observable signatures are controlled by the local baryon density and the ionisation fraction of the gas. In practice, this implies that potential signatures must be sought across a wide range of environments: large-scale structure and galaxy clusters, galactic haloes, stellar and planetary atmospheres, and the interstellar medium. $\bar{\rm AQN}$s could also modify the thermal and ionisation history of the early Universe. As a continuous source of ionising photons, they may contribute during the epoch of reionisation. An additional radiation component of this type could delay the cooling of primordial gas into molecular hydrogen, thereby affecting the onset of early star formation and possibly the formation of the first black holes. 

The breadth of these potential consequences might suggest that $\bar{\rm AQN}$ signatures should be ubiquitous and therefore already detected. However, this intuition is misleading: the presence of many possible channels does not imply that each produces a large or easily identifiable signal. Not all astrophysical environments are conducive to producing detectable imprints. For instance, massive O and B stars possess radiative envelopes, and their outer atmospheres are dominated by strong, fast stellar winds. Any localised energy deposition from an $\bar{\rm AQN}$ would be rapidly advected away, preventing the build-up of a sustained coronal signature, leaving no clear $\bar{\rm AQN}$-induced observational imprint. By contrast, low- and intermediate-mass stars with convective envelopes host magnetically confined coronae. In these systems, $\bar{\rm AQN}$ energy injection could trigger magnetic reconnection events and contribute to coronal heating.

A comparatively more accessible environment is the Earth’s atmosphere, where both ground-based detectors and balloon-borne experiments (e.g. \cite{2015ExA....40....3A}) can be deployed.   Although a single 100 g $\bar{\rm AQN}$ carries an energy comparable to several megatons of TNT, even in this environment observational signatures are typically subtle. This is because a very large amount of energy would be deposited along an extended trajectory, in contrast to the case of point-like explosions. $\bar{\rm AQN}$s may interact with terrestrial thunderstorms, which are characterised by strong, large-scale electric fields capable of accelerating charged particles toward the aggregates. This enhanced baryon flux onto the $\bar{\rm AQN}$ surface would increase the annihilation rate and lead to rapid, localised energy release, potentially producing spallation-like particle cascades. \cite{2025Symm...17...79Z} argues that such interactions could generate bursts of relativistic particles, gamma rays, and radio emission, thereby offering a possible connection between $\bar{\rm AQN}$s and transient high-energy atmospheric phenomena.

\cite{2020PhRvD.101d3512L} has demonstrated that the detection of a daily modulation in the flux of high-velocity axions could serve as a signature of $\bar{\rm AQN}$s that have recently traversed the Earth. These axions are naturally generated through the interaction of the axion domain wall with terrestrial matter during transit. Searches for such daily modulations, induced by AQNs crossing the Earth, have already commenced \cite{2025PhRvD.111h2009C,2025PhRvD.112l1305K}.

Finally, searching for $\bar{\rm AQN}$ signatures frequently demands substantial expertise, significant computational resources, and the analysis of large observational datasets. One example is detailed in \cite{2024A&A...691A..38S}, who develop an observational strategy to detect $\bar{\rm AQN}$ signatures in nearby, well-characterised galaxy clusters such as Virgo and Fornax. They conclude that a  multi-wavelength approach is essential. Any candidate $\bar{\rm AQN}$ emission must be carefully separated from dominant astrophysical backgrounds, particularly the thermal bremsstrahlung from the intracluster medium in X-rays and the non-thermal synchrotron radiation from cosmic rays in radio. This requires robust component separation and consistent modelling across different frequency bands. Another example is the search for the so-called “dark glow”, which we now discuss.

\medskip

{\bf Key takeaway.} The search for observational evidence of AQNs mainly focuses on diffuse sky background maps. Signals previously discarded as noise in past observations could contain valuable information, and may warrant renewed attention.

\subsection{Prediction: the dark glow}
\label{sec:dark_glow}

Figure~\ref{fig:sky_monopole_with_AQN} demonstrates that the $\bar{\rm AQN}$ signal can generate a diffuse uniform glow across the sky, potentially extending into the UV.  One therefore expects that the $\bar{\rm AQN}$ emission should display anisotropies over a broad range of frequencies, from the radio to the UV. Fig.~\ref{fig:SPT_upper_bound} shows the anisotropy spectrum, $C_\ell$, at $\nu = 95\,\mathrm{GHz}$ for three different values of the mean $\bar{\rm AQN}$ mass. The principal challenge is that the predicted signal, the so called dark glow, is extremely faint, and therefore these fluctuations cannot be detected individually.

\begin{figure}[ht]
\centering
  \includegraphics[width=14 cm]{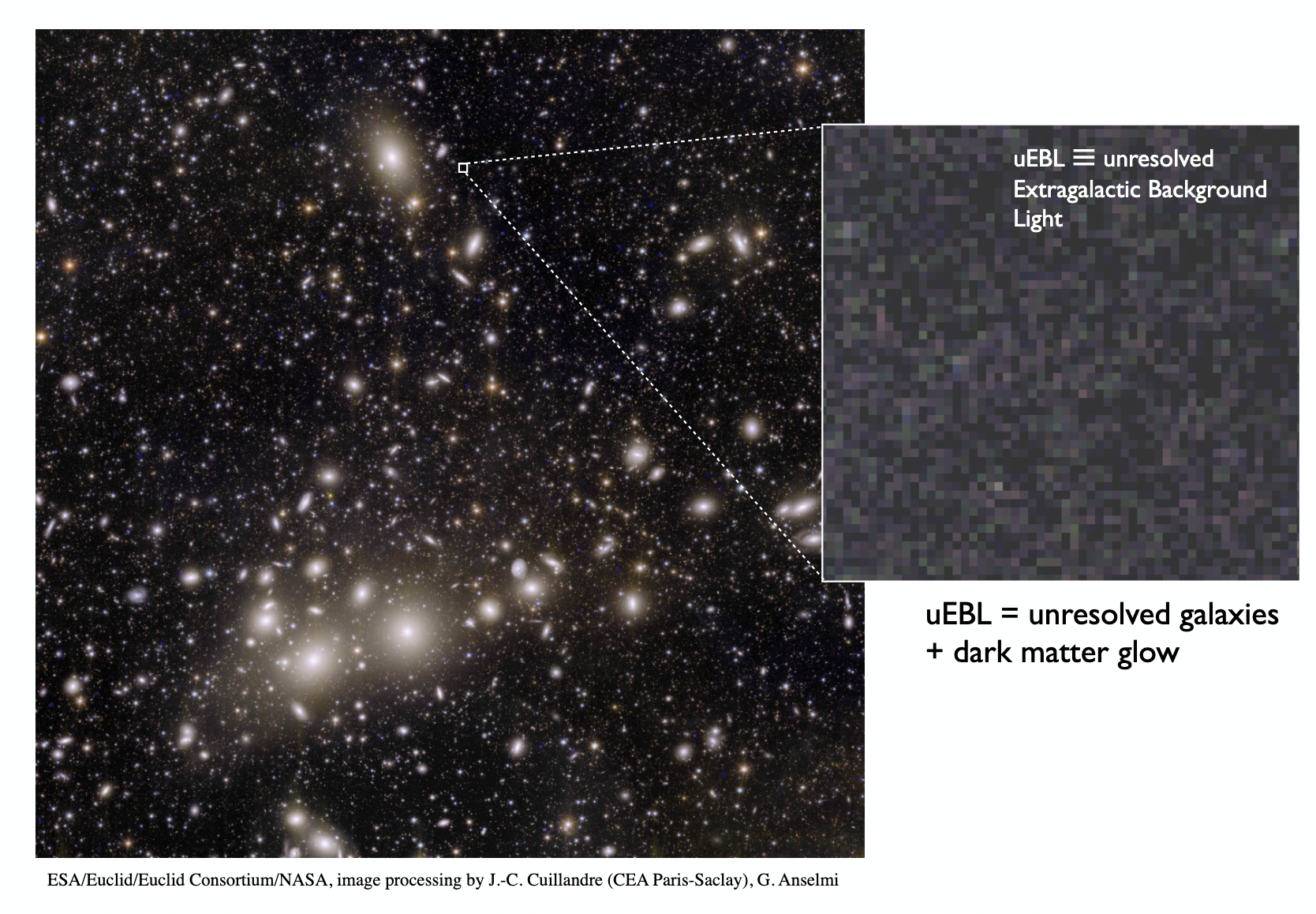}
  \vspace{-0cm}
  \captionof{figure}{(Left) Euclid image of the Extragalatic Background Light (EBL). (Right) Zoom in on the unresolved EBL (uEBL), obtained after removing  resolved sources. The uEBL consist of unresolved galaxies and the dark-matter glow.}
  \label{fig:uEBL.png}
\end{figure}

The new space observatories Euclid and James Webb Space Telescope provide an exceptionally stable environment for ultra-faint photometry, enabling exploration of the extremely low surface-brightness Universe.
Fig.~\ref{fig:uEBL.png} displays an image of the Perseus cluster obtained with the Euclid space telescope, with an inset showing a magnified region of the background sky where no discrete sources are detected. Thanks to the stability of space-based imaging, it may be possible to search for fluctuations in the sky background that lie below the nominal sky level itself.

\cite{2024arXiv240612122M} demonstrated that extending the SPT anisotropy spectrum to optical wavelengths yields a signal that lies within the nominal detection threshold of Euclid \cite{2025A&A...697A..11C}.
The grey region in Fig.~\ref{fig:dark_glow_detection_threshold} represents the expected $\bar{\rm AQN}$ signal at $\ell = 3000$ as a function of frequency, for $\langle m_{\bar{AQN}} \rangle$ in the range $10$–$100$ g. The coloured rectangles denote the projected $1\,\sigma$ uncertainties achievable with SPT-3G, COSMOS-Web, and the Euclid Wide Survey.
It can be seen that the predicted root-mean-square dark glow amplitude falls within the anticipated sensitivity of each of these instruments.

\begin{figure}[ht]
\centering
  \includegraphics[width=13 cm]{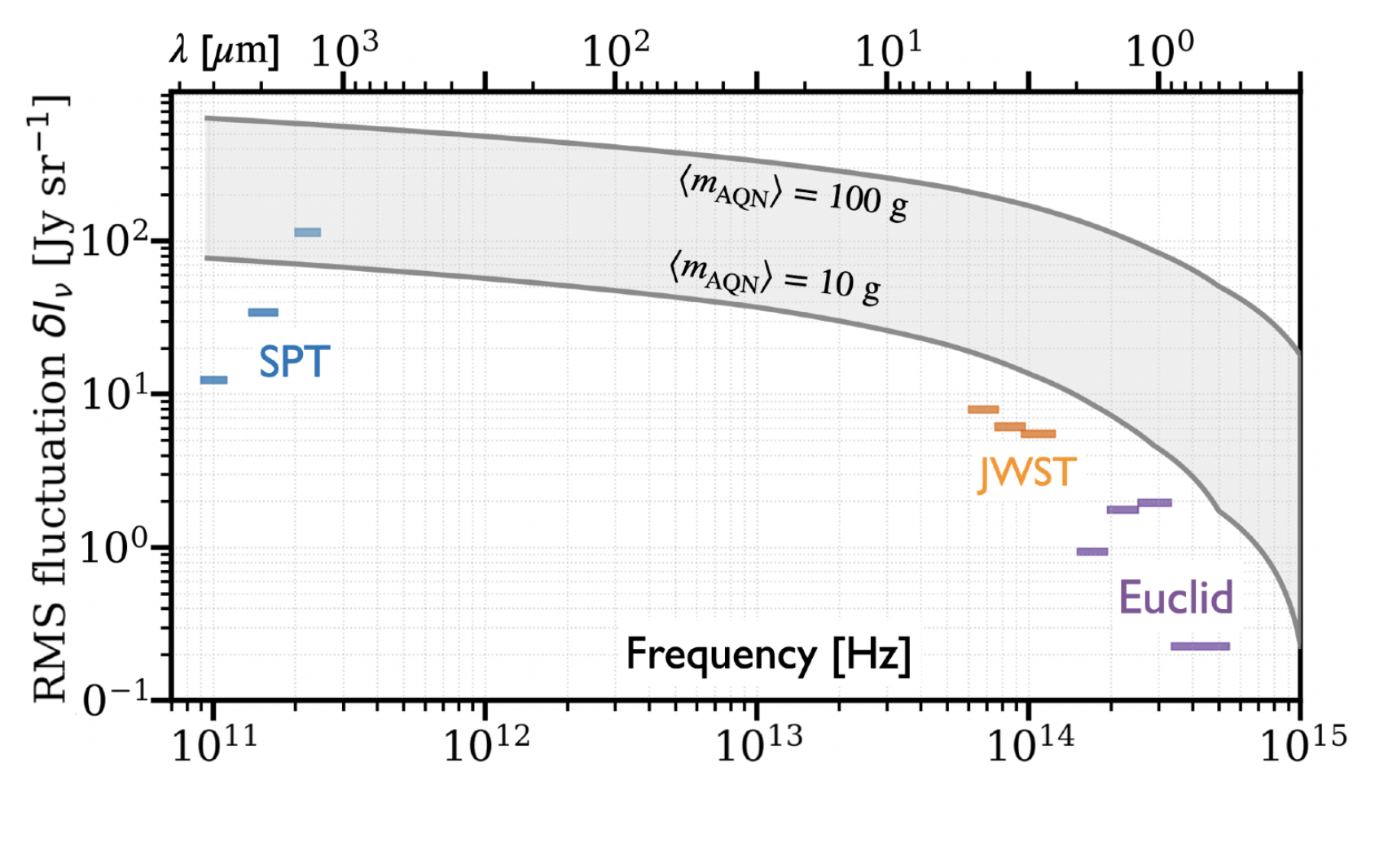}
  \vspace{-0cm}
  \captionof{figure}{Simulated RMS of $\bar{\rm AQN}$ density fluctuations at $\ell=3000$ for $\langle m_{\bar{AQN}}\rangle\in[10-100]\rm g$, cumulative across all redshifts. The coloured rectangles indicate a  $1\;\sigma$ uncertainties for the Euclid Wide and JWST's COSMOS-Web surveys for $\langle m_{\bar{AQN}}\rangle=10\;\rm g$, and the
  $1\;\sigma$ uncertainties of the Cosmological Infrared Background component from SPT-SZ \cite{2021ApJ...908..199R}.}
  \label{fig:dark_glow_detection_threshold}
\end{figure}

\bigskip
\indent {\bf Components separation }

Fig. \ref{fig:dark_glow_detection_threshold} shows the potential for a statistical detection of the $\bar{\rm AQN}$ glow, not an individual fluctuation detection. A statistical detection relies on components separation techniques and cross-correlations. The emission from unresolved galaxies originates at intermediate to high redshifts, typically $z \sim 1\text{--}3$, whereas the dark glow arises from energy injection by $\bar{\rm AQN}$s throughout cosmic history and contributes to anisotropies and spectral distortions over a very broad redshift range ($z \gtrsim 5$).

From observations of the Extragalactic Background Light (EBL), one can remove the resolved sources in order to isolate the unresolved Extragalactic Background Light (uEBL), as illustrated in Fig. ~\ref{fig:uEBL.png}. 
The uEBL has never been used as a scientific observable in the optical/NIR; standard pipelines treat it as a nuisance and discard it, however it could be transformed it into a window for observational cosmology, reaching z > 6 and enabling probes of the Universe’s history up to the Epoch of Reionisation. 

Contrary to unresolved galaxies, the intensity of the dark-glow signal increases with redshift (Majidi et al). Using the intensity fluctuations within the uEBL, the dark-glow signal can be distinguished from unresolved galaxy light by its distinctive dependence on redshift $z$, electromagnetic frequency $\nu$, and harmonic wavenumber $l$. Detecting a significant dark glow at high-redshift would indicate that dark matter interacts with ordinary matter, a paradigm shift in itself that would strenghten the case for macroscopic DM. Futhermore, a ($z$, $\nu$, $\ell$) dependence of the observed signal that matches the predictions of the QCD-AQN model would constitute compelling evidence that AQNs are a leading dark-matter candidate.

\section{AQNs in the early Universe}
\label{sec: formation}

The QCD-AQN framework is rooted in Quantum Chromodynamics, the quantum field theory that describes the strong nuclear force. Formulated in the early 1970s \cite{WeinbergQFT2}, QCD is based on the non-Abelian gauge group $SU(3)_{\rm colour}$. It describes quarks interacting through the exchange of massless gauge bosons called gluons. Asymptotic freedom \cite{1973PhRvL..30.1343G,PhysRevLett.30.1346} implies that quarks and gluons are never observed in isolation but are bound into hadrons. QCD is one of the most thoroughly experimentally tested components of the Standard Model of particle physics and is regarded as a firmly established and quantitatively verified theory of the strong interaction \cite{PDG2024}.

\begin{figure}[ht]
\centering
  \includegraphics[width=12 cm]{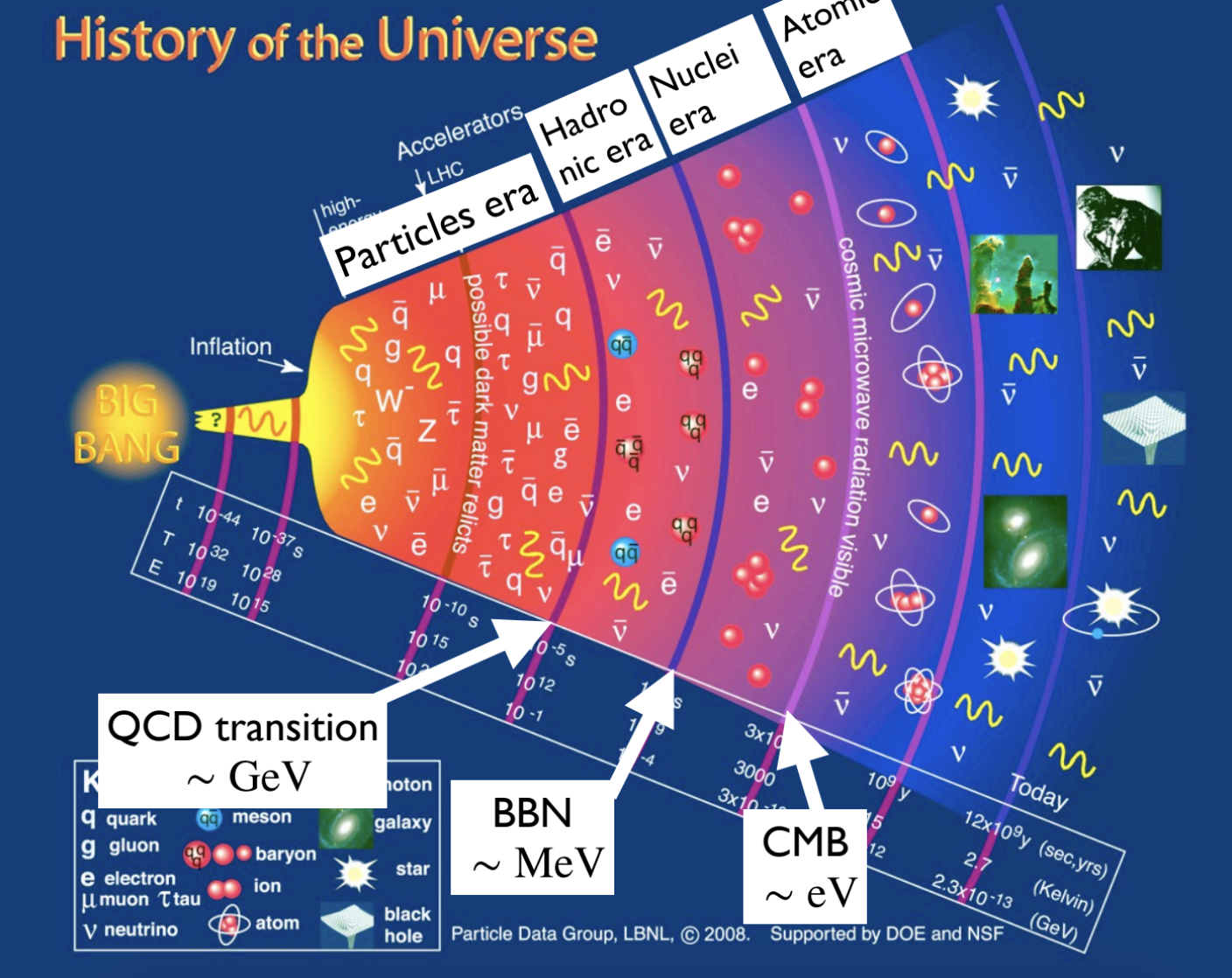}
  \vspace{-0cm}
  \captionof{figure}{History of the Universe (from Particle Data Group, 2008)), annotated to highlight QCD transition, BBN and CMB.}
  \label{fig:history_QCD_transition}
\end{figure}

\subsection{AQN formation}
\label{sec:AQN_formation}

The formation of AQNs, first hypothesised by Zhitnitsky in 2003 \cite{2003JCAP...10..010Z}, begins at the QCD transition in the early Universe (Fig. \ref{fig:history_QCD_transition}), at temperatures of order $ T \sim 170 \mathrm{MeV} $ ( $ t \sim 10^{-5}\,\mathrm{s}$), when quarks and gluons become confined into hadrons. Fig.\ref{fig:QCD_phase_diagram} illustrates several possible AQN formation pathways within the QCD phase diagram: either via the hadron gas phase or through the quark–gluon plasma, before completing their evolution in the colour-superconducting phase.

The formation of AQNs requires the presence of an axion field, which constitutes \textbf{the sole extension of the Standard Model of particle physics employed in the QCD-AQN framework.} This hypothesis is well motivated, as it is presently the only viable solution to the strong CP problem\footnote{"CP" symmetry means the combined  charge conjugation (C) and parity (P) symmetry.}.

\begin{itemize}
    \item {\bf The strong CP problem}
\end{itemize}

The "strong CP problem" can be stated as follows: Why does QCD appear to conserve CP symmetry, even though the theory permits a term that would produce large CP violation? In the QCD Lagrangian, the CP-violating term can be written as :

\begin{equation}
\mathcal{L}_\theta
=
\theta \,\frac{g_s^{\,2}}{32\pi^2}\,
G^a_{\mu\nu}\,\widetilde{G}^{\mu\nu}_a,
\label{eq:CPterm}
\end{equation}
where $G^a_{\mu\nu}$ is the gluons field and $g_s$ is the strong force coupling constant. $\theta$ is a free parameter which is strongly constraints by experiments, the strongest upper bound being $|\theta|\lesssim 10^{-10}$ from the absence of detection of a neutron electric dipole moment \cite{2020PhRvL.124h1803A}. This constitutes the so-called strong CP problem: within the Standard Model, there is no symmetry that naturally enforces such an extraordinarily small value of $\theta$.
\begin{figure}[t]
\centering
\includegraphics[width=15 cm]{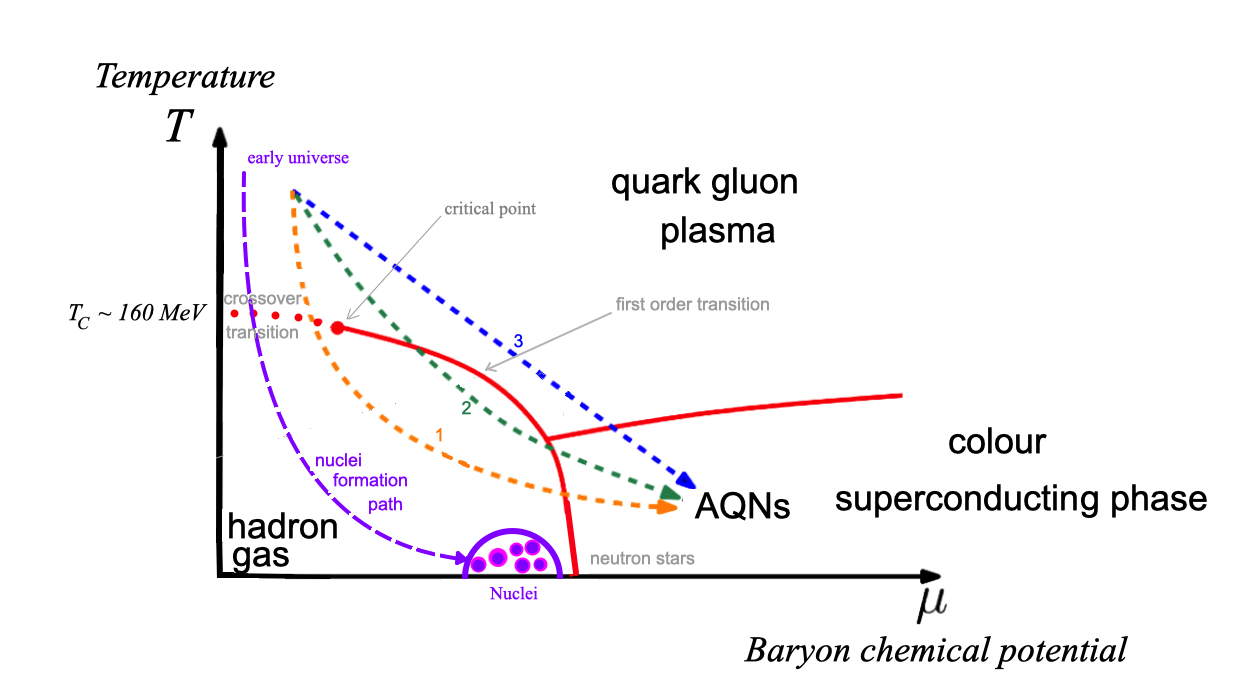}
  \vspace{-0cm}
  \captionof{figure}{The QCD phase diagram and its three phases: quark-gluon plasma, hadron gas and colour superconductor. Possible formation pathways for AQNs are numbered 1,2,3. The nuclei formation pathway (dashed purple line) passes through the crossover transition and proceeds within the hadron gas phase. Adapted from \cite{2018PhRvD..97d3008G} }
  \label{fig:QCD_phase_diagram}
\end{figure}

\begin{itemize}
    \item {\bf The Peccei–Quinn mechanism}
\end{itemize}

The original Peccei–Quinn (PQ) proposal \cite{1977PhRvD..16.1791P} correctly identified a dynamical solution to the strong CP problem by introducing a global symmetry $U(1)_{\rm PQ}$ with a new complexe scalar field $\phi$:

\[
\phi(x)\equiv (f_a+\rho(x))e^{i\alpha},
\]
where $\alpha$ is a new global symmetry and $f_a=\langle\phi\rangle$ is the axion decay constant.
The dynamical solution consists in assuming a spontaneous symmetry breaking at scale $f_a$ and promote $\alpha$, the phase of $\phi$, to a dynamical field $a/f_a$. A new term appears in the Lagrangian:

\[
\mathcal{L}_{aG\tilde G}
=
\frac{a(x)}{f_a} \,\frac{g_s^{\,2}}{32\pi^2}\,
G^a_{\mu\nu}\,\widetilde{G}^{\mu\nu}_a.
\]
The CP violating term in (Eq. \ref{eq:CPterm}) can now be expressed as an effective angle $\theta_{\rm eff}$:

\[
\theta_{\rm eff}\equiv \theta+\frac{a(x)}{f_a}.
\]
In QCD, the axion mass $m_a$ is given by:

\[
m_a^2=\frac{\chi_{\rm top}(T)}{f_a^2},
\]
where $\chi_{\rm top}(T)$ is the QCD topological susceptibility\footnote{The QCD topological susceptibility measures how strongly the QCD vacuum responds to the CP-violating parameter $\theta$; physically, it quantifies the strength of non-perturbative topological fluctuations that generate the axion potential and mass.}. In the limits of low $T\sim 0$ and high temperature $T\gg \Lambda_{\rm QCD}$, the topological susceptibility can be calculated analytically:

\begin{equation}
\left\{
\begin{aligned}
\chi_{\rm top}(T\sim0)
&\propto
m_\pi^2 f_\pi^2,
\\[6pt]
\chi_{\rm top}(T\gg\Lambda_{\rm QCD})
&\propto
\Lambda_{\rm QCD}^4
\left(\frac{\Lambda_{\rm QCD}}{T}\right)^n,
\qquad \text{with } n\sim 7\text{--}8,
\end{aligned}
\right.
\label{eq:chi_tot}
\end{equation}
where $\Lambda_{\rm QCD}\sim 100-200\,{\rm MeV}$, $m_\pi$ is the pion mass and $f_\pi$ is the pion decay constant\footnote{The pion properties encode the structure of the QCD vacuum and sets the scale of the QCD topological susceptibility.}. As a result, the mass of the axion field varies greatly over cosmic time:

\begin{subequations}\label{eq:a_mass}
\begin{align}
m_a(T\sim 0) &\approx 
5.7\,\mu\mathrm{eV}
\left(
\frac{10^{12}\,\mathrm{GeV}}{f_a}
\right),
\\[6pt]
m_a(T\gg \Lambda_{\rm QCD}) &\propto 
\frac{1}{f_a}
\left(
\frac{\Lambda_{\rm QCD}}{T}
\right)^{\!n/2}
\xrightarrow[T \to \infty]{} 0 .
\end{align}
\end{subequations}

Originally, the Peccei--Quinn proposal identified the axion symmetry-breaking scale $f_a$ with the electroweak scale, $f_a \sim 250\,\mathrm{GeV}$. This led to a large axion mass $m_a\sim 100\,{\rm keV}$, as shown by Eq.(\ref{eq:a_mass}), and therefore strong couplings to Standard Model particles. It predicted observable effects in particle experiments and significant signatures in stellar cooling that were in strong conflict with collider constraints and astrophysical observations. Nevertheless, although the original PQ axion scale turned out to be phenomenologically excluded, the axion mechanism itself remains a viable solution to the strong CP problem \cite{2008LNP...741....3P}.

\begin{itemize}
    \item {\bf Axion dynamics and the onset of AQN formation}
\end{itemize}

DFSZ and KSVZ\footnote{Dine–Fischler–Srednicki–Zhitnitsky \cite{DFSZ1,DFSZ2} and  Kim–Shifman–Vainshtein–Zakharov \cite{KSVZ1,KSVZ2}.} showed that one can realise the PQ symmetry breaking differently, allowing the breaking scale $f_a$ to be arbitrarily large. 
With such a large value, the axion field becomes {\it invisible}, and the cosmological impact of the axion field can be re-evaluated. After PQ breaking, the axion remains essentially massless, until, at $T\sim 1\,{\rm GeV}$,  non-perturbative QCD effects generate a potential $V(a)$ for the axion:

\begin{equation}
    V(a)\sim \chi_{\rm top}(T)\left(1-\cos{\left(\frac{a}{f_a}+\theta\right)}\right),
\label{eq:a_potential}
\end{equation}
where $\chi_{\rm top}(T)$ is given by Eq.(\ref{eq:chi_tot}). As a result, the axion field $a$ begins to relax towards the minimum of $V(a)$, thereby cancelling the $\theta$ term and driving  $\theta_{\rm eff}\rightarrow 0$, hence resolving the strong CP problem. 

It is at this stage, $T\sim 1\;{\rm GeV}$, that the AQN formation begins. In the limit of small mass density perturbations, one finds that the homogeneous axion field $a$ satisfies the equation of a harmonic oscillator damped by Hubble friction:

\begin{equation}
    \ddot{a} + 3H \dot{a} + m_a^2(T)\, a = 0.
\label{eq:a_osc}
\end{equation}
From Eq.(\ref{eq:a_mass}), one can see that at $T\gg 1\;{\rm GeV}$, the axion mass is extremely small $m_a\sim 0$, hence the axion field is {\it frozen} because the Hubble friction term dominates. At $T\sim 1\;{\rm GeV}$, calculations show that $m_a\sim 3H$ and the field starts to oscillate globally. As the Universe continues expanding, the restoring force dominates over the Hubble damping: the field can now roll and oscillate about the minimum of the potential while $\theta_{\rm eff}\rightarrow 0$. At $T\sim 1\;{\rm GeV}$, the effective angle reaches $\theta_{\rm eff}\simeq 10^{-3}-10^{-2}$.

Figure~\ref{fig:AQN_formation_3D_track}  illustrates the evolution and oscillatory behaviour of $\theta_{\rm eff}$ as the Universe cools. Note that the third axis defines a new region of the QCD phase diagram that remains entirely unknown.

\begin{figure}[t]
\centering
  \includegraphics[width=11 cm]{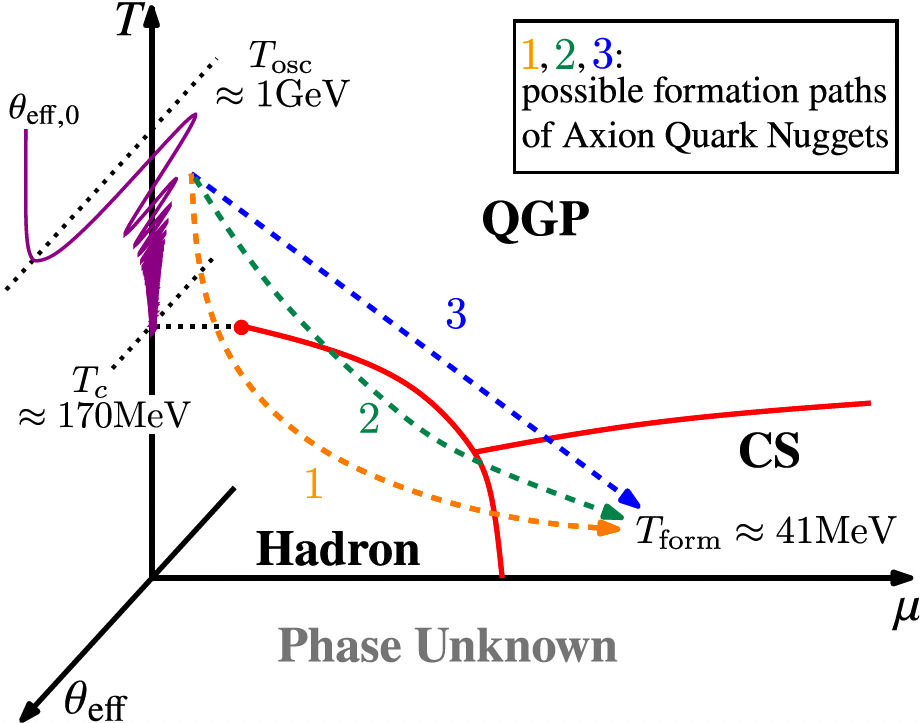}
  \vspace{-0cm}
  \captionof{figure}{The oscillations of $\theta_{\rm eff}$ between $T\sim 1$ GeV and $T\sim 170 $ MeV and three possible AQN formation pathways. The AQNs condensation starts aroung 200 MeV in the quark-gluon phase (QGP). The baryon number of the aggregates grows  as they migrate through the diagram. The AQNs formation is complete $\sim 41$ MeV.  From \cite{2018PhRvD..97d3008G} }
  \label{fig:AQN_formation_3D_track}
\end{figure}

\newpage

\begin{itemize}
    \item {\bf The axion field configuration at the QCD epoch}
\end{itemize}

What is the spatial configuration of the axion field across the Universe at $T\sim 1\;{\rm GeV}$? The answer depends on whether the PQ symmetry breaking happens before or after inflation. 

If the PQ symmetry breaks before inflation\footnote{Only this scenario is explored in this subsection, for the sake of simplicity. The second scenario will be explained in (\ref{sec:two_inflations}).}, then at $T\sim 1\;{\rm GeV}$, $\theta_{\rm eff}\simeq 10^{-3}-10^{-2}$ assumes the same {\it frozen} value everywhere in the Universe. 
As the Universe expands and cools, topological regions with $\theta_{\rm eff}\rightarrow \theta_{\rm eff}+2\pi$ are formed (see Eq.(\ref{eq:a_potential})). These regions assume the same vacuum expectation value $V(\theta_{\rm eff})=V(\theta_{\rm eff}+2\pi)$, but they are topologically distinct.  ${\rm N_{DW}}=1$ domain walls form and \cite{VilenkinShellard1994}, 
\cite{1998PhRvD..59b3505C} showed that approximately $10\%$ of them form closed configurations. Domain walls normally decay when intersected by strings, but for sufficiently small closed walls the probability of such an intersection is low, allowing them to survive. These small, closed domain walls can subsequently collapse and give rise to AQNs.

\begin{itemize}
    \item {\bf Axion-$\eta'$ domain walls and dynamical quark trapping}
\end{itemize}

In 2003, \cite{2003JCAP...10..010Z} proposed that the collapse of the above-mentioned closed domain walls can be halted if a sufficiently large number of quarks become trapped inside them. However, the axion field alone cannot confine quarks, and there is no strict prohibition preventing quarks from crossing the domain-wall boundaries. This raises the question of how quarks can remain trapped within the walls.

As shown in \cite{2016PhRvD..94h3502L} and \cite{2017PhRvD..96f3514G}, axion domain walls are not purely axionic objects. It is a well-established property of QCD that the axion coupling to the gluonic operator $G\tilde G$ implies that any spatial variation of $\theta_{\rm eff}$ necessarily excites the $\eta'$ meson\footnote{This is the QCD response to the anomalous $U(1)_A$ symmetry}. It means that axion domain walls boundaries where $\theta_{\rm eff}$ jumps by $2\pi$, also contain a layer of $\eta'$ meson, implying that the walls acquires a nontrivial CP-odd structure through coupled axion–$\eta'$ dynamics. At the onset of domain wall formation, equal numbers of quarks and antiquarks are present both inside and outside the wall, which they can cross freely. The axion–$\eta'$ structure subsequently favours one species over the other. As a result, some domain walls acquire an excess of quarks, while others acquire an excess of antiquarks. At a later stage, the quarks (or antiquarks) confined within a closed domain wall become compressed into a dense colour-superconducting (CS) phase, whose bulk free energy can be lower than that of the surrounding vacuum. Fermi pressure and the CS pairing counteract the collapse. Once the CS phase forms, it becomes energetically costly for quarks to leave, and the system behaves as if the quarks are “trapped.” The trapping is therefore dynamical and thermodynamic, not due to an impenetrable microscopic barrier.

During this dynamical phase, the oscillating axion background acts as a CP-odd bias, effectively behaving like a chemical potential for baryon number: $\Delta \mu_B\sim \dot \theta_{\rm eff}$. Integrated coherently over many oscillations during the collapse, this bias leads to a macroscopic baryon number accumulation inside the AQN. Once the configuration stabilizes, the baryon number is conserved, even as $\theta_{\rm eff}\rightarrow 0$.

\begin{itemize}
    \item {\bfseries Baryon number separation}
\end{itemize}

At $T\sim 1\;{\rm GeV}$, when the axion oscillations begin  and acts as an effective baryonic chemical potential, $\Delta \mu_B\sim \dot \theta_{\rm eff}$, the sign of $\dot \theta_{\rm eff}$ is the same everywhere within a given CP-odd region. 
The collapse of different domain walls happens at different times and the collapse dynamics are stochastic. Hence, some bubbles will start collapsing when $\dot\theta_{\rm eff}>0$ and some when $\dot\theta_{\rm eff}<0$. Once the sign inside a bubble is set, each oscillation cycle will generate a very small $\Delta \mu_B$ enhacing the baryon number $|B|$ of the bubble. The larger the $|B|$ , the more energetically favourable it becomes for quarks (antiquarks) to remain confined within the AQN ($\rm \bar{AQN}$). This is a runaway selection effect: Once a sign is selected by the initial bias, it reinforces itself, and each bubble ultimately acquires a large $|B|$.\footnote{The asymmetry only affects a thin layer at the surface of the closed domain wall that bounds the AQN, which thickness is determined by the QCD physical scale ${\Lambda_{\rm QCD}}^{-1}$: The baryon number separation is essentially a surface effect rather than a volume effect.}

\begin{itemize}
    \item {\bf Color superconductivity and the freeze-out of AQN formation}
\end{itemize}

When the AQN core enters a CS phase, quarks (antiquarks) near the Fermi surface form Cooper pairs and the excitation spectrum develops an energy gap:
\[
\Delta_{CS} \sim \Lambda_{\rm QCD}.
\]
Consequently, exciting higher-energy states becomes  more difficult at temperature $T\ll \Delta_{CS}$. 

Various studies indicate that $\Delta_{\rm CS}\sim 40-60\,{\rm MeV}$ \cite{1999NuPhB.537..443A,2001afpp.book.2061R,2002PhRvD..66g4017L,2005PhRvD..72c4004R,2008RvMP...80.1455A}.
Hence, when the cosmic temperature drops below $T\lesssim 40\,{\rm MeV}$, creating or removing quark excitations requires an amount of energy greater than the thermal bath can supply. Quark excitation inside the AQN becomes rare, quark evaporation is suppressed and quark accretion becomes inefficient. As a result, baryon exchange across the AQN boundary becomes exponentially suppressed, and the baryon number freezes. This marks the end of the active phase of AQN formation. From that point onward, AQNs with baryon number in the range $|B| \sim 10^{22}\text{--} 10^{28}$ are remarkably stable \cite{2019PhRvD..99k6017G}, enough to survive to the present epoch.

\subsection{Matter-antimatter asymmetry}
\label{subsection:asymmetry}

\begin{itemize}
    \item {\bf Why $\Omega_{\rm {DM}} \,\sim \Omega_b  $}
\end{itemize}

The final asymmetry between matter and antimatter AQNs is of order unity because the axion field oscillates a very large number of times, $\sim 10^7$, during the epoch when closed domain walls form and baryon number separation occurs. 
If the initial CP violating parameter was exactly zero ($\theta_{\rm eff}=0$) at $T\sim 1\,\rm GeV$, equal numbers of AQNs and  $\rm \bar{AQN}$s would form and no visible baryons would remain. With the small nonzero CP violation at $T\sim 1\,{\rm GeV}$, the oscillating axion background slightly favours one sign. Over many collapse events, a small statistical bias gradually builds up and ultimately becomes of order unity. This baryon number separation mechanism implies that the total baryon number of the Universe remains zero at all times:
\begin{equation}
    B_{\rm universe} 
    = B_{\rm visible} +B_{\rm AQN} 
    +  {\bar{B}}_{\rm \bar{AQN}}   = 0
\label{eq:neutral_charge}
\end{equation}
The resulting effect is an order-unity asymmetry between the total baryon number stored in matter  AQNs and antimatter $\bar{\rm AQN}$s, which can be parametrised as
\begin{equation}
{\bar B}_{\rm \bar{AQN}} = c(T_{\rm form})\, B_{\rm AQN},
\qquad
|c(T_{\rm form})| \sim \mathcal{O}(1).
\end{equation}
Hence, assuming the asymmetry favours antimatter (i.e. $|B_{\rm AQN}| < |B_{\bar{\rm AQN}}|$), a population of free baryons will remain unbound. This naturally leads to $\Omega_{DM} \simeq \Omega_{b}$. 

\begin{itemize}
    \item {\bf Masses of matter and antimatter AQNs}
\end{itemize}

Matter and antimatter AQNs originate from the same population of closed axion domain walls produced with random orientations of the axion field. The sign of the baryon number accumulated on a given wall is determined by the local value of  $\dot\theta_{\rm eff}$ at the time of formation, but this does not  affect the probability that a closed wall forms. As a result, the formation mechanism produces comparable numbers of matter and antimatter AQNs:
\[
n_{\rm AQN} \sim n_{\bar{\rm AQN}}.
\] The baryon number asymmetry manifests itself through an asymmetric mass distribution of aggregates, with antimatter $\bar{\rm AQN}$s being, on average, more massive than matter AQNs:
\[
\langle m_{\rm AQN} \rangle < \langle m_{\bar{\rm AQN}} \rangle
\]

\begin{itemize}
    \item {\bf Mass densities ratio}
\end{itemize}

A precise determination of the ratio 
$\Omega_b : \Omega_{\rm AQN} : \Omega_{\bar{\rm AQN}}$ 
would require non-perturbative QCD calculations, which are currently beyond reach. For the purpose of a visual illustration, let us assume that all other DM components are subdominant (including conventional axion production) and that $\Omega_{\rm DM} \simeq 5\,\Omega_{\rm visible}$.  
Equation~(\ref{eq:neutral_charge}) then implies relative density fractions of approximately $1:6$ for visible matter, $2:6$ for matter AQNs, and $3:6$ for antimatter AQNs. This situation is illustrated in Fig.~\ref{fig:3_types_of_matter}.  
The main purpose of this illustration is to emphasise that:

(ii) the number densities of the two populations are of the same order;

(i) the mean baryon number stored in antimatter $\bar{\rm AQN}$s exceeds that stored in matter AQNs, hence $\rm\bar{AQN}$ are represented with a larger radius..

\begin{figure}[t]
\centering
  \includegraphics[width=13 cm]{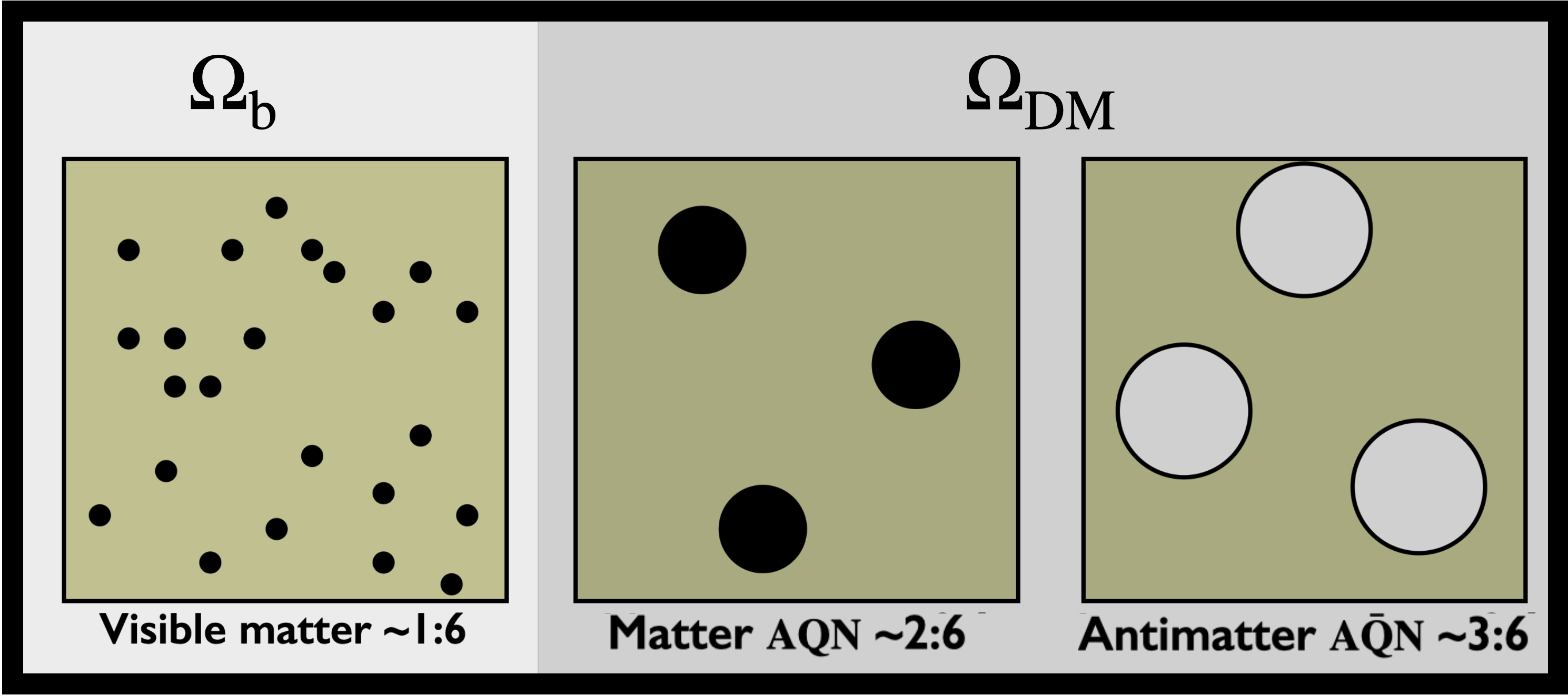}
  \vspace{-0cm}
  \captionof{figure}{Visible matter, AQNs and antimatter ${\rm A\bar{Q}N}$s before BBN. The aggregates form in a roughly 2:3 ratio because the axion-induced CP violation at the QCD epoch biases quark versus antiquark capture during formation, sequestering more antibaryon number in antimatter AQNs and thereby producing the observed visible baryon excess.
}
  \label{fig:3_types_of_matter}
\end{figure}

\subsection{AQNs remarkable stability }
\label{sec: AQN_stability}

\cite{2019PhRvD..99k6017G} investigated the stability and survival of AQNs during the early stages of the Universe. At high temperatures, $T \gg 1\,{\rm MeV}$, most collisions between protons and AQNs lead to elastic scattering : the resulting mass loss from annihilation processes remains negligible.
\newpage
\begin{itemize}
    \item {\bf AQNs during BBN}
\end{itemize}
\label{sec:AQN_stability}

AQNs formation ends at $T_{\rm form}\sim 40\,{\rm MeV}$, far before the BBN energy scale $T_{\rm{BBN}}\sim 1\,{\rm MeV}$. By the time nucleosynthesis begins, AQNs are already macroscopic, tightly bound objects in a CS phase, with a binding energy per baryon of order $\Lambda_{\rm QCD} \sim 100\,\mathrm{MeV}$, far exceeding BBN thermal energies. Their interaction cross section per baryon is geometrically suppressed, preventing significant participation in nuclear reactions. They neither disrupt light-element abundances (except for heavy nuclei, see Section 4.5) nor inject sufficient energy to alter neutron-proton freeze-out or the standard BBN expansion history \cite{2021MPLA...3630017Z}.

One may ask how antiquark aggregates can survive in such a hot, relativistic plasma with a large particle number density, composed of electrons, positrons, and, most importantly, a dense population of free baryons at high temperature. Electron-positron pairs remain in thermal equilibrium down to temperatures of order $T \sim 20\,{\rm keV}$. It was shown in \cite{2019PhRvD..99b3517F} that the large positron number density inhibits the capture of baryons (protons and helium nuclei) by $\bar{\rm AQN}$, thereby preventing rapid annihilation. The positrons remain sufficiently abundant for long enough to form an effective shielding layer around the aggregate.

\begin{itemize}
    \item {\bf AQNs between BBN and recombination}
\end{itemize}

AQNs remain stable throughout the period between BBN and the CMB epoch. During this interval, the plasma is both dense and fully ionised from the perspective of an AQN. Interactions between AQNs and such a medium are analysed in \cite{2025arXiv251205401M}, while the more general conditions prevailing from the post-BBN era up to recombination are discussed in \cite{2019PhRvD..99k6017G}.

In contrast to the BBN epoch, the Coulomb capture rate of ambient protons in a 
$T \sim \mathrm{eV\text{--}keV}$ plasma is significantly enhanced. Because nuclei move more slowly than during BBN, the efficiency of Coulomb capture increases. The resulting large capture rate produces a screening layer around the AQNs, which prevents any runaway 
annihilation. For typical baryon numbers $|B| \sim 10^{22}\text{--}10^{28}$, the corresponding annihilation rate --- and thus the associated mass loss --- remains negligibly small.

\newpage

\begin{itemize}
    \item {\bf Post recombination}
\end{itemize}

For the remainder of cosmic history, thermal and collisional processes in most astrophysical environments are far too weak to disrupt AQNs. Their enormous binding energy, combined with their extremely low number density, renders catastrophic collisions exceedingly unlikely. Only a minuscule fraction of antimatter aggregates lose a significant amount of energy through interactions with ambient baryons, while the probability of AQN–$\rm\bar{AQN}$ collisions is entirely negligible.

After recombination the baryon number distribution of the AQNs will essentially be fixed until the present day, but it may develop localized anisotropy in regions of particularly high matter density such as stars or planets.
\smallskip

{\bf Key takeaway.} Both matter and antimatter AQNs survive from the end of their formation at $T \sim 40 \, \rm{MeV}$ through BBN, the CMB epoch, and the subsequent stages of structure formation, all the way to the present day, without experiencing any significant mass loss \cite{2019PhRvD..99k6017G}.

\subsection{Photon to baryon ratio, \texorpdfstring{$\eta$}{eta}}

The baryon-to-photon ratio $\eta \equiv n_{\rm baryon}/n_{\rm photon}$ is measured independently by BBN \cite{Cooke:2017cwo} and the CMB \cite{Planck:2018vyg}. The respective constraints are:
\begin{eqnarray}
    \eta_{\rm BBN}&\simeq& 6.0\times 10^{-10}\nonumber\\
    \eta_{\rm CMB}&\simeq& 6.1\times 10^{-10}.\nonumber
\end{eqnarray}

The agreement between the two measurements constitutes a remarkable success of Big-Bang cosmology. It also places a strong constraint on any entropy injection (i.e. changes in the photon number) occurring between BBN and recombination, which cannot exceed $\delta n_\gamma/n_\gamma\sim 5\%$ at the $2\sigma$ level. This constraint is consistent with the QCD-AQN framework (see ~\ref{sec:AQN_stability}).

Within the standard Big-Bang picture, however, the origin of the observed value of $\eta$ remains unexplained. Using the well-known cross-sections for baryon–antibaryon annihilation, the freeze-out temperature is estimated to be of order $T_{p\bar p}\sim 20\,{\rm MeV}$\footnote{This corresponds to the temperature when the $p-\bar p$ collision rate drops below the Hubble expansion rate}, yielding the prediction $\eta\sim 10^{-20}$, which fails both to reproduce the observed baryon asymmetry and to account for the absence of large quantities of antimatter in the Universe \cite{KolbTurner1990}. Hence, numerous scenarios have been proposed to account for the baryon asymmetry and the observed value of $\eta$. Many of these invoke physics beyond the Standard Model of particle physics, frequently associated with CP violation at the electroweak symmetry-breaking scale.

\smallskip
In the QCD–AQN framework, primordial photons arise from the same thermal processes that operate in the standard hot Big-Bang scenario. At temperatures well above the QCD scale, the Universe is filled with a dense, strongly interacting plasma in thermal equilibrium. Photons are continuously produced and absorbed through rapid pair annihilations, thereby maintaining a blackbody radiation field.

During the QCD epoch, when axion domain walls collapse, a large fraction of baryon– antibaryon pairs continue to annihilate. These annihilations predominantly produce pions, with neutral pions subsequently decaying into photons (e.g. $\pi^0 \rightarrow \gamma\gamma$). However, since this takes place at temperatures of order $T \sim 100\text{--}200\,\mathrm{MeV}$,
the interaction rates are extremely high, and the injected electromagnetic energy thermalises very efficiently. The net effect is not a distortion of the photon spectrum, but rather a redistribution of energy between matter and radiation.

Baryon–antibaryon annihilation ceases when AQN formation ends at $T\sim T_{\rm form}$, as there are no longer free antibaryons available to annihilate \cite{2003JCAP...10..010Z}. The AQN core becomes sequestered from the surrounding plasma. Meanwhile, the baryons that were not sequestered into matter AQNs remain as free particles in a plasma devoid of antibaryons. The resulting baryon–to– photon ratio is then determined by a Boltzmann suppression factor,

\begin{equation}
\eta \sim e^{-m_p/T_{\rm form}},
\end{equation}
where $T_{\rm form}$ corresponds to the binding-energy scale of the colour superconducting phase inside the aggregate. For $\eta \simeq 6.0 \times 10^{-10}$, it yields
\[
T_{\rm form} \approx 40\,\mathrm{MeV},
\]
which is of the same order as the results quoted in~ \ref{sec:AQN_formation}.

\medskip 
{\bf Key takeaways}. The QCD-AQN framework provides a physical motivation for 
$\eta \sim 10^{-10}$ without invoking fine-tuning or introducing new baryon-number--violating physics. 

\subsection{Primordial ${}^{7}\mathrm{Li}$ abundance}

There is a discrepancy between the primordial abundance of ${}^{7}\mathrm{Li}$ predicted by standard BBN and that inferred from observations of old, metal-poor halo stars. Using the baryon density precisely determined from CMB measurements, standard BBN predicts a primordial abundance of ${}^{7}\mathrm{Li/H} \simeq (4$--$5)\times10^{-10}$. In contrast, spectroscopic observations of warm Population~II stars consistently find a significantly lower and nearly constant abundance, ${}^{7}\mathrm{Li/H} \simeq (1$--$2)\times10^{-10}$. Meanwhile, the predicted abundances of other light elements, notably deuterium and ${}^{4}\mathrm{He}$, are in good agreement with observations. Proposed resolutions broadly fall into three categories: astrophysical solutions involving lithium depletion in stellar interiors, nuclear physics solutions invoking unaccounted-for reaction channels during BBN, and new-physics solutions that modify early-Universe conditions, such as late-time energy injection or non-standard particle content. To date, none of these explanations has achieved broad consensus.
\smallskip

In the QCD–AQN framework, antimatter is sequestered into AQNs prior to BBN and does not participate in nuclear reactions during this epoch. AQNs are non-relativistic and extremely rare, and consequently make a negligible contribution to the radiation density. They do not affect the cosmic expansion rate or the effective number of relativistic species, $N_{\rm {eff}}$. Energy injection from AQNs during BBN is likewise negligible, leading to no significant photodissociation of light nuclei. 

Under these conditions, BBN might proceed with a free baryon-to-photon ratio that is lower than the value inferred at later times from the CMB. This selective reduction primarily impacts the synthesis of lithium.
\cite{2019PhRvD..99b3517F} showed that the relative depletion of nuclei -- quantified by the number of trapped and captured ions of atomic number Z -- can be expressed as

\begin{equation}
\frac{\delta n_Z}{n_Z} \simeq \frac{4\pi R_{\rm cap}^3}{3}\,n_{\rm AQN}\, \exp\!\left[\frac{(Z-1)\,\alpha\,Q(r)}{r\,T}\right],
\end{equation}
where $(4\pi R_{\rm cap}^3/3)\,n_{\rm AQN}$ represents the effective geometrical capture volume associated with the presence of AQNs. For deuterium and helium, this expression yields $\delta n/n \ll 1$, indicating that their primordial abundances remain essentially unchanged, in agreement with observational constraints. In contrast, for lithium ($Z=3$) one finds $\delta n/n \sim 1$, implying a significant depletion. The depletation is amplified by the fact that most primordial ${}^{7}\mathrm{Li}$ is initially synthesised in the form of ${}^{7}\mathrm{Be}$, for which the depletion factor is even larger.

\medskip
{\bf Key takeway.} The QCD-AQN framework may offer a plausible explanation for the  ${}^{7}\mathrm{Li}$ discrepancy.

\subsection{AQN mass distribution}
\label{sec:AQN_mass_range}

The AQN formation process yields a distribution of masses which cannot be precisely computed from first-principles QCD, for several reasons.

(i) AQNs formation is a highly non-linear, out-of-equilibrium process involving axion domain wall formation, collapse, and baryon number accretion in an expanding cosmological background. QCD at finite temperature and finite baryon chemical potential in this regime is intrinsically non-perturbative and cannot be solved analytically.  

(ii) The relevant scales span many orders of magnitude: microscopic QCD dynamics ($\Lambda_{\mathrm{QCD}}^{-1}$) must be coupled to the axion scale $m_a^{-1}$ and the macroscopic horizon-scale correlations set by the axion field. This multi-scale problem is beyond current lattice QCD capabilities.  

(iii) The final AQN mass depends on stochastic properties of the axion domain wall network (correlation length, percolation, collapse time), which are determined by cosmological initial conditions rather than by QCD microphysics.

In other words, QCD fixes the characteristic density and stability properties of the AQNs but it does not allow a prediction of the AQN mass distribution. From a theoretical standpoint, only an order-of-magnitude estimate can be obtained.

\subsubsection{AQN mass range at formation freeze-out}
\label{sec:formation_mass_range}

Since the mass density of the AQN core is comparable to nuclear density\footnote{The mass density of quark matter in the CS phase slightly exceeds nuclear density, $\rho_{\rm CS}\simeq (2\text{–}5)\rho_{\rm nucl}$ \cite{2008RvMP...80.1455A}.}, the AQN mass function is directly determined by the AQN size distribution, i.e. the baryon number distribution. 

\begin{itemize}
    \item {\bf Characteristic baryon number and radius}
\end{itemize}
The size of an AQN is set by the initial size of the collapsing $N_{DW}$ = 1 domain walls from which it originates which, in turn, is set by the defect correlation length $\xi_{\rm DW}$, which scales inversely with the axion mass, $\xi_{\rm DW} \sim m_a^{-1}$ \cite{PhysRevD.30.2036}. The evolution of $\xi_{\rm DW}(T)\sim m_a^{-1}(T)$ from $T_{\rm osc}\sim 1\,{\rm GeV}$ to $T_c\sim 170\,{\rm MeV}$ 
was computed in \cite{2019PhRvD..99k6017G} 
on the basis of detailed modelling rather than purely analytical arguments. For $m_a\sim 10^{-4}\,{\rm eV}$, the correlation length of DW is $\xi_{\rm DW}\sim 0.1-1\,{\rm cm}$, which is much smaller than the horizon size at QCD transition ($\sim 10\,\rm km$) but much larger than the QCD physical scale, $\Lambda_{\rm QCD}^{-1}\sim $fm.

As we saw in Section~\ref{sec:AQN_formation}, the formation of AQNs follows a sequence of damped oscillations of axion domain walls. Each collapsing domain-wall bubble undergoes radial oscillations whose amplitude gradually decreases due to friction. This friction arises from several thermal and dynamical processes related to the QCD scale : proton–antiproton annihilation within the core, thermal exchange with the surrounding QG plasma, pressure differences between the interior and exterior, and repeated “refill” events in which baryons leave and re-enter the system.
As the oscillation proceeds, the chemical potential $\mu$ of the AQN core} builds up with a definite sign $B>0$ or $B<0$. This growing chemical potential causes the equilibrium radius to shrink until the oscillation damps around a final radius $R_{\rm form}$ (see Fig. \ref{fig:T_form}).

\begin{figure}[ht]
\centering
  \includegraphics[width=14 cm]{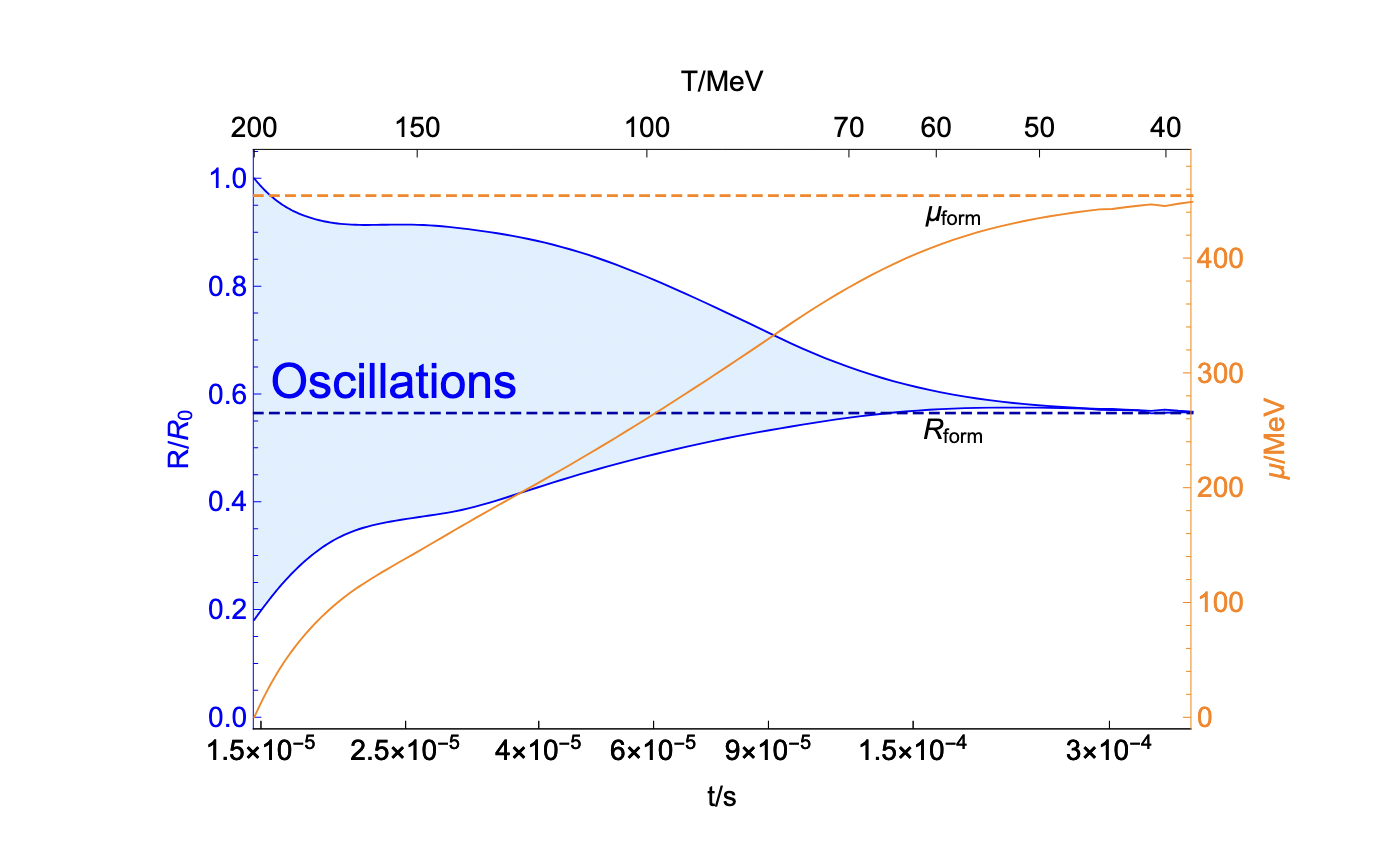}
  \vspace{-0cm}
  \captionof{figure}{Evolution of the AQN radius, expressed as a fraction of the inital radius $R_0$, between 200 MeV and 40 Mev. Matter AQNs and antimatter AQNs complete their formation at $\sim 40$ MeV. The solid orange line represents the
lower envelope of the $\mu$ oscillation.  It reaches $\mu_{\rm{form}} 
\sim 450 \, \rm{MeV}$, consistent with
formation of a CS phase\cite{2019PhRvD..99k6017G,2021MPLA...3630017Z}.}
  \label{fig:T_form}
\end{figure}

The characteristic AQN mass is set by the number of oscillation cycles occurring as the temperature decreases from $T\sim 170\,{\rm MeV}$ to $T\sim 40\,{\rm MeV}$ , which is of order $\sim 10^7$. When the AQN formation ends at $t\sim 10^{-4}\,\rm s$ ($T\sim 40\,{\rm MeV}$), the typical baryon number is determined by:

\begin{equation}
|B|\sim 4\pi \xi_{\rm DW}^2 \times \Lambda_{\rm QCD}^{-1}\times \rho_{\rm nucl} \times m_p^{-1}\sim 10^{24}
\label{eq:Bestim}
\end{equation}

where $4\pi \xi_{\rm DW}^2$ is the surface area of the axion domain wall, $(\Lambda_{\rm QCD}^{-1}\times \rho_{\rm nucl}) $  the surface mass density of the region in which baryon number separation occurs\footnote{This estimate is based on the "thin-wall" approximation which was used in AQN formation papers. The wall can be much thicker, but in that case the baryon are also diluted such that Eq.(\ref{eq:Bestim}) still holds.}, and $m_p$  the proton mass. Eq.~(\ref{eq:Bestim}) yields
$|B| \sim 10^{24}$, which corresponds to a typical CS core of size $R_{\rm AQN}\sim 10^{-5}\,{\rm cm}$ and a mass of the order of gram.

\begin{itemize}
    \item {\bf Mass lower bound}
\end{itemize}

The theoretical lower bound arises from AQN cosmological survival. During their formation, AQNs must withstand thermal evaporation, annihilation with ambient baryons, and diffusion across the axion domain wall before baryon exchange freezes out at $T_{\rm form} \sim 40\,\mathrm{MeV}$. If the baryon number is too small, surface effects dominate over bulk binding, leading to efficient evaporation and eventual destruction. Some AQNs do not have time to grow, either because they started their formation not very long before $T\sim T_{\rm form}$ or because they started to grow in a small domain wall, or both. As a consequence, they cannot reach the nominal value $\mu\sim 450\,{\rm MeV}$ as shown on Fig. \ref{fig:T_form}, and do not survive.
Detailed analysis based on the axion domain wall size distribution and the AQN formation growth by \cite{2006PhRvD..74d3515Z} indicate a minimum baryon number of order $|B|_{\min} \sim 10^{22}-10^{23}$, corresponding to a mass $m_{\min} \sim 10^{-2}-10^{-1}\,\mathrm{g}$. 
 \begin{itemize}
     \item {\bf Mass upper bound}
 \end{itemize}
 
The  theoretical upper bound  arises from the fact that the domain walls size distribution is exponentially supressed at large scale and that the formation time for AQNs is limited.  Together, these two limitations lead to an upper bound $|B|_{\max} \sim 10^{28}$ \cite{2006PhRvD..74d3515Z}.

\medskip
 
Hence $|B| \sim 10^{22}\text{--}10^{28}$, with the precise distribution depending on the axion mass $m_a$ and the details of the collapse process. 
Converting to mass through $m \simeq m_p B$, this corresponds to a formation mass range of approximately
\begin{equation}
m_{AQN} \sim 10^{-2}\text{--}10^{4}\,\mathrm{g}
\label{eq:mass_theory}
\end{equation}

\subsubsection{A mass distribution constrained by observations}

The mass distribution of AQNs is driven by the combination of the following constraints: 

(i) The axion domain wall size, $R$, follows a Gaussian distribution $\propto \exp(-(R/\xi_{\rm DW})^2)$ 

(ii) AQN formation is active during the period $T\simeq [200-40]\,{\rm MeV}$. 

Taken together, these two considerations lead to a power law AQN mass distribution:
\begin{equation}
    f(m_{\rm AQN}) \propto m_{\rm AQN}^{-\alpha}, \,
\label{eq:f_mass_distribution}
\end{equation}

which is bounded by the theoretical $[m_{\min},m_{\max}]\simeq [10^{-2}, 10^{4}] \rm{g}$ limits given above.

\medskip
$\bar{\rm{AQN}}$s mass constraints from observations have been discussed in Section~\ref{sec:AQN_and_Obs}. They are generally given in terms of  the mean $\bar{\rm{AQN}}$s mass, $\langle m_{\bar{\rm{AQN}}s} \rangle $:
\[
    \langle m_{\bar{\rm{AQN}}s} \rangle
    = \int_{m_{\min}}^{m_{\max}} \mathrm{d}m_{\bar{\rm{AQN}}s}\,
      m_{\bar{\rm{AQN}}s}\, f(m_{\bar{\rm{AQN}}s}) \, ,
\]

These observational constraints were summarised in (\ref{sec:synthesis_mass}): they are within the $[10^{-2}, 10^{4}] \rm{g}$ theoretical limits.

The value of $\alpha$ cannot be predicted from first principles, but it can be confronted to observations.
 To calculate the amount of energy injected in the Solar corona, ~\cite{2018PhRvD..98j3527R} assumed the power law given by Eq.(\ref{eq:f_mass_distribution}) with a mass range 
$m_{\bar {\rm AQN}}\in [10^{-2},10^{4}]\,{\rm g}$, which corresponds to a characteristic value $|B| \sim 10^{25}$. The mean $\bar{\rm AQN}$ mass depends on the full distribution, while the annihilation events rate is controlled by the number density of $\bar{\rm AQN}$s. By comparison with the observed statistics of nanoflares in the corona, the authors show that the data are consistent with an $\bar{\rm AQN}$ mass function with \(\alpha \sim 2\), for which the energy contribution per logarithmic interval of mass is approximately constant. For a typical $\bar{\rm AQN}$ window $10^{23} \lesssim |B| \lesssim 10^{28}$, the broadly viable region is approximately
\[
1 \lesssim \alpha \lesssim 2.5.
\]
This analysis does not provide a strong constraint on either the slope $\alpha$ or the mean $\bar{\rm AQN}$ mass, because all $\rm\bar{AQN}$s are expected to annihilate once they reach an altitude of about $\sim 2000\,{\rm km}$ above the photosphere, independently of the assumed slope of the mass distribution. The allowed parameter space is therefore characterised by a wide mass interval, with a typical average mass of order $\sim 10$ grams, but with significant uncertainty in the precise form of the distribution. This uncertainty mainly reflects the fact that the formation and growth of AQNs take place during the QCD epoch, in a strongly coupled and non-perturbative regime, where the microphysical evolution of closed domain walls and the accumulation of baryon number cannot presently be calculated from first principles.

\begin{table}[t]
\centering
\small
\setlength{\tabcolsep}{4pt}
\begin{tabular}{p{3.2cm} p{3.5cm} p{5.2cm}}
\hline
\textbf{Observation} & \textbf{Common overinterpretation} & \textbf{What is actually constrained} \\
\hline
Bullet Cluster 
& Proof of collisionless particle DM 
& Low self-interaction cross section per unit mass ($\sigma/m$) at large scales; microscopic nature not specified \\

CMB acoustic peaks 
& Requires non-baryonic particle DM 
& Cold, pressureless component before recombination; negligible photon coupling \\

Large-scale structures 
& Confirms cold particle DM 
& Collisionless gravitational clustering \\

Direct-detection null results 
& Strong constraints on particle DM 
& Specific scattering cross sections; does not exclude macroscopic DM \\
\hline
\end{tabular}
\caption{Observations often interpreted as evidence for elementary particle dark matter. The final column lists the actual physical quantities constrained.}
\label{tab:DM_misinterpretations_compact}
\label{tab:no_proof}
\end{table}

\subsection{Two inflation scenarios}
\label{sec:two_inflations}

The spatial structure of the domain walls responsible for AQN formation depends on whether Pecci-Quinn symmetry breaking occurs before or after inflation, leading to a different domain-wall topology.

\begin{itemize}
    \item {\bf Pre-inflation scenario}
\end{itemize}

Here, since the axion field across the observable Universe is homogenised by inflation, the matter--antimatter asymmetry arises naturally from AQN formation through baryon number separation. All free baryons ($B=1$) are leftover baryons from AQN formation. Therefore $\Omega_{\rm DM} \sim \Omega_b$ with {\bf no need for baryogenesis.} 

\begin{itemize}
    \item {\bf Post-inflation scenario}
\end{itemize}

Here the sign of $\theta_i$ is random in each causal horizon volume. At PQ symmetry breaking, the Universe is composed of large domains where  $\theta_i>0$  or $\theta_i<0$. Hence, adopting the formation scenario described in (\ref{sec:AQN_formation}), the Universe is a mosaic of regions dominated by AQNs with leftover antibaryons, and regions dominated by $\bar{\rm {AQN}}$s with leftover baryons. In each of these regions, baryon number conservation implies that
\[
B_{\rm universe}=B_{\rm visible}+B_{\rm AQN}+\bar{B}_{\bar{\rm AQN}}=0 .
\]
The relation $\Omega_{\rm DM} \sim \Omega_b$ applies both locally within each domain, and globally. There, a conventional baryogenesis may be required.

\subsection{Broadening the dark-matter paradigm}

Over the past four decades, DM as an elementary particle has emerged as the prevailing paradigm for a combination of theoretical, phenomenological, and experimental reasons. Extensions of the Standard Model of particle physics -- such as supersymmetry or hidden-sector frameworks -- predict new stable particles, making it economical to associate one of them with DM. The so-called “WIMP miracle” further reinforced this idea: a weak-scale particle with typical electroweak interactions yields the observed relic abundance through thermal freeze-out. In parallel, cold, collisionless particle DM provides an excellent fit to large-scale structure formation and the CMB acoustic peaks. Moreover, particle DM candidates opened the possibility of laboratory detection through direct searches, indirect signals, or collider production. 

Over time, the search for elementary particle DM generated considerable sociological momentum, occasionally leading to inherent overinterpretations of some observational evidence. Selected examples are summarised in Table~\ref{tab:no_proof}. But today broadening the DM paradigm is not a matter of debating astronomical observations: The fundamental question is whether an alternative DM framework can offer an appeal comparable to that of elementary particle candidates. As an illustration, Table \ref{tab:WIMP_vs_AQN} contrasts the QCD–AQN framework with the WIMP paradigm. 

\begin{table}[t]
\centering
\small
\setlength{\tabcolsep}{4pt}
\begin{tabular}{p{3.2cm} p{5.8cm} p{5.8cm}}
\hline
\textbf{Aspect} & \textbf{WIMPs (Weakly Interacting Massive Particles)} & \textbf{QCD--AQN Framework} \\
\hline

Prior theoretical motivation 
& Arises in many extensions of the Standard Model (e.g. SUSY, hidden sectors); historically supported by the ``WIMP miracle.'' 
& Arises as a testable extension of the Standard Model rooted in QCD and axion domain-wall dynamics; no dark sector\\
\hline
Explanatory scope 
& Explains cosmological DM abundance and structure formation; does not naturally account for the observed coincidence $\Omega_{\rm DM} \sim \Omega_b$ without additional assumptions. 
& $\Omega_{\rm DM} \sim \Omega_b$ through shared QCD-scale; baryon-to-photon ratio $\eta$; possible lithium-7 tension; no baryogenesis if PQ occurs after inflation \\
\hline
Observational evidence 
& Strong evidence for cold, collisionless DM; but no direct detection.
& Behave as cold and effectively collisionless DM. Consistent with cosmological constraints; multiple faint signatures across environments . \\
\hline
Free parameters 
& At least two parameters ($m$ and $\sigma$), often more depending on the specific model. 
& Dominated by a single macroscopic parameter, $\langle m_{\rm AQN} \rangle$. \\
\hline
Characteristics 
& Elementary particle; microscopic, weak-scale interactions; small interaction cross section and large number density; mass typically 1\,GeV--100\,TeV depending on model. 
& Macroscopic composite objects; gram-scale masses from QCD formation; very small number density; observationally, mean mass $\sim 5$--$200\,\mathrm{g}$ . \\
\hline
Stability across cosmic history 
& Stability is imposed by symmetry (e.g. R-parity, Kaluza–Klein parity).
& Stabilised by QCD pressure and axion domain-wall dynamics; macroscopic binding ensures survival from the QCD transition to the present epoch. \\
\hline
Direct detection experiments 
& LUX-ZEPLIN (LZ), XENONnT, PandaX, SuperCDMS, EDELWEISS, DarkSide, and DEAP-3600. No confirmed direct detection.
& No dedicated experiments; \\
\hline
Indirect detection facilities 
& Fermi Gamma-ray Space Telescope, H.E.S.S., MAGIC, VERITAS, IceCube, ANTARES, AMS-02, and DAMPE
& Possible Dark Glow  detection with Euclid + JWST + SPT.\\
\hline
Collider searches
& Large Hardron Collider: ATLAS and CMS
& Not applicable \\

\hline
\end{tabular}
\caption{Comparison between the WIMP paradigm and the QCD--AQN framework.}
\label{tab:WIMP_vs_AQN}
\end{table}

After several decades of sustained experimental efforts to detect WIMPs without definitive success, the time may be ripe to encourage a more diversified theoretical landscape. As shown in Table~\ref{tab:dm_candidates_comparison}, one compelling direction is axion physics, the topic of our next section. 

\section{The QCD axion }
\label{section:axion}

The QCD axion{\footnote{In this lecture, {\it  axion} refers to the QCD axion, with the DFSZ and KSVZ models serving as the two primary benchmarks. For a review of QCD axions and their cosmological implications, see \cite{2016PhR...643....1M}.}} and the QCD-AQN framework both originate from the same well-motivated extension of the
Standard Model: the Peccei–Quinn solution to the strong CP problem. Yet they represent distinct dark matter candidates. 

Axions are ultralight bosons. Their interactions are extremely weak but non-zero, allowing them to evade laboratory detection, while astrophysical and cosmological observations already place strong constraints on the allowed mass range. The axion remains experimentally testable.

In contrast, AQNs are rare macroscopic composite objects. Their interactions with ordinary matter are infrequent but not negligible, occurring primarily through baryon annihilation at their surface, allowing them to evade conventional particle DM searches while remaining astrophysically testable.

\begin{figure}[ht]
\centering
  \includegraphics[width=13 cm]{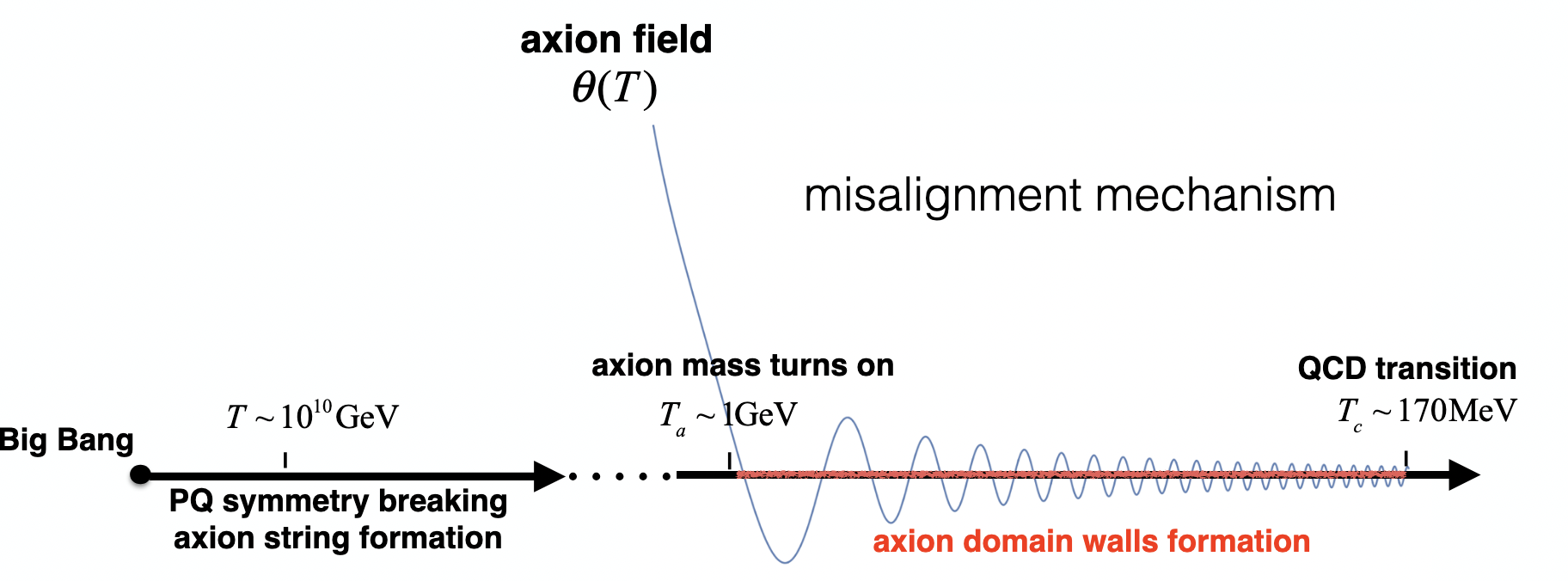}
  \vspace{-0cm}
  \caption{Timeline of axion formation from the Big Bang to the QCD transition.}
  \caption*{\footnotesize\emph{Source:} \url{https://indico.cern.ch/event/707123/contributions/3143173/attachments/1729576/2794761/1_IPA_presentation_Shuailiang_Ge.pdf}}
  \label{fig:misalignment}
\end{figure}

Axions are formed during the PQ symmetry breaking ($T \sim f_a \sim 10^{9}$-$ 10^{12} \, {\rm GeV}$) while AQNs are formed during the QCD transition ($ 170 \,{\rm MeV} \gtrsim T \gtrsim 40 \, {\rm MeV}$) (see \ref{sec:AQN_formation}). The misalignment mechanism is central to both the axion-field dynamics
and the baryon-number separation in the AQN framework
(Fig.~\ref{fig:misalignment}). The key difference is that AQN formation additionally requires the coupling of domain walls with $\eta'$ mesons.

Both DM candidates behave as cold, collisionless DM without requiring fine-tuned annihilation cross-sections. They cluster gravitationally, do not disrupt BBN, leave the CMB moslty unaffected, and reproduce the observed large-scale structure.

Axions couple to photons, nucleons, and electrons in well-defined ways set by QCD. The coupling to photons is given by
\[  
\mathcal{L}_{a\gamma\gamma}
= -\frac{1}{4}\, g_{a\gamma\gamma}\, a\, F_{\mu\nu}\,\tilde F^{\mu\nu} \, ,
\]
where the axion-photon coupling is 
\[  
g_{a\gamma\gamma} \simeq \frac{\alpha}{2\pi f_a} \, .
\]
Since $f_a \sim 10^9\text{--}10^{12}\,\mathrm{GeV} $,
the axion--photon coupling is extremely suppressed, which yields an axion lifetime
\[
\tau_a \gtrsim 10^{25}\,\mathrm{s}
\]
much larger than the age of the Universe. Axions and AQNs share a remarkable stability on cosmological timescales.

\subsection{QCD axion mass range}

The axion mass $m_a \simeq 5.7\,\mu\mathrm{eV}\,(10^{12}\,\mathrm{GeV}/f_a)$ leads to a present day axion relic density parameter $\Omega_a$:

\begin{equation}
\Omega_a h^2 \simeq (0.12)\,\theta_i^2 \,\left(\frac{f_a}{10^{12}\,{\rm GeV}}\right)^{7/6},
\label{eq:axion_relic_density}
\end{equation}
where $\theta_i$ is the initial misalignment angle. Since $\Omega_{\rm DM} h^2 \simeq 0.12$ today \cite{2020A&A...641A...6P}, one finds that if the axion is the sole DM component, it saturates dark matter at $m_a\simeq 5.7\,\mu\mathrm{eV}$,  hence lower masses are excluded for cosmological reasons.

On the other hand, stellar energy-loss arguments, derived from observations of red giants, horizontal-branch stars, white dwarfs, neutron stars and supernova 1987A, impose an upper bound on the axion mass,
$m_a \lesssim 10^{-3}\,\mathrm{eV}$, depending on the assumed axion couplings. These considerations restinct the broad theoretical window to a mass window spanning 
\[
m_a \sim 10^{-6}\text{--}10^{-3}\,\mathrm{eV},
\]
which still leaves a wide region of parameter space in which AQNs and axions can coexist. The idea that both DM candidates contribute concurrently to the dark matter budget will be discussed in Section~ \ref{sec:axion_in_AQN}. 

\subsection{Other types of axions}

The QCD axion is sometimes confused with axion-like particles (ALPs) or ultra-light axions (ULAs). To clarify:

\begin{itemize}
    \item Axion-like particles: ALPs are pseudo-scalar bosons for which the mass
$m_a$ and decay constant $f_a$ are not related by the QCD dynamics. They arise generically from the
breaking of global symmetries but do not necessarily possess a QCD
anomaly and therefore need not couple to the gluon field. ALP cannot solve the strong CP problem. They typically interact very weakly with photons, fermions, or gluons, with couplings suppressed by a large energy scale, and may span a vast
mass range, from sub-$10^{-22}\,\mathrm{eV}$ to the eV scale and beyond.

    \item Ultra-light axions: ULAs are pseudo-scalar fields with extremely small masses, typically $m_a \sim 10^{-22}\,\mathrm{eV}$ . They are a subset of ALPs with a specific importance in cosmology because of its possible connection with dark energy \cite{2017PhRvD..95d3541H}. ULAs formation via the misalignment mechanism can yield the observed relic abundance for suitable decay constants $f_a$. Their de~Broglie wavelength can be of order kiloparsecs, so that DM behaves as a coherent classical field rather than as individual particles. In this regime, quantum pressure arising from the field gradients counteracts gravitational collapse below a characteristic Jeans scale, thereby suppressing small-scale structure formation. 
\end{itemize}

\subsection{Current and upcoming experiments}

A broad and rapidly expanding experimental programme is currently underway to search for the QCD axion. The most mature and sensitive searches are haloscope experiments, which aim to detect axions constituting the local Galactic DM halo through their conversion into microwave photons in a resonant cavity permeated by a strong magnetic field. The Axion Dark Matter eXperiment (ADMX) has already reached sensitivity sufficient to probe QCD axions in the mass range
$m_a \sim 2- 4\,\mu\mathrm{eV}$. Next-generation haloscopes, including HAYSTAC, CAPP, ORGAN, and QUAX, are extending coverage to both higher and lower masses using improved cryogenic amplifiers, quantum-limited readout techniques, and novel resonator designs. Looking further ahead, large-scale projects such as MADMAX, based on dielectric haloscopes, aim to explore the higher-mass regime (
$m_a \sim 40 \text{--} 400\,\mu\mathrm{eV}$).

Complementary approaches target axions produced in astrophysical or laboratory environments. Helioscope experiments search for axions emitted by the Sun via their conversion into X-rays in strong laboratory magnetic fields. The CERN Axion Solar Telescope (CAST) has provided leading constraints on axion–photon couplings, while the next-generation International Axion Observatory (IAXO) is expected to improve sensitivity by more than an order of magnitude, probing deeply into the QCD axion parameter space. Laboratory-based “light-shining-through-a-wall” experiments, such as ALPS II, provide purely terrestrial tests of axion–photon conversion, independent of cosmological assumptions.

Axions can also be probed through their couplings to nuclear spins and electrons. Experiments such as CASPEr exploit nuclear magnetic resonance techniques to search for axion-induced oscillating electric dipole moments \cite{JacksonKimball:2017elr}, while QUAX and related setups target axion–electron couplings using magnetic resonance in solid-state systems.

\begin{figure}[t]
\centering
  \includegraphics[width=12 cm]{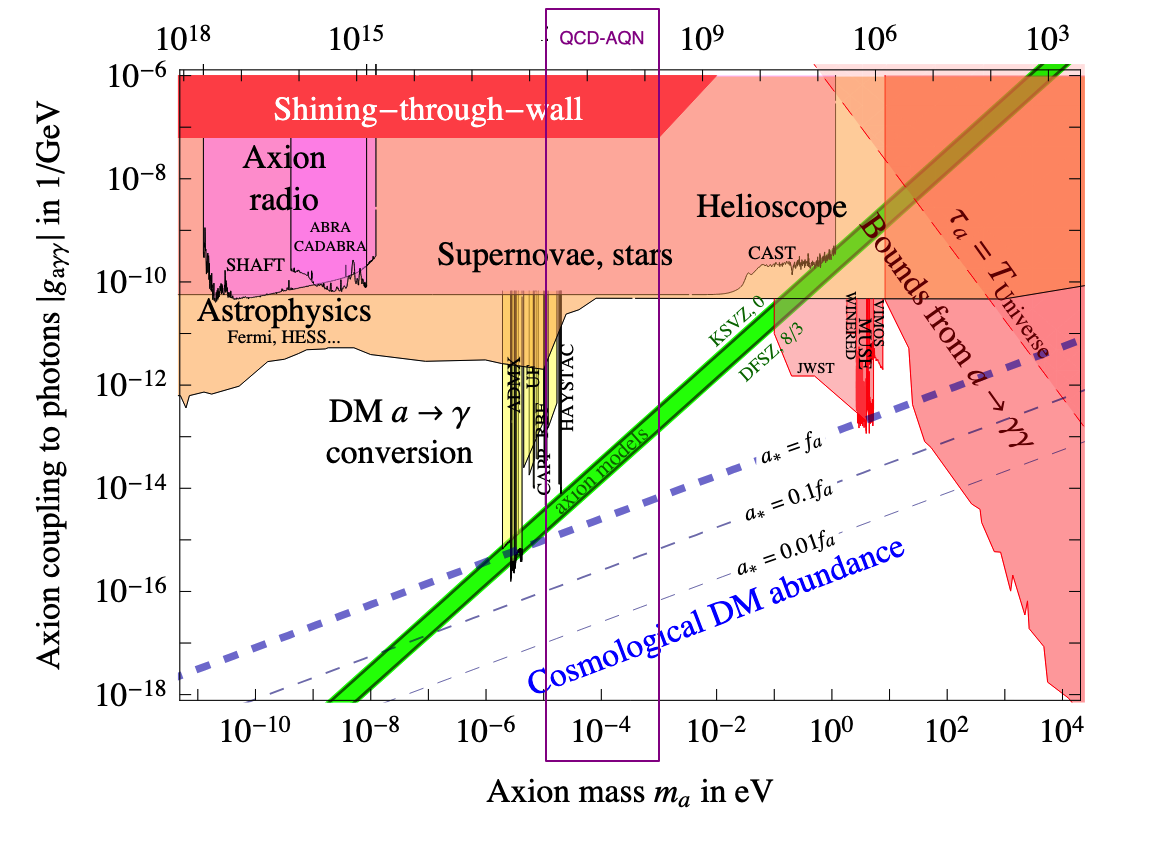}
  \vspace{-0cm}
  \captionof{figure}{Parameter space for the experimental axion search. Shown are: the range favoured by axion models (green band); the initial axion vacuum expectation value $a_*$ that reproduces the observed DM abundance (blue dashed lines); the exclusion bounds from non-observation of axion decays (coloured regions). The search window compatible with the QCD-AQN framework is the purple rectangle. Adapted from {\cite{2024arXiv240601705C}}}
  \label{fig:axion_mass.png}
\end{figure}

So far, most axion detection efforts have focused on the $m_a\sim \, 10^{-6 }\, $ --  $10^{-5} \, \mathrm{eV}$ window (see Fig.~\ref{fig:axion_mass.png}).

\subsection{The importance of inflation in axion cosmology}
\label{sec:axion_and_inflation}

The cosmological implications of the QCD axion depend on whether Peccei–Quinn symmetry breaking occurred before or after cosmic inflation. The epoch of the breaking remains unknown, as it depends on the PQ energy scale, which is not determined by the theory and is not yet constrained by observations. The two scenarios must hence be explored; they correspond to distinct cosmological histories for the axion field and lead to different phenomenological predictions. In particular, we will focus on two aspects: the allowed axion mass range and the upper bound on the inflationary energy scale, $H_I$.

\subsubsection{Axion mass and inflation}

The key parameter controlling the axion abundance is the initial misalignment angle: 

\[
\theta_i = a_i/f_a
\]
If PQ symmetry breaks before inflation, $\theta_i$ is homogeneous across the observable Universe, whereas if PQ symmetry breaks after inflation, it varies randomly between causally disconnected regions.

\begin{itemize}
    \item {\bf If PQ symmetry breaks before inflation}
\end{itemize}

In this case, the subsequent exponential expansion stretches a single causal region to encompass the entire observable Universe. As a result, the axion field takes a single, perfectly homogeneous value across our entire cosmological horizon. The relic axion abundance, $\Omega_a$, then depends on
this unique $\theta_i$ and on $f_a$. 

In this pre-inflation scenario, cosmological observations impose strong constraints. During inflation, quantum fluctuations of any light scalar field are generated with an amplitude proportional to the inflationary Hubble
scale, $H_I$. If the axion exists during inflation, these fluctuations produce spatial variations in the axion field that translate into fluctuations in the DM density. Such perturbations correspond to isocurvature
modes, which are distinct from the adiabatic perturbations observed in the CMB. Observations show that the primordial perturbations are almost entirely adiabatic, placing tight limits on any isocurvature component. 

\begin{figure}[t]
\centering
\includegraphics[width=16cm]{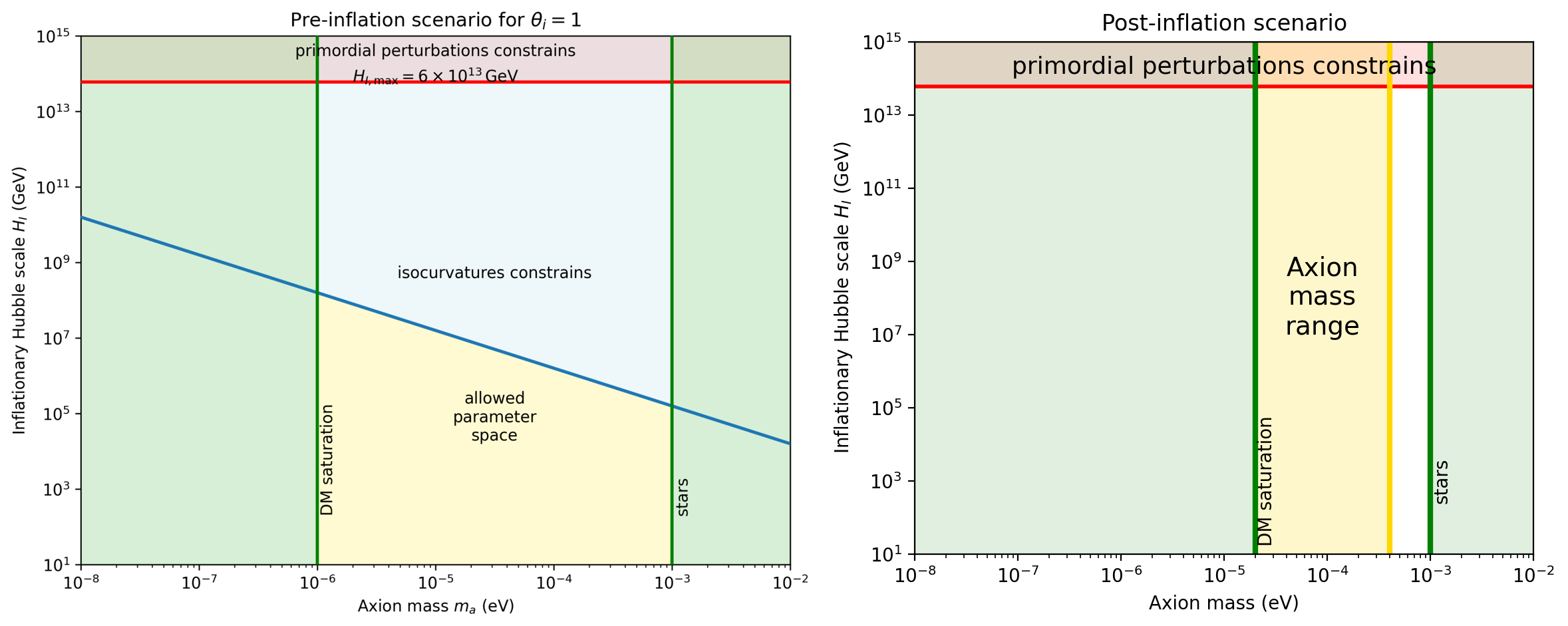}
\caption{Inflationary Hubble scale $H_I$ vs axion mass $m_a$ for the pre- and post-inflation scenarios.  The allowed parameter space is highlighted in yellow.
Left: pre-inflation scenario (for $\theta_i=1$, $\Omega_{\rm DM} = \Omega_a$), showing constraints on $H_I$ from isocurvatures observations, DM saturation and stellar energy loss arguments. 
Right: post-inflation scenario with its narrow axion mass range.}
\label{fig:axion_pre_post_inflation}
\end{figure}

Here the axion mass is only weakly theoretically constrained and may lie within the broad theoretical window set by the decay-constant range
\[
10^{-10}\,\mathrm{eV} \lesssim m_a \lesssim 10^{-2}\,\mathrm{eV}.
\]
with the caveat that the actual viable region depends on $\theta_i$ and $H_I$, as illustrated in Fig.~\ref{fig:axion_pre_post_inflation}.

\newpage

\begin{itemize}
    \item {{\bf If PQ symmetry breaks after inflation}}
\end{itemize}

A different cosmological history arises in the post-inflation scenario. 
In this case, inflation does not homogenise the axion field across the observable Universe. Instead, $\theta_i$
takes different values in different causal regions. The axion field therefore becomes a spatially varying quantity whose initial value is effectively random from one horizon volume to another. When the PQ
symmetry breaks, the network of global axion strings that forms eventually decays, producing additional axions that contributes to $\Omega_a$, potentially overclosing the Universe if the axion is too light.

Early simulations suggested a narrow mass range around $m_a \sim 10^{-5}$--$10^{-4}\,\mathrm{eV}$. However, significant progress in large-scale simulations has shown that the predicted mass range depends sensitively on the axion radiation spectrum emitted by the string network and on the dynamics of the final string collapse. Large lattice simulations performed by \cite{2024JCAP...10..043S} emphasised previously underestimated systematic effects, including sensitivity to initial conditions and discretisation artefacts. Their analysis favours a comparatively high axion mass range
\[
m_a \simeq 95\text{--}450~\mu{\rm eV}.
\]
Recent work has clarified that the dominant uncertainty in the post-inflation prediction is not the zero-temperature equation (Eq.\ref{eq:a_mass}a) but rather the real-time dynamics of cosmic strings during the QCD transition.  Taking current numerical uncertainties into account, the axion mass lies in the approximate range
\[
m_a \sim 25\text{--}400~\mu{\rm eV}.
\]

Several key characteristics of the two scenarios are contrasted in Table~\ref{tab:PQ_inflation}.

\subsubsection{Inflationary Hubble scale upper bound}

The current observational upper bound on  $H_I$ is derived from limits on the tensor-to-scalar ratio of primordial perturbations, $r$, measured in the CMB. Current data constrain this ratio to $r < 0.032$ at $95\%$ confidence \cite{2022PhRvD.105h3524T}, which translates into:
\begin{equation}
H_I\simeq (10^{14}\,{\rm GeV})\,\sqrt{\frac{r}{0.01}}\lesssim 6\times 10^{13}\,{\rm GeV}
\label{eq:limit_on_H1}
\end{equation}

\begin{itemize}
    \item {\bf In the pre-inflation scenario}

$H_I$ must also be sufficiently small to prevent excessive axion isocurvature fluctuations $\delta a \sim \frac{H_I}{2\pi}$. In this case, CMB observations lead to a much tighter upper bound on $H_I$:
\begin{equation}
    H_I\lesssim (1.6\times 10^7\,{\rm GeV})\,\theta_i\left(\frac{10\,{\rm \mu eV}}{m_a}\right)\frac{\Omega_{\rm DM}}{\Omega_a}.
    \label{eq:isocurvature_H1_bound}
\end{equation}
Assuming $\Omega_{\rm DM}=\Omega_a$, this does not lead to an observational conflict with (Eq.~\ref{eq:limit_on_H1}) (see Fig.~\ref{fig:axion_pre_post_inflation}).

\hspace*{1.5em} Several inflation models are considered theoretically unattractive if $H_I$ is forced to be very low, as this typically requires a more finely tuned inflaton potential and can complicate the reheating process after inflation. It also strongly suppresses primordial tensor modes, eliminating the possibility of detecting inflationary gravitational waves (and B-modes in the CMB). For these reasons, some model builders favour an axion mass in the $\mu{\rm eV}$ range. 

\hspace*{1.5em} A growing body of literature discusses mechanisms capable of relaxing the isocurvature constraint, e.g. \cite{2007PhRvD..75j3507B,2025JHEP...12..028G}. The effective axion decay constant during inflation, $f_a^{\rm eff}$, may be substantially larger than the QCD epoch value $f_a$, 
such that at the end of the inflationary era, $f_a$ could reach lower values consistent with a high $m_a$, without generating significant isocurvature fluctuations. The main conclusion is that a relatively large present-day $m_a$ does not preclude the detection of primordial B-modes and remains compatible with a high $H_I$. However, for such a large axion mass,  $\Omega_a$ cannot account for the entirety of $\Omega_{\rm DM}$ (Eq.\ref{eq:axion_relic_density}). Nothing in principle prevents the ratio $\Omega_a/\Omega_{\rm DM}$ from evolving at early times. This is consistent with the AQN formation scenario, as we shall see in (\ref{sec:axion_in_AQN}), where  AQNs can co-exist with QCD axions.

\textbf{    \item In the post-inflation scenario}

The axion field does not exist as a physical degree of freedom during inflation, and therefore inflationary fluctuations do not generate axion isocurvature perturbations. The CMB constraints on isocurvature modes do not apply.
\end{itemize}

Appart from the upper bounds described above, the inflationary Hubble scale $H_I$ remains unconstrained observationally, and current data do not require inflation to occur at a high energy scale.

\begin{table}[t]
\centering
\small
\begin{tabular}{p{4cm} p{5cm} p{5cm}}
\hline
Aspect & PQ breaking \textbf{before inflation} & PQ breaking \textbf{after inflation} \\
\hline
Theoretical $ f_a$ parameter space & Large: $10^{9} \lesssim f_a \lesssim 10^{12}\,\mathrm{GeV}$ (model dependent) & Small: $10^{11} \lesssim f_a \lesssim 10^{12}\,\mathrm{GeV}$ \\
$\theta_i$ & $-\pi \le \theta_i \le \pi$ & Random variable with $\langle \theta_i^2 \rangle = \frac{\pi^2}{3}$. \\
Axion mass for DM saturation & $m_a \sim 10^{-6}\,\mathrm{eV}$  & 
$m_a = 25.2 \pm 11.0\,{\rm eV}$;  $m_a \simeq 95\text{--}450~\mu{\rm eV}$ \cite{2024JCAP...10..043S}\\
\hline
\end{tabular}
\caption{Comparison of  pre- and post-inflation scenarios aspects \cite{2020PhRvL.124p1103B}. }
\label{tab:PQ_inflation}
\end{table}

\bigskip

\subsection{Dark matter as a two-component system}
\label{sec:axion_in_AQN}

Some critics of the QCD axion argue that the model has become overly focused on the narrow mass range around $m_a \sim 10^{-6}\,\mathrm{eV}$. This region has been the target of several experiments, yet no confirmed detection has emerged. The same authors claim that the parameter space has been repeatedly probed without success and that the continued emphasis on this range reflects theoretical bias. Exploring much broader mass ranges often implies that axions would constitute only a fraction of the dark matter rather than its entirety, which departs from the conventional approach of constructing DM models that account for the full DM abundance. 

The QCD-AQN framework provides a strong motivation for exploring the region $m_a \gtrsim \mu{\rm eV}$, regardless of the time when PQ breaking occurs. The formation of AQNs is accompanied by the generation of a residual population of free QCD axions:
\[
\Omega_{\rm DM}
=\Omega_{\rm a} +
\Omega_{\rm AQN}
+\Omega_{\bar{\rm AQN}}
\]
where both AQN components are generated at the QCD epoch, while the axion mass turns on at $\sim$ 1 GeV and spans a wide interval:
\[
10^{-5} \lesssim m_a \lesssim 10^{-3}\,{\rm eV},
\]
compatible with cosmology and AQN formation. 

\cite{2018PhRvD..97d3008G} perform quantitative calculations of the relative abundances of axions, AQNs, and left over baryons. They show that AQNs would constitute the dominant component of dark matter\footnote{Unless $\Omega_{\rm DM} \simeq \Omega_a$}: 
\[
\Omega_{\rm AQN}+ \Omega_{\bar{\rm AQN}} \gg \Omega_a.
\]
Interestingly, they find that isocurvature perturbations are significantly suppressed even for large values of $m_a$, which enlarges the allowed parameter space for the inflationary Hubble scale and strengthens the motivation for experimental axion searches across the QCD-AQN window shown in Fig.~\ref{fig:axion_mass.png}.
 
\medskip

In Sections \ref{sec: formation} and \ref{section:axion} we examined the ability of QCD to account for dark matter within a theoretically well-motivated framework that does not require extensions of the Standard Model beyond the QCD axion. The purpose of our final section is to illustrate that QCD may offer a conceptually simple, theoretically grounded, and explanatory framework for dark energy.  

\section{QCD and dark energy}
\label{sec:QCDDE}

Could the fundamental nature of dark energy (DE) be rooted in non-perturbative QCD effects?
\cite{2009PhRvD..80f3001U,2010PhLB..688....9U,2010NuPhB.835..135U} propose that the origin of DE may reside in the topological structure of the QCD vacuum. This idea is further developped in \cite{2013AnPhy.336..462Z}, where it is argued that the vacuum energy responsible for cosmic acceleration originates from tunnelling between distinct topological sectors of QCD in an expanding Universe\footnote{See \cite{2015PhRvD..92d3512Z,2025arXiv250614182V} for references and a brief overview of this idea.}.

This idea has recently been implemented within a Big Bang cosmological framework \cite{2025arXiv250614182V}. In this scenario, the DE contribution to the Friedmann equation is not strictly constant, but instead scales with the Hubble parameter, thereby offering an explanation for why DE becomes cosmologically relevant only at late times. The mechanism relies exclusively on non-perturbative QCD dynamics in expanding spacetime and does not introduce any additional fundamental fields. It therefore constitutes a topological, rather than dynamical, explanation of the origin of dark-energy.

Although this line of investigation was motivated in part by recent DESI results{\footnote{Recent analyses from the Dark Energy Spectroscopic Instrument (DESI) suggest that DE may not be strictly constant in time, in contrast with the standard assumption of a cosmological constant $\Lambda$. When DESI measurements of galaxy clustering are combined with observations of the CMB, weak gravitational lensing, and Type~Ia supernovae, the resulting datasets show deviations from the minimal $\Lambda$CDM model at the level of approximately $2.8$--$4.2\,\sigma$. While this significance is insufficient to claim a discovery by the conventional $5\,\sigma$ standard, it is strong enough to motivate theoretical and observational scrutiny. DE may have been more influential in the past and has since weakened, or its equation-of-state parameter $w$ may have evolved and possibly crossed the so-called phantom divide at $w=-1$. }, it is not intended as a direct explanation of those observations, which may still be affected by internal tensions arising from residual systematics and the combination of heterogeneous data sets.

\subsection{Overview}
Within this scenario, dark energy is identified with the time-dependent QCD vacuum energy, rather than with an arbitrary constant added by hand to the action ($S$). 

\medskip
\begin{itemize}
    \item {\bf Time-dependent vacuum energy}
\end{itemize}

Following the conceptual suggestion by Zel’Dovich~\cite{1967JETPL...6..316Z} and the formalisation by \cite{2010PhLB..688....9U}, the calculation of the DE density proceeds by adopting the Minkowski spacetime as the static flat-space reference background. Any vacuum energy that exists with the same value in flat space is regarded as gravitationally irrelevant; only variations induced by expansion and/or curvature relative to the Minkowski space contribute to the effective DE density. The meaningful quantity is therefore not the absolute vacuum energy -- which is divergent in quantum field theory -- but the difference between its value in an expanding Universe and in flat spacetime\footnote{The Minkowski spacetime corresponds to the limiting case $H \to 0$, with no expansion and no gravitational dynamics.}. 

Hence:
\begin{equation}
\rho_{\rm DE} = E^{\rm FLRW}_{\rm vac} - E^{\rm Minkowski}_{\rm vac},
\label{eq:rhode_as_difference}
\end{equation}
where the FLRW\footnote{Friedmann–Lemaître–Robertson–Walker} background metric depends explicitly on the Hubble parameter \(H\). Any residual contribution that survives when \(H \neq 0\) -- \emph{i.e.} terms proportional to \(H\), \(H^2\), or other invariants -- defines the gravitationally active component.

\medskip
\begin{itemize}
    \item {\bf QCD vacuum energy}
\end{itemize}

\cite{2015PhRvD..92d3512Z} argue that the QCD topological sector in an expanding FLRW Universe produces such a vacuum energy difference.  
Here $E^{\rm Minkowski}_{\rm vac}$ is noticably negative \cite{1974PhRvD...9.3471C,1974PhRvD..10.2599C}:
\[
E^{\rm Minkowski}_{\rm vac}=-\Lambda_{\rm QCD}^4.
\]  
The calculation of \(E^{\rm FLRW}_{\rm vac}\) in a general FLRW Universe cannot be performed in full generality. However, it was evaluated in \cite{2015PhRvD..92d3512Z} for the specific case of a de~Sitter Universe driven by a constant  value \(\bar H\) of $H(t)$:
\begin{equation}
    E^{\rm FLRW}_{\rm vac}=-\Lambda_{\rm QCD}^4\left(1-c_{\bar H}\frac{\bar H}{\Lambda_{\rm QCD}}\right),
    \label{eq:QCDvac}
\end{equation}
where $c_{\bar H}$ is a dimensionless coefficient of order $\sim m_q/\Lambda_{\rm QCD}\sim 0.01-0.02$\footnote{$m_q\sim m_u\sim m_d$ is the quark mass. $m_q/\Lambda_{\rm QCD}$ is a QCD related factor, which consistently accompanies tunnelling transitions in QCD.}.  Inserting in Eq.(\ref{eq:rhode_as_difference}) yields a positive vacuum energy:
\begin{equation}
\rho_{\rm DE}=c_{\bar H} \bar H \Lambda_{\rm QCD}^3.
\label{eq:rho_DE}
\end{equation}
In the pure de Sitter limit, the Friedmann equation becomes:
\begin{equation}
\bar H^2=\frac{8\pi G}{3}\rho_{\rm DE}.
\end{equation}
Using Eq.(\ref{eq:rho_DE}) and introducing natural units $G=M_{\rm PL}^{-2}$ we obtain:
\[
\bar{H} = c_{\bar H} \, \frac{8\pi \Lambda_{\mathrm{QCD}}^{3}}{3 M_{\mathrm{PL}}^{2}},
\qquad
\rho_{\mathrm{DE}} = c_{\bar H}^{2} \, \frac{8\pi \Lambda_{\mathrm{QCD}}^{6}}{3 M_{\mathrm{PL}}^{2}}.
\]
Using $\Lambda_{\rm QCD}\sim 100\,{\rm MeV}$ and $c_{\bar H}=0.01-0.02$, we calculate:
\begin{equation}
\left\{
\begin{aligned}
    \bar H &\simeq 1.7-3.4\times 10^{-33}\,{\rm eV}
    \simeq 80-160\,{\rm km\,s^{-1}\,Mpc^{-1}} \\
    \rho_{\rm DE} &\simeq 4-16\times 10^{-11}\,{\rm eV^4}
    \simeq 9-36\times 10^{-30}\,{\rm g\,cm^{-3}} \\
    t_0 &\simeq \frac{1}{\bar{H}}
    \simeq 12-6\,{\rm Gyr}
\end{aligned}
\right.
\label{eq:scales}
\end{equation}

to be compared to the observed values:
\[
\left\{
\begin{aligned}
H_0 &\sim 70\,\mathrm{km\,s^{-1}\,Mpc^{-1}} \\
\rho_{\rm DE} &\sim 10^{-11}\,\mathrm{eV}^4 \\
t_0 &\sim 14\,\mathrm{Gyr}
\end{aligned}
\right.
\]
The calculated values have the same order of magnitude as those observed -- a non-trivial result as all cosmological scales in this framework are expressed in terms of $\Lambda_{\rm QCD}$. This proximity suggests that the late-time de Sitter–like expansion of the Universe may originate from a small time dependence of QCD tunnelling processes in an expanding Universe.

\subsection{A large scale effect}

\subsubsection{Global gauge topology and macroscopic quantum effects}
At first sight, Eq.(\ref{eq:rho_DE}) appears puzzling: how can a local quantum field theory such as QCD generate a contribution to the vacuum energy that manifests itself on cosmological, i.e. extremely large, distance scales? 

In quantum field theory, global properties of gauge fields can produce observable effects that depend on the topology of the gauge configuration rather than on the local field strengths. In quantum electrodynamics, for instance, the Aharonov–Bohm and Aharonov–Casher effects illustrate this phenomenon: a particle may acquire a measurable phase shift\footnote{This is a pure quantum effect which has been measured and does not exist in the classical limit of electrodynamic.} even when travelling in a region where the electromagnetic fields vanish. The resulting phase depends only on the enclosed flux and therefore persists over arbitrarily large separations.  Both effects have been experimentally confirmed and are well-established quantum phenomena. 

In QCD, the classical vacuum is degenerate: it is composed of an infinite set of topologically distinct sectors, each labelled by an integer winding number. Tunnelling transitions between these sectors can happen, the gauge configurations describing the tunnelling events are called instantons. By analogy with the Aharonov–Bohm and Aharonov–Casher effects, such tunnelling events can give rise to macroscopic consequences, even though the underlying theory is local, thereby generating large-scale effects in the Universe.

These considerations remove the apparent contradiction: a local quantum field theory such as QCD can generate contributions to the vacuum energy on cosmological scales, because its vacuum structure is topologically non-trivial and physical observables may depend on global, rather than purely local, properties of the gauge configuration\footnote{Computations for exactly solvable models support this claim (see [114] and references therein).}.

\subsubsection{Time-dependent expansion rate}

 The vacuum energy that enters the Friedmann equation, expressed in Eq.(\ref{eq:rhode_as_difference}), is not associated with a local stress–energy density that can be evaluated point by point. Rather, it represents a global property of the spacetime background. More generally, such differences in vacuum energy cannot be computed in a purely local manner for an arbitrary, time-dependent geometry. \textit{This strongly suggests that vacuum energy—and by extension dark energy—may be intrinsically non-local}, in close analogy with the quantum phase effects discussed earlier.

 \cite{2015PhRvD..92d3512Z} conjectured that, in the de~Sitter limit, the QCD vacuum energy
$E^{\rm Minkowski}_{\rm vac} = -\Lambda_{\rm QCD}^4$ acquires a correction term linear in the Hubble parameter \(H\) \cite{1974PhRvD...9.3471C,1974PhRvD..10.2599C}. 
One could expect such corrections to scale quadratically,
\(\propto H^2 \Lambda_{\rm QCD}^2\,\sim 10^{-52}\,{\rm eV}^4\), as required by the general covariance for local propagating fields \cite{2000PhLB..475..236S,2002JHEP...02..006S}. 
However, this reasoning does not apply to instantons, which are not propagating degrees of freedom but rather topological configurations of the QCD vacuum. Since here the effect originates from non-local topological structure rather than from local field fluctuations, a linear term \(\propto H \Lambda_{\rm QCD}^3 \sim \,10^{-6} \rm {eV}^4\) is not excluded. 
Moreover, because:
\[
H \Lambda_{\rm QCD}^3 \gg H^2 \Lambda_{\rm QCD}^2,
\]
such a linear correction can generate a much larger large-scale contribution, thereby producing macroscopic effects despite the microscopic origin of vacuum energy.
\medskip

{\bf Key takeaway.} Dark energy may arise from the global topological structure of the QCD vacuum, allowing a local theory to generate a macroscopic, non-local cosmological effect.

\subsection{Present-day interpolation}

Tunneling events between topological sectors can arise only in the adiabatic regime, when the Hubble parameter changes slowly relative to the expansion rate. This condition is automatically satisfied in all computations, where parameters such as $\kappa$ in \cite{2015PhRvD..92d3512Z} or the Hubble parameter are assumed to be constant. This requirement can be expressed as  
\begin{equation}
    |\dot{H}| \ll H^2,
    \label{eq:adiabatic_condition}
\end{equation}
which ensures that the background geometry changes sufficiently slowly for the topological configurations to be well defined. This condition is naturally satisfied in the de~Sitter limit, where \(H = \bar{H}\) is constant.

By contrast, in the very early Universe, as the scale factor \(a \to 0\), the adiabatic condition is not fulfilled. This does not pose a problem, however, since DE is dynamically irrelevant in a Universe dominated by matter and radiation. A more subtle issue concerns how the Universe will transition from the matter dominated era to the DE dominated era. To address this \cite{2025arXiv250614182V}  generalised the strictly de~Sitter expression \(\rho_{\rm DE}=c_{\bar H}\bar H \Lambda_{\rm QCD}^3\) to the time-dependent form \(\rho_{\rm DE}=c_{H} H \Lambda_{\rm QCD}^3\), with the aim of determining when the adiabatic condition (\ref{eq:adiabatic_condition}) could be marginally satisfied.

With this time-dependent DE term, the Friedmann equation for a spatially flat Universe can be rewritten as a quadratic equation for the Hubble parameter $H$:
\[
H^2 - \bar H\,H
- H_i^2 \left[
\Omega_{m,i}\left(\frac{a_i}{a}\right)^3
+ \Omega_{r,i}\left(\frac{a_i}{a}\right)^4
\right]
= 0 \, ,
\]
 where the index $i$ denotes an arbitrary reference epoch in the past at which DE was negligible. The solution is given by:
\begin{equation}
H(a)
= \frac{\bar H}{2}
\left[
1 + \sqrt{
1 + B\left(\frac{a_i}{a}\right)^3
+ C\left(\frac{a_i}{a}\right)^4
}
\right] \, ,
\label{eq:H_QCD}
\end{equation}
where the dimensionless coefficients are defined as
\[
B \equiv 4\left(\frac{H_i}{\bar H}\right)^2 \Omega_{m,i} \, ,
\qquad
C \equiv 4\left(\frac{H_i}{\bar H}\right)^2 \Omega_{r,i} \, .
\]
From the above equations, following the derivations in \cite{2025arXiv250614182V}, one can derive the expressions for $\dot H$ and $H$, and the adiabatic condition (Eq.\ref{eq:adiabatic_condition}) becomes:
\[
    \left(1+w\frac{\rho_{\rm DE}}{\rho_{\rm DE}+\rho_{\rm M}}\right)\ll \frac{3}{2}
\]
where $\rho_{\rm M}$ is the matter mass density. 
At \(z=0\),
\[
    \left(1 + w\,\frac{\rho_{\rm DE}}{\rho_{\rm DE} + \rho_{\rm M}}\right) \simeq 0.5.
\]
This indicates that the adiabatic condition is only partially satisfied at the present epoch. Consequently, the use of the time-dependent expression \(\rho_{\rm DE} = c_{\bar H} H \Lambda_{\rm QCD}^3\) cannot be regarded as fully justified today.

In order to overcome this limitation, \cite{2025arXiv250614182V} introduced a {\it switch} function $\beta(a)$, which role is to describe the activation of the instantons large scale effects, i.e. activate the QCD vacuum energy contribution at a given epoch with a controlled rate:
\[ 
    \rho_{\rm DE}=\beta(a) c_{\bar H} H(a) \Lambda_{\rm QCD}^3,
\]
where $0\le \beta(a)\le1$ and  $\beta(a)$ should vanish at large redshift:
\[
\left\{
\begin{aligned}
\beta(a \rightarrow \infty) &= 1 \\
\beta(z \gtrsim 1\text{--}2) &\sim 0.
\end{aligned}
\right.
\]
The switch function \(\beta(a)\) could, in principle, be derived from first principles. In practice, however, such a calculation is not feasible, as it would require solving non-perturbative QCD in a non-trivial background corresponding to an expanding Universe characterised by the Hubble parameter $H$. It is therefore not introduced as an ad hoc modification of the theory, but rather as a physically motivated quantity that will be treated as a parametrised function.

With this prescription,  the Friedmann equation becomes 
\begin{equation}
\boxed {
\frac{d a}{d\tau}
= \beta(\tau)\,\frac{a}{2}
\left[
1 + \sqrt{
1
+ \frac{B}{\beta^2(\tau)}\left(\frac{a_i}{a}\right)^3
+ \frac{C}{\beta^2(\tau)}\left(\frac{a_i}{a}\right)^4 
}\right] }\, ,
\label{eq:enhanced_friedmann}
\end{equation}
where we have introduced the dimensionless time variable

\[
\tau \equiv \frac{8\pi G}{3}\,\Lambda_{\rm QCD}^3\,c_H\, t \, .
\]
and where the switch function $\beta$ must be determined empirically from observations.

\subsection{Effective DE equation of state}

Since this dark-energy model does not rely on the dynamics of a new fluid component, one may ask what effective equation-of-state parameter $w$ would be inferred if DE were described in the conventional form:
\begin{equation}
{\rm P_{DE}}=w\rho_{\rm DE}
\end{equation}

For a given choice of switch function $\beta$, the Hubble parameter $H=\dot a/a$ can be obtained from Eq.~\ref{eq:enhanced_friedmann}.
By combining the Friedmann equation with the acceleration equation, one can then derive the corresponding effective $w$:

\begin{equation}
1+w=\frac{-2\dot{H}-3H_i^2\Omega_{m,i}^2\left(\frac{a_i}{a}\right)^3-3H_i^2\Omega_{r,i}^2\left(\frac{a_i}{a}\right)^4}{3c_H\bar{H}H}.
\label{eq:w_QCD}
\end{equation}
It was shown in \cite{2025arXiv250614182V} that Eq.(\ref{eq:w_QCD}) does not impose any constraints on $w(z)$. 

Unlike the Chevallier–Polarski–Linder parametrisation \cite{2001IJMPD..10..213C}, this QCD switch–function approach is physically motivated and does not require the introduction of an additional dark sector. 

\medskip
{\bf Predictions}

This proposal makes several distinctive predictions:

    (i) The equation of state of DE, $w_{\rm DE}$, is generally greater than -1 today, approaching the de Sitter limit -1 asymptotically in the future.
    
    (ii) $w_{\rm DE}(z) $ can cross the phantom boundary (w = -1) at redshifts z > 0, potentially multiple times, qualitatively in line with recent measurements from the DESI survey ( DESI Collaboration et al., 2025).
    
    (iii) Departures from the standard $\Lambda$CDM expansion history are expected.

\smallskip
Preliminary cosmological constraints -- obtained from a combination of observational probes including baryon acoustic oscillations, type Ia supernovae, CMB anisotropies, and large-scale-structure data --are encouraging \cite{LeeInPrep2026}.

\medskip
While the present section on dark energy lies outside the main focus of this lecture, namely QCD-driven dark matter, it reinforces the idea that the strongly coupled non-perturbative QCD regime may provide a unified framework for cosmology. 

\section{Conclusion}

The QCD-AQN  framework offers a phenomenologically rich approach to dark matter rooted in QCD dynamics and the Peccei–Quinn mechanism. It links the origin of dark matter to the matter–antimatter asymmetry without introducing an entirely new particle sector beyond the axion, a particle currently being sought by state-of-the-art experiments.  It proposes a shift from dark matter as weakly interacting particles to macroscopic, strongly bound aggregates whose interactions with baryonic matter generate observable electromagnetic signatures.

The explanatory scope of QCD-AQN is particularly compelling. It naturally accounts for the observed relation $\Omega_{\rm DM} \simeq 5\,\Omega_b$ by linking the dark matter abundance directly to baryonic physics at the QCD epoch: dark matter and visible baryons originate from the same QCD-driven dynamics, which also sets the baryon-to-photon ratio $\eta \sim 10^{-10}$. The apparent matter--antimatter asymmetry is reinterpreted as a baryon number separation. Furthermore, QCD-AQN may offer new avenues for alleviating the $^{7}\mathrm{Li}$ discrepancy. Several puzzles of early-Universe cosmology are seen as interconnected consequences of QCD-scale physics.

A distinctive feature of the QCD-AQN framework is its multi-channel predictivity: the same macroscopic parameter governing the AQN mass distribution simultaneously affects diffuse Galactic emission and spectral distortions of the CMB, as well as characteristics signature in dense objects, including the Sun and the Earth. Rather than being tuned independently for each phenomenon, these observables emerge from a unified physical mechanism. 

The framework is governed by a single free parameter~-- the mean mass of the AQN mass distribution~-- which is currently constrained by independent astrophysical considerations to lie in the range $\sim 1\text{--}100\,\mathrm{g}$. These bounds emerge consistently across multiple, otherwise unrelated observations, with the independently inferred mass windows showing a remarkable degree of agreement with theoretical constrains.

If the upcoming observational tests (511 keV bulge emission, 21 cm emission, cosmic radio background) are successfully passed, or if the predicted dark glow is detected, confidence in the QCD-AQN framework will grow. It would open a new and rich line of investigation, connecting a wide range of well-established areas of physics. Rather than requiring exotic extensions of the Standard Model, QCD-AQN suggests that new phenomena may emerge from known physics operating in unusual regimes. This perspective invites a collective effort to search for observational signatures across many environments. In this sense, the exploration of the QCD-AQN framework may develop into a playful yet rigorous field of research, where discarded sky backgrounds, diffuse emissions, and unexplained excesses become valuable laboratories for testing fundamental physics. Whether AQNs, together with a population of QCD axions, ultimately make up all of the dark matter remains uncertain; nevertheless, the framework provides a compelling illustration of how QCD physics can leave observable imprints in cosmology, possibly extending beyond dark matter to the domain of dark energy.

\section*{Acknowledgements}
These lecture notes were prepared for Les Houches 2025 Summer School. I am thankful to the organizers for the invitation and for creating such a stimulating scientific environment. I am particularly grateful to Benoîte Pfeiffer for her significant contribution to the writing of this lecture. Her careful reading, insightful comments, and constructive suggestions substantially improved both the clarity and the structure of the material. Her engagement with the subject and thoughtful discussions were invaluable in refining several key arguments and strengthening the overall presentation. I would like to thank Eric Zhitnitsky and Xunyu Liang for our lively and constructive discussions about the AQNs framework since 2017. Students in our AQN group at UBC have contributed significantly to several results presented in this lecture, in particular Shuailiang Ge, Fereshteh Majidi, Egor Peshkov, Nayyer Raza, Ben Scully, Michael Sekatchev, Md Shahriar Rahim Siddiqui, Saloni Soni; it is a pleasure to thank them for their commitment and hard work. Finally, I would like to thank Karim Benabed, Robert Branderberger, Jean-Charles Cuillandre, Bryce Cyr, Cail Daley, Klaus Dolag, Hendrik Hildebrant, Ryley Hill, Gary Hinshaw, Adrian Liu, Jayant Murthy, Douglas Scott, Joe Silk, and Julian Sommer for insightful discussions on specific aspects of this work, which proved particularly helpful when exploring new directions and new tests of the AQN framework. This work was supported by the Natural Sciences and Engineering Research Council of Canada.

\bibliography{master_bibliography.bib}


\end{document}